\documentclass[12pt,a4paper]{article}
\usepackage[T1]{fontenc}
\usepackage{lmodern,microtype}
\usepackage[margin=27mm]{geometry}
\usepackage{amsmath,amssymb,amsthm,mathtools}
\usepackage{enumitem,booktabs,array,xcolor,needspace,placeins}
\usepackage{hyperref}

\usepackage[round,authoryear,sort]{natbib}
\usepackage[labelsep=period,font=small]{caption}
\hypersetup{pdftitle={Extremal Perfect Monotone Equilibria in Bayesian Games},
pdfsubject={Perfect equilibria and strategic complementarities in Bayesian games}}
\setlist{topsep=4pt,itemsep=3pt}
\theoremstyle{plain}
\newtheorem{theorem}{Theorem}
\newtheorem{proposition}{Proposition}
\newtheorem{lemma}{Lemma}
\newtheorem{corollary}{Corollary}
\newtheorem{claim}{Claim}

\theoremstyle{definition}
\newtheorem{definition}{Definition}
\newtheorem{assumption}{Assumption}
\newtheorem{example}{Example}

\theoremstyle{remark}
\newtheorem{remark}{Remark}

\newcommand{\BR}{\operatorname{BR}}
\newcommand{\R}{\mathbb R}

\newcommand{\st}{\preceq_{\mathrm{st}}}
\newcommand{\argmax}{\operatorname*{arg\,max}}
\newcommand{\supp}{\operatorname{supp}}
\newcommand{\PP}{\mathcal P}
\newcommand{\EE}{\mathcal E}
\newcommand{\dd}{\,d}

\title{Extremal Perfect Monotone Equilibria in \\ Bayesian Games with Strategic Complementarities}
\author{Wei He\thanks{\raggedright Department of Economics, The Chinese University of Hong Kong, Hong Kong (\texttt{hewei@cuhk.edu.hk}).}
\and Yeneng Sun\thanks{\raggedright Departments of Economics and Mathematics, National University of Singapore, Singapore (\texttt{ynsun@nus.edu.sg}).}
\and Hanping Xu\thanks{\raggedright Institute for Advanced Study in Mathematics, Harbin Institute of Technology, Harbin (\texttt{xuhanping@u.nus.edu}).}}
\date{\today}
\begin{document}
\pagenumbering{arabic}
\maketitle
\begin{abstract}

We study sharp bounds on equilibrium behavior in Bayesian games when players may make mistakes. Under strategic complementarities, the set of behavioral perfect equilibria has a largest and a smallest element, both in pure strategies. We ask when these bounds are monotone and how they depend on the way mistakes are made. For monotone supermodular games, we consider limits of equilibria under increasing perturbations, in the sense that conditional tremble distributions have full support and increase stochastically with type. We show that the largest and smallest perfect equilibria exist, which are pure and monotone. Under the general perfection notion, the extremal perfect equilibria continue to be monotone in the setting with finite actions. With infinite actions, the extremal perfect equilibria can be nonmonotone, and the set of perfect monotone equilibria can even lack extrema. 

\end{abstract}

\newpage
\tableofcontents
\newpage

\section{Introduction}\label{sec:introduction}

Bayesian games with strategic complementarities are widely used to study various economic applications, including investment, technology adoption, and coordination under private information. Under standard supermodularity and regularity conditions, these games have a largest and a smallest Nash equilibrium; see \cite{milgromRoberts1990} and \cite{vives1990}. Those extremal equilibria provide bounds on players' choices and permit comparative statics without selecting among multiple equilibria. With suitable conditions on payoffs and beliefs, these strategies are also increasing in private types; see \cite{vzv}.

The predictions are less convincing when the extremal equilibria involve weakly dominated actions. A player may choose such an action because she is indifferent when opponents follow their equilibrium strategies, yet strictly prefer another action once they make mistakes with small probability. Indeed, an extremal Bayesian Nash equilibrium can fail both the basic criteria of admissibility and trembling-hand perfection.\footnote{See Section~\ref{sec:general-definition} and Example~\ref{ex:regional-chain}.} To address this issue, we aim to study whether the set of perfect equilibria has a largest and a smallest element, whether these equilibria are monotone in private information, and how they change with payoffs and beliefs.

Towards this end, we study  monotone supermodular Bayesian games. Specifically, we impose two assumptions: first, players' actions exhibit complementarity; second, higher types have a greater direct gain from higher own actions and expect stochastically higher others' types. Under these assumptions, when opponents use increasing strategies, both the direct payoff effect and the information conveyed by type favor a higher own action. We allow action spaces to be compact metric lattices, including discrete, continuous, and multidimensional choices. We also permit subjective interim beliefs and both discrete and continuous type distributions.\footnote{For extremal Bayesian Nash equilibria without an order on types, see \citet{vanZandt2010}.}

To start with, we first model how a player may make mistakes. A natural benchmark is a type-independent tremble: the player draws from the same action distribution regardless of her information. However, a player who makes mistakes need not ignore her private information. A firm may misjudge an investment project while remaining more likely to invest when its private return is high. For example, suppose it occasionally compares its return plus an assessment error with a fixed cost, rather than evaluating the full strategic problem. Any realization of the error that leads to investment at a lower return also leads to investment at a higher return. When the error distribution is independent of type, the probability of investment increases with the return. We allow such mistakes, as well as type-independent trembles, through \emph{increasing perturbations}; that is, conditional on a tremble, higher types choose stochastically higher actions.

Under increasing perturbations, an increasing intended strategy remains increasing in distribution after mistakes are included. Other players can therefore continue to infer higher actions from higher types, provided the intended strategies are increasing. Under strategic complementarities, they have an incentive to respond with higher actions themselves. In each perturbed game, players choose their intended behavior optimally given opponents' realized behavior. An equilibrium is \emph{perfect under increasing perturbations} if it can be obtained as a limit of these equilibria as tremble probabilities vanish.\footnote{The conditional tremble distributions have full support and may vary along the supporting sequence. Neither the equilibria in that sequence nor the limiting equilibrium is required to be monotone.}

Theorem~\ref{thm:main} establishes a largest and a smallest equilibrium under this refinement, both pure and monotone. They bound every behavioral equilibrium supported by increasing perturbations, including equilibria that are not monotone. Since the two bounds are themselves perfect under the refinement, they are also the largest and smallest perfect monotone equilibria. 
One can therefore obtain sharp bounds without selecting an equilibrium or specifying a particular distribution of mistakes.

Methodologically, the difficulty comes from the comparison across different perturbations. Solving a single perturbed game gives bounds conditional on its error distributions. 
However, different mistakes can support different limiting behavior, so finding extremal equilibria for each perturbed game does not suffice. We construct common upper and lower perturbations for any two given perturbations. These induce stochastically ordered realized behavior at every intended action and type, without requiring a larger maximum tremble probability. Strategic complementarities then order the associated extremal equilibria. We use these comparisons to obtain bounds across perturbations and show that the bounds themselves can be approached by perturbed-game equilibria as mistakes vanish.\footnote{Analogous existence, monotonicity, and comparative-statics results hold when conditional tremble distributions are required to be type independent but may vary along the sequence. Prescribing one profile of tremble distributions for the entire sequence is a further restriction; Section~\ref{oa:fixed} shows that it can eliminate largest and smallest elements.}

We next ask which conclusions remain valid under a broader perfection criterion. Instead of focusing on increasing perturbations, we require the candidate behavior to be approximated by best responses to completely mixed profiles converging to it. The supporting profiles need not be induced by perturbed-game equilibria, nor need they be increasing in type. By only requiring players' actions to be strategic complementarities, we prove in Theorem~\ref{thm:general-global} that pure largest and smallest perfect equilibria still exist. However, these extremal equilibria may fail to be monotone.

To study the monotonicity property, we further  require that higher types have a greater direct gain from higher own actions and expect stochastically higher opponents' types. Even so, with the general perturbations, higher types need not expect higher actions from opponents whose supporting behavior is nonmonotone. With finite actions, we can construct increasing supporting profiles for the global extrema and establish the monotonicity, regardless of whether the type spaces are finite (Theorem~\ref{prop:general-finite}). With infinite actions, however, nearby best responses may converge to an action that is never itself optimal against the supporting profiles. We provide examples to demonstrate the issues that the global extrema are nonmonotone, and the nonempty set of perfect monotone equilibria has neither a largest nor a smallest element. These failures occur within the framework of monotone supermodular games.

We further investigate how the extremal equilibria change in response to beliefs and payoffs. Under increasing perturbations, both extrema increase with an upward shift in interim beliefs. Under broader perfection, we establish this comparison for finite actions, but it can fail with infinite actions. Payoff comparisons are less restrictive: when a higher parameter raises the gain from a higher action against any fixed opponents' behavior, both extrema increase under either refinement.


The two refinements can still agree on the bounds even though they allow different supporting behavior. Every equilibrium under increasing perturbations is perfect under the broader criterion, so the former bounds lie between the latter. When actions and private types are finite, the pairs coincide. With finitely many types, the best-response conditions supporting each extremum hold simultaneously at almost every type once mistakes are sufficiently small. The same extremum can then be supported by exact equilibria of increasing perturbed games. With continuously distributed types, each fixed type may eventually prefer its prescribed action while some other types still prefer to deviate at every positive tremble probability. Indeed, they can differ even when each player has only two actions, types are independent, and there is a common prior.


More generally, we establish the pure monotone bounds under increasing perturbations under complementarity conditions on interim expected payoffs. A separate stochastic-order restriction on beliefs is unnecessary for this result. This matters when a higher type has opposing effects on the return to an action: a firm with a better project may also expect higher input costs, yet still have a larger net gain from investment. Rather than require the direct payoff effect and the change in beliefs to be favorable separately, we impose the required conditions on their combined effect after taking expectations. The resulting bounds continue to cover all behavioral equilibria supported by increasing perturbations. 


\medskip

\textbf{Literature review.} 
The existence of monotone Bayesian equilibria has been established by \citet{athey}, \citet{mcadams}, and \citet{reny} under conditions on a player's incentives against monotone opponents' strategies. \citet{heSunXu2026} establish the existence of perfect monotone equilibria. None of these papers establish extremal monotone perfect equilibria, which is the focus of this paper. In addition, these existence results may allow strategic substitutes.\footnote{To illustrate, consider two firms whose entry decisions are strategic substitutes. Their private types are independent and uniformly distributed on $[0,1]$. Firm $i$ chooses whether to enter, $a_i\in\{0,1\}$, and earns $u_i(a_i,a_j,t_i,t_j)=a_i(1+t_i-3a_j)$. There are three equilibria: in the first two equilibria, one firm enters and the other never does; in the third, each enters above type $1/2$. All three equilibria are perfect and monotone under both perfection criteria here, but they are pairwise incomparable. As a result, there is no largest or smallest perfect equilibrium. We provide the proof in Section~\ref{app:entry-continuous}.} Our extremal results use strategic complementarities to bound the full behavioral perfect-equilibrium set, rather than restrict the comparison to monotone strategies.

The study of trembling-hand perfection for finite games goes back to \citet{selten}. \citet{simonStinchcombe1995} study extensions to infinite games.  \citet{bfv16} and \citet{cn21} develop perfection in Bayesian games without imposing strategic complementarities. Our comparison of refinements concerns the bounds on equilibrium behavior, including the possibility that two criteria admit monotone extrema but yield different bounds even with binary actions.

Under complete information, \citet{ks97} study undominated equilibria in games with strategic complementarities, and \citet{cnm15} establish existence and compactness of the set of pure perfect equilibria under sequential better-reply security. \citet{ckk} obtain pure maximal and minimal perfect equilibria and comparative statics in the weak set order for pseudo-supermodular games as well as largest and smallest equilibria in each constrained game. We obtain global largest and smallest perfect equilibria in Bayesian games on action lattices, and obtain monotone equilibria in monotone supermodular Bayesian games.

Section~\ref{sec:model} presents the model. Section~\ref{sec:results} studies increasing perturbations, and Section~\ref{sec:general-comparison} studies broader perfection and compares the refinements. Section~\ref{sec:discussion} gives the extension to interim complementarities. Section~\ref{sec:conclusion} concludes. The Appendix contains the proofs and the analysis of prescribed tremble distributions.

\section{The model}\label{sec:model}

There are $N$ players, indexed by $i\in I=\{1,\ldots,N\}$. Player $i$ chooses an action from a nonempty compact metric lattice $(A_i,d_i)$, endowed with its Borel $\sigma$-algebra. 
For any $a_i,a_i'\in A_i$, write $a_i\vee a_i'$ for their least upper bound and $a_i\wedge a_i'$ for their greatest lower bound, both of which exist as $A_i$ is a lattice.\footnote{On a subset of Euclidean space closed under coordinatewise maxima and minima, they are the coordinatewise maximum and minimum.} 
We call a nonempty subset closed under these operations a sublattice. The operations are continuous, and $A_i$ is compact in the topology induced by $d_i$. There is a smallest action $\underline a_i=\bigwedge A_i$ and a largest action $\overline a_i=\bigvee A_i$.\footnote{A compact metric lattice is complete: every nonempty subset $S$ has a least upper bound $\bigvee S$ and a greatest lower bound $\bigwedge S$. Its order is also closed, so $\{(a,b)\in A_i\times A_i\mid a\le b\}$ is closed.} Let $A=\prod_{i\in I}A_i$ and $A_{-i}=\prod_{j\ne i}A_j$, with the product topologies and coordinatewise orders. Joins and meets of profiles are taken coordinatewise.

Player $i$ observes a private type $t_i\in T_i$. We allow a general measurable space $(T_i,\mathcal T_i)$, with  a partial order and a reference probability $\lambda_i$. Residual uncertainty is represented by a state $t_0\in T_0$, where $(T_0,\mathcal T_0)$ is a measurable space with a partial order. Write $T=T_0\times\prod_{i\in I}T_i$ and $T_{-i}=T_0\times\prod_{j\ne i}T_j$, with the product orders and $\sigma$-algebras.\footnote{The reference probabilities need not be the marginals of a common prior that generates the players' beliefs. Unless stated otherwise, the phrases ``almost every type'' and ``almost everywhere'' refer to the reference probability of the relevant player. On the type spaces, we do not impose any condition on topology, or require $\lambda_i$ to be atomless or have a density.}

At type $t_i$, player $i$ holds a belief $p_i(\cdot\mid t_i)$ about $t_{-i}$. We work directly with these interim beliefs rather than assume that they come from a common prior. Each $p_i(\cdot\mid t_i)$ is a probability measure on $T_{-i}$, and $t_i\mapsto p_i(B\mid t_i)$ is measurable for every measurable $B\subseteq T_{-i}$. 
We identify strategies that agree $\lambda_i$-almost everywhere.  A change to an opponent's strategy on a $\lambda_j$-null set should then leave a player's interim incentives unchanged at almost every own type. As a result, we impose the following compatibility condition: for every $i$, every $j\ne i$, and every $N_j\in\mathcal T_j$ with $\lambda_j(N_j)=0$,
\[
\int_{T_i}p_i(\{t_j\in N_j\}\mid t_i)\lambda_i(\dd t_i)=0.
\]
The condition holds when beliefs come from a common prior with marginals $\lambda_i$, but subjective beliefs are permitted as well.\footnote{Equivalently, the marginal on $T_j$ of player $i$'s belief, averaged over $t_i$ with respect to $\lambda_i$, is absolutely continuous with respect to $\lambda_j$. 
The pointwise formulation in \citet{vzv} does not require this condition.} To compare beliefs, we use first-order stochastic dominance. For probability measures $P,P'$ on $T_{-i}$, write $P\st P'$ if $\int\psi\dd P\le\int\psi\dd P'$ for every bounded measurable increasing function $\psi$ on $T_{-i}$.

A pure strategy is a measurable map $\alpha_i:T_i\to A_i$. A behavioral strategy $r_i$ allows the player to randomize at each type: $r_i(\cdot\mid t_i)$ is a Borel probability measure on $A_i$, and $t_i\mapsto r_i(B\mid t_i)$ is measurable for each Borel set $B\subseteq A_i$. We denote the Borel probability measures on $A_i$ by $\mathcal M(A_i)$, which is endowed with the topology of weak convergence.\footnote{We write $\mu_n\Rightarrow\mu$ if $\int\phi\dd\mu_n\to\int\phi\dd\mu$ for every bounded continuous function $\phi$ on $A_i$. The same convention applies on product action spaces. The Prohorov metric $\rho_i$, defined in Section~\ref{sec:general-definition}, metrizes weak convergence on $\mathcal M(A_i)$.} A pure profile $\alpha=(\alpha_i)_{i\in I}$ can be regarded as the degenerate behavioral profile $\delta_\alpha=(\delta_{\alpha_i(t_i)})_{i\in I}$. A behavioral strategy is completely mixed if $r_i(\cdot\mid t_i)$ has full support on $A_i$ at every type.\footnote{The join and meet of pure strategies can be defined pointwise: $(\alpha_i\vee\beta_i)(t_i)=\alpha_i(t_i)\vee\beta_i(t_i)$ and $(\alpha_i\wedge\beta_i)(t_i)=\alpha_i(t_i)\wedge\beta_i(t_i)$. 
}

We call $\alpha_i$ monotone if there is a measurable $\widetilde\alpha_i$ equal to $\alpha_i$ almost everywhere such that $\widetilde\alpha_i(t_i)\le\widetilde\alpha_i(t_i')$ whenever $t_i\le t_i'$.\footnote{In this case, we call $\widetilde\alpha_i$ an increasing representative of $\alpha_i$. The use of representatives allows us to  treat strategies that differ only on a null set as the same strategy.} A pure profile is monotone if each component is monotone.

For action distributions $\mu,\nu\in\mathcal M(A_i)$, write $\mu\st\nu$ if
\[
\int\phi\dd\mu\le\int\phi\dd\nu
\]
for every bounded Borel measurable increasing function $\phi$ on $A_i$. We use the same order on distributions over product action spaces.\footnote{On compact metric lattices, continuous increasing functions suffice. Also, $\mu\st\delta_b$ if and only if $\mu(\{a\mid a\le b\})=1$,  $\delta_b\st\mu$ if and only if $\mu(\{a\mid b\le a\})=1$, and $\delta_a\st\delta_b$ if and only if $a\le b$; see Lemma~\ref{lem:stoch}.} For behavioral profiles, $r\st h$ means $r_i(\cdot\mid t_i)\st h_i(\cdot\mid t_i)$ for every player and almost every type. For pure profiles, $\alpha\le\beta$ means $\alpha_i(t_i)\le\beta_i(t_i)$ almost everywhere for each player, or equivalently $\delta_\alpha\st\delta_\beta$. We write $\alpha<\beta$ if $\alpha\le\beta$ and the profiles are distinct modulo null sets. A behavioral strategy is stochastically increasing in type if $r_i(\cdot\mid t_i)\st r_i(\cdot\mid t_i')$ whenever $t_i\le t_i'$.

Player $i$ receives payoff $u_i(a,t)$. We assume that $u_i$ is bounded and jointly measurable on $A\times T$, and continuous in $a$ for each $t$. 
Against opponents' behavioral strategies $r_{-i}$, the player's interim payoff from action $a_i$ at type $t_i$ is
\begin{equation}\label{eq:V}
V_i(a_i,t_i;r_{-i})=
\int_{T_{-i}}\int_{A_{-i}}
u_i(a_i,a_{-i},t_i,t_{-i})
\bigotimes_{j\ne i}r_j(\dd a_j\mid t_j)\,
p_i(\dd t_{-i}\mid t_i).
\end{equation}
Players' randomizations are independent conditional on the type profile, though the types need not be independent. Write 
$$\BR_i(t_i;r_{-i})=\argmax_{a_i\in A_i}V_i(a_i,t_i;r_{-i}).
$$ 
A behavioral profile $r$ is a Bayesian Nash equilibrium if $\supp r_i(\cdot\mid t_i)\subseteq\BR_i(t_i;r_{-i})$ for every player and almost every type. For a pure strategy Bayesian Nash equilibrium $\alpha$, $\alpha_i(t_i)\in\BR_i(t_i;\delta_{\alpha_{-i}})$ almost everywhere for every player.

We consider games in which a player has a greater incentive to choose a higher action when her opponents choose higher actions. When her own action has several components, the payoff exhibits complementarities among those components. Formally, $u_i$ is supermodular in own action if
\[
u_i(a_i\vee a_i',a_{-i},t)+u_i(a_i\wedge a_i',a_{-i},t)
\ge u_i(a_i,a_{-i},t)+u_i(a_i',a_{-i},t)
\]
for every $a_i,a_i'\in A_i$, $a_{-i}\in A_{-i}$, and $t\in T$. This condition is automatic on a totally ordered action space. For another ordered argument $z$, the function has increasing differences in $(a_i,z)$ if $u_i(a_i',z)-u_i(a_i,z)$ is increasing in $z$ whenever $a_i'\ge a_i$, holding the remaining arguments fixed.

\begin{assumption}\label{ass:main}
For every player $i$ and every $t\in T$, $u_i$ is supermodular in $a_i$ and has increasing differences in $(a_i,a_{-i})$.
\end{assumption}

A Bayesian game satisfying Assumption~\ref{ass:main} is supermodular. Under this assumption, against any opponents' behavioral strategies, a player's interim best response set is a nonempty compact sublattice. It has largest and smallest elements, both of which increase when opponents' strategies increase in the stochastic order at every type. Thus, we can compare best responses across opponents' behavior without yet imposing any condition involving types.

For best responses also to increase with private information, we impose conditions on both payoffs and beliefs.

\begin{assumption}\label{ass:types}
For every player $i$,
\begin{enumerate}[label=(\alph*)]
\item $u_i(a_i,a_{-i},t)$ has increasing differences in $(a_i,t)$ for every $a_{-i}\in A_{-i}$;
\item $p_i(\cdot\mid t_i)$ is increasing in $t_i$ in the stochastic order.
\end{enumerate}
\end{assumption}

Part~(a) concerns the entire type profile, including the residual state. A higher own type raises the payoff gain from a higher action at any fixed opponents' behavior. Under part~(b), the player also expects stochastically higher opponents' types and residual states. If opponents use increasing strategies, she therefore expects higher actions from them as well. Under the two assumptions, her largest and smallest best responses to increasing opponents' strategies are increasing in her type. Constant interim beliefs satisfy part~(b). Following \citet{vzv}, we refer to games satisfying Assumptions~\ref{ass:main} and~\ref{ass:types} as monotone supermodular games.


\section{Perfect equilibria under increasing perturbations}\label{sec:results}

\subsection{Increasing perturbations}\label{sec:definition}

Extremal equilibria provide upper and lower predictions for behavior, but these predictions may be fragile. In particular, an extremal Bayesian Nash equilibrium may involve weakly dominated actions. The monotonicity property alone does not resolve this concern. Thus, we ask whether extremal equilibrium predictions can be sustained under the widely-adopted trembling-hand refinement. It raises a further modeling choice in Bayesian games: how should a player’s mistakes depend on her private information?

We focus on trembles that retain the direction of the response to private information. A player may make mistakes and depart from optimal strategic behavior, while continuing to respond to the incentives associated with her type. The simplest case is a tremble that does not depend on type. It describes a departure from the intended strategy in which the player makes mistakes independent of her private information. We shall include this benchmark, but also accommodate imperfect decision procedures whose mistakes may still depend on type. Specifically, we allow the conditional distribution of a trembling player’s action to depend on her type, subject to the requirement that higher types choose stochastically higher actions. We call perturbations with this property \textit{increasing perturbations}.

Consider a firm deciding whether to invest, $A_i=\{0,1\}$. Its private type $t_i\in\R$ is the return on investment, and $c_i$ is a fixed cost. Ordinarily, the firm takes strategic interaction into account. With a small probability, it instead uses a simplified assessment and invests whenever $t_i+Z_i\ge c_i$. The assessment error $Z_i$ has a continuous distribution function $F_i$ with full support on $\R$, independent of the firm's type and of other firms' errors. Conditional on using this rule, the firm invests with probability
\[
\mu_i(\{1\}\mid t_i)=1-F_i(c_i-t_i)\in(0,1).
\]
Both actions have positive probability at every type. For any given error, a firm that invests at one private return also invests at a higher return. Its probability of investment therefore increases with type, although it is not making an optimal strategic choice.\footnote{This is a binary random-parameter model of stochastic choice; see \citet{apesteguiaBallester2018}. For a related example on a finite, totally ordered action set, suppose the player maximizes $w_i(a_i,t_i)+Z_i(a_i)$ when she trembles. If $w_i$ is measurable and has increasing differences in action and type, and the shock vector is independent of type, the largest maximizer is increasing in type for every shock realization. The resulting action distributions are stochastically increasing.}

In this example, for any fixed realization of the error, a higher type leads to a weakly higher action.
The same reasoning applies to a decision rule that is increasing in type for each realization of a type-independent random input. 
By working with increasing perturbations, we abstract from imposing a particular decision procedure or distribution of assessment errors, and putting stochastic monotonicity directly on these action distributions.\footnote{Allowing increasing perturbations is more general than requiring perturbations to be mixtures over monotone strategies. Such mixtures induce stochastically increasing conditional action distributions, but the converse can fail on partially ordered spaces; see \citet[Example~1.1]{fillMachida2001}. }


\begin{definition}
An increasing perturbation $q=(\varepsilon,\mu)$ specifies, for each player $i$, a tremble probability $\varepsilon_i\in(0,1)$ that is constant across types and a transition probability $\mu_i$ from $T_i$ to $A_i$. Each $\mu_i(\cdot\mid t_i)$ has full support on $A_i$, and
\[
t_i\le t_i'
\quad\Longrightarrow\quad
\mu_i(\cdot\mid t_i)\st\mu_i(\cdot\mid t_i').
\]
\end{definition}

We use ``perturbation'' to mean an increasing perturbation in this section. Write $q_i=(\varepsilon_i,\mu_i)$ and $\|q\|=\max_i\varepsilon_i$. Players may differ in their tremble probabilities, and these probabilities may differ across perturbations. 

If player $i$ intends to use $r_i$, the actual action distribution under perturbation is
\begin{equation}\label{eq:residual}
r_i^{q_i}(\cdot\mid t_i)
=\varepsilon_i\mu_i(\cdot\mid t_i)+(1-\varepsilon_i)r_i(\cdot\mid t_i).
\end{equation}
She follows $r_i$ with probability $1-\varepsilon_i$ and otherwise draws an action from $\mu_i(\cdot\mid t_i)$. When $r_i$ is stochastically increasing in type, so is $r_i^{q_i}$.
Thus, if an opponent intends to choose higher actions at higher types, she continues to do so in distribution after mistakes are introduced.

Let $r^q=(r_i^{q_i})_{i\in I}$ and $r_{-i}^{q_{-i}}=(r_j^{q_j})_{j\ne i}$. For a pure intended action and a pure strategy, respectively, write
$$a_i^{q_i}(\cdot\mid t_i)
=\varepsilon_i\mu_i(\cdot\mid t_i)+(1-\varepsilon_i)\delta_{a_i}, \qquad 
\alpha_i^{q_i}(\cdot\mid t_i) =(\alpha_i(t_i))^{q_i}(\cdot\mid t_i).
$$
In the perturbed game $G(q)$, players choose intended actions and receive payoffs from the realized actions. The other primitives are unchanged. Player~$i$'s payoff is 
\begin{equation}\label{eq:perturbedpayoff}
u_i^q(a,t)=\int_A u_i(x,t)
\bigotimes_{j\in I}a_j^{q_j}(\dd x_j\mid t_j).
\end{equation}

Throughout, superscripts $p$ and $b$ refer to pure and behavioral strategies. Denote the pure and behavioral Bayesian Nash equilibria of $G(q)$ by $E^p(q)$ and $\EE(q)$. A player's own tremble contributes a payoff term independent of her intended choice. She chooses an intended best response to opponents' realized behavior:
\begin{equation}\label{eq:perturbedoptimality}
\begin{aligned}
r\in\EE(q)\quad\Longleftrightarrow\quad
&\supp r_i(\cdot\mid t_i)\subseteq\BR_i(t_i;r_{-i}^{q_{-i}})\\
&\text{for every }i\text{ and almost every }t_i.
\end{aligned}
\end{equation}
For a pure profile $\alpha$ in $E^p(q)$,  $\alpha_i(t_i)\in\BR_i(t_i;\alpha_{-i}^{q_{-i}})$ almost everywhere for every player. 

\begin{definition}\label{def:perfect}
A behavioral profile $r$ is perfect under increasing perturbations if there are perturbations $q^n$ with $\|q^n\|\to0$ and equilibria $r^n\in\EE(q^n)$ such that
\[
(r_i^n)^{q_i^n}(\cdot\mid t_i)\Rightarrow r_i(\cdot\mid t_i)
\]
for every player $i$ and almost every type $t_i$. Denote this set by $\PP_{\mathrm{inc}}^b$, and let $\PP_{\mathrm{inc}}^p=\{\alpha\mid\delta_\alpha\in\PP_{\mathrm{inc}}^b\}$ be  the set of pure strategy perfect equilibria.
\end{definition}

The tremble distributions may vary along the sequence. 
The supporting equilibria along the sequence may be behavioral even when the limit is pure. 
Since tremble probabilities vanish, intended strategies and realized action distributions have the same conditional weak limits. Every strategy profile in $\PP_{\mathrm{inc}}^b$ is a Bayesian Nash equilibrium. 


\subsection{Extremal equilibria}\label{sec:main-result}

Fixing a perturbation, one picks a particular model of mistakes. For each fixed increasing perturbation, the perturbed game has a largest and a smallest Bayesian Nash equilibrium. Different perturbations may nevertheless support different limiting behavior.  However, allowing the perturbation to vary admits different equilibrium selections, and those extremal equilibria may not be comparable across perturbations. In the following theorem, we show that the globally largest and smallest perfect equilibria exist across all equilibria and all increasing perturbations. Importantly, both the globally largest and smallest perfect equilibria are attained by pure monotone equilibria.

\begin{theorem}\label{thm:main}
Under Assumptions~\ref{ass:main} and~\ref{ass:types}, the set $\PP_{\mathrm{inc}}^b$ has a largest and a smallest element, both induced by pure monotone strategy profiles.
\end{theorem}

Write $\alpha_{\mathrm{inc}}^+$ and $\alpha_{\mathrm{inc}}^-$ as the largest and smallest perfect equilibria, which are increasing pure-strategy profiles. 
These two monotone pure-strategy profiles provide sharp bounds on the set of all perfect equilibria under increasing perturbations that may be neither pure nor monotone. In particular, every $r\in\PP_{\mathrm{inc}}^b$ satisfies
\[
\delta_{\alpha_{\mathrm{inc},i}^-(t_i)}
\st r_i(\cdot\mid t_i)
\st\delta_{\alpha_{\mathrm{inc},i}^+(t_i)}
\]
for every player and almost every type.  As $\alpha_{\mathrm{inc}}^-$ and $\alpha_{\mathrm{inc}}^+$ are themselves perfect and monotone, they are also the smallest and largest perfect monotone equilibria. In particular, there is a unique behavioral equilibrium under the refinement exactly when the two pure profiles coincide almost everywhere. Furthermore, each bound can be approached by extremal pure equilibria of perturbed games, though the upper and lower bounds may require different supporting sequences.\footnote{There are perturbations $q^{n,+}$ and $q^{n,-}$ with $\|q^{n,+}\|,\|q^{n,-}\|\to0$ such that the largest pure equilibria of $G(q^{n,+})$ converge almost everywhere to $\alpha_{\mathrm{inc}}^+$, and the smallest pure equilibria of $G(q^{n,-})$ converge almost everywhere to $\alpha_{\mathrm{inc}}^-$. Both limits have representatives that are increasing at every type.}

The theorem, and all other results in this section and Section~\ref{sec:discussion}, also hold if tremble distributions are required to be type independent; that is, $\mu_i^n(\cdot\mid t_i)=\mu_i^n$ for every $i,n,t_i$, with the distributions still allowed to vary with $n$. One may further require one profile of tremble distributions for the entire sequence. Under that restriction, largest and smallest elements may fail to exist, as Proposition~\ref{prop:failure} shows.


To understand why Theorem~\ref{thm:main} does not require a prescribed distribution of mistakes, it is useful to separate comparisons within a perturbed game from comparisons across perturbations. For every perturbation $q$, following the standard argument of strategic complementarities, the game $G(q)$ has a largest and a smallest pure Bayesian Nash equilibrium, denoted by $\bar\alpha(q)$ and $\underline\alpha(q)$. Both $\bar\alpha(q)$ and $\underline\alpha(q)$ are increasing in type; see \citet{vives1990} and \citet{vzv}. However, existence of bounds for each perturbation does not by itself give bounds across all perturbations. Different mistakes need not induce ordered behavior, and simply raising a tremble probability need not shift realized actions in one direction. In the following proposition, we construct a common upper perturbation and a common lower perturbation for any two given perturbations. The comparison holds for every action and every type, without increasing the maximum tremble probability for the original pair.

\begin{proposition}\label{prop:common}
For any perturbations $q^1,q^2$, there are perturbations $q^+,q^-$ with $\|q^+\|=\|q^-\|=\max\{\|q^1\|,\|q^2\|\}$ such that
\[
a_i^{q_i^-}(\cdot\mid t_i)
\st a_i^{q_i^k}(\cdot\mid t_i)
\st a_i^{q_i^+}(\cdot\mid t_i)
\]
for every player $i$, type $t_i$, intended action $a_i$, and $k=1,2$.
\end{proposition}

Strategic complementarities then give
\[
\bar\alpha(q^+)\ge\bar\alpha(q^1)\vee\bar\alpha(q^2),
\qquad
\underline\alpha(q^-)\le\underline\alpha(q^1)\wedge\underline\alpha(q^2)
\]
at every type. For any bound on tremble probabilities, we can place two largest perturbed-game equilibria below a third without exceeding that bound. The analogous comparison holds from below for smallest equilibria. We take bounds over these families and let the maximum permitted tremble probability tend to zero. The limiting profiles belong to $\PP_{\mathrm{inc}}^p$ and bound every element of $\PP_{\mathrm{inc}}^b$.\footnote{For each bound on tremble probabilities, the largest perturbed-game equilibria form an upward directed family and the smallest form a downward directed family. We use their measurable essential bounds. The common upper and lower perturbations remain type independent when the original perturbations are type independent. 
}

Suppose that a sequence of pure perfect equilibria under increasing perturbations converges in measure to a pure strategy profile: the measure of types at which prescribed actions differ appreciably from the limit tends to zero. The limiting profile remains perfect under increasing perturbations.\footnote{Formally, $\alpha^k$ converges in measure to $\alpha$ if, for each player $i$ and each $\eta>0$, $\lambda_i(\{t_i\mid d_i(\alpha_i^k(t_i),\alpha_i(t_i))>\eta\})\to0$.}

\begin{proposition}\label{prop:closure}
The set $\PP_{\mathrm{inc}}^p$ is closed under convergence in measure of pure strategy profiles, and hence under pointwise convergence almost everywhere.
\end{proposition}

The result does not require compactness of the pure strategy space or a topological density condition on types. It is proved together with Proposition~\ref{prop:interim}.


In the end, we study comparative statics for the extremal equilibria. We consider games with the same action spaces, type spaces, and reference probabilities. We study two cases. In the first case, we show that both extremal equilibrium predictions increase under an upward shift in interim beliefs. Second, we assume that payoffs vary with a parameter $\theta$ in a partially ordered set, and show that both extremal equilibrium predictions increase in $\theta$.

Write $\alpha_{\mathrm{inc}}^+(p)$ and $\alpha_{\mathrm{inc}}^-(p)$ for the largest and smallest equilibria under increasing perturbations at belief profile $p$. When only payoffs vary with a parameter $\theta$ in a partially ordered set, use $\alpha_{\mathrm{inc}}^+(\theta)$ and $\alpha_{\mathrm{inc}}^-(\theta)$. The following corollary establishes that both extremal predictions move in the direction of the change in incentives.

\begin{corollary}[Comparative statics]\label{cor:comparative}
\begin{enumerate}[label=(\roman*)]
\item If the games with beliefs $p,p'$ satisfy Assumptions~\ref{ass:main} and~\ref{ass:types} and $p_i(\cdot\mid t_i)\st p_i'(\cdot\mid t_i)$ for every $i,t_i$, then
\[
\alpha_{\mathrm{inc}}^+(p)\le\alpha_{\mathrm{inc}}^+(p'),
\qquad
\alpha_{\mathrm{inc}}^-(p)\le\alpha_{\mathrm{inc}}^-(p')
\quad\text{almost everywhere}.
\]
\item If Assumptions~\ref{ass:main} and~\ref{ass:types} hold at every $\theta$ and each $u_i^\theta$ has increasing differences in $(a_i,\theta)$, then $\theta\le\theta'$ implies
\[
\alpha_{\mathrm{inc}}^+(\theta)\le\alpha_{\mathrm{inc}}^+(\theta'),
\qquad
\alpha_{\mathrm{inc}}^-(\theta)\le\alpha_{\mathrm{inc}}^-(\theta')
\quad\text{almost everywhere}.
\]
\end{enumerate}
\end{corollary}



\section{Perfect equilibria}\label{sec:general-comparison}


In  Section~\ref{sec:results}, we consider the class of increasing perturbations. In this section, we shall relax this constraint and allow for all perturbations. We show that under strategic complementarities (under Assumption~\ref{ass:main}), the set of all perfect equilibrium still admits a largest and a smallest extremal perfect equilibrium, which can be taken in pure strategies. In monotone supermodular games (under Assumptions~\ref{ass:main} and~\ref{ass:types}), these extremal perfect equilibria are monotone with finite actions.  However, with infinite actions, the extremal perfect equilibria can be nonmonotone. Indeed, the set of perfect monotone equilibria can itself lack extrema.


\subsection{Perfect equilibrium and existence}\label{sec:general-definition}

For a strategy profile, we ask whether the action prescribed at almost every type can be approximated by best responses to completely mixed opponents whose behavior converges to that profile.\footnote{For pure strategies, the definition follows from \citet[Definition~11]{heSunXu2026}. Here we extend the notion for behavioral strategies.} 

Against opponents' behavioral strategies $g_{-i}$, player $i$ can use any distribution supported on her best response set. These behavioral best responses form
\[
\mathcal B_i(t_i;g_{-i})
=\{\nu_i\in\mathcal M(A_i)\mid\nu_i(\BR_i(t_i;g_{-i}))=1\}.
\]
For a nonempty $C\subseteq\mathcal M(A_i)$, let $\rho_i(\nu_i,C)=\inf_{\xi_i\in C}\rho_i(\nu_i,\xi_i)$, where $\rho_i$ is the Prohorov metric.\footnote{For a Borel set $B\subseteq A_i$, put $B^\eta=\{a_i\in A_i\mid\inf_{b_i\in B}d_i(a_i,b_i)<\eta\}$, with $\varnothing^\eta=\varnothing$. The Prohorov metric is
\[
\rho_i(\mu,\nu)=\inf\left\{\eta>0\,\middle|\,
\begin{array}{l}
\mu(B)\le\nu(B^\eta)+\eta\ \text{and}\ \nu(B)\le\mu(B^\eta)+\eta\\
\text{for every Borel set }B\subseteq A_i
\end{array}\right\}.
\]
It metrizes weak convergence on $\mathcal M(A_i)$.}

\begin{definition}\label{def:general}
A behavioral profile $r$ is perfect if there are completely mixed behavioral profiles $g^n$ such that, for every player $i$ and almost every type $t_i$,
\begin{equation}\label{eq:general-behavioral-criterion}
g_i^n(\cdot\mid t_i)\Rightarrow r_i(\cdot\mid t_i),
\qquad
\rho_i\bigl(r_i(\cdot\mid t_i),\mathcal B_i(t_i;g_{-i}^n)\bigr)\longrightarrow0.
\end{equation}
Denote this set by $\PP^b$, and let $\PP^p=\{\alpha\mid\delta_\alpha\in\PP^b\}$ include the pure strategy profiles.
\end{definition}

Every perfect profile is a Bayesian Nash equilibrium by Lemma~\ref{lem:optimality}. A pure perfect equilibrium is a perfect monotone equilibrium if each component has an increasing representative.\footnote{Henceforth, ``perfect'' without qualification refers to Definition~\ref{def:general}; ``perfect under increasing perturbations'' refers to Definition~\ref{def:perfect}.} 
For a pure profile $\alpha$, the requirements in~\eqref{eq:general-behavioral-criterion} are equivalent to
\begin{equation}\label{eq:general-criterion}
g_i^n(\cdot\mid t_i)\Rightarrow\delta_{\alpha_i(t_i)},
\qquad
d_i\bigl(\alpha_i(t_i),\BR_i(t_i;g_{-i}^n)\bigr)\longrightarrow0
\end{equation}
for every player and almost every type. With finite actions, $\alpha_i(t_i)$ must itself be a best response for all sufficiently large indices at each such type. This excludes weakly dominated actions in the interim game. With infinite actions, a player can have best responses arbitrarily close to $\alpha_i(t_i)$ even though that action is never optimal against the supporting profiles. A weakly dominated limiting action may then be perfect, as Example~\ref{ex:regional-chain} will illustrate.

We first establish global bounds without imposing the conditions in Assumption~\ref{ass:types}.

\begin{theorem}\label{thm:general-global}
Under Assumption~\ref{ass:main}, the set $\PP^b$ has a largest and a smallest element, both induced by pure strategy profiles.
\end{theorem}

Write $\alpha^+$ and $\alpha^-$ for the corresponding pure profiles. Every $r\in\PP^b$ satisfies
\[
\delta_{\alpha_i^-(t_i)}\st r_i(\cdot\mid t_i)\st\delta_{\alpha_i^+(t_i)}
\]
for every player and almost every type. No restriction to pure or monotone equilibria is needed for the comparison.

For the largest equilibrium, we begin with pure profiles that are best responses to completely mixed behavior concentrated near them. Given two such profiles, we construct a third above both, supported in the same way. We control how far the new supporting behavior is from the new pure profile. A convergent sequence of these profiles gives a pure perfect equilibrium, and the same comparison places every behavioral perfect equilibrium below it. The smallest equilibrium is obtained by reversing the action orders. 


\subsection{Monotonicity and comparative statics}\label{sec:general-monotone}

Theorem~\ref{thm:general-global} does not indicate whether  $\alpha^-$ and $\alpha^+$ are monotone. We show that under Assumption~\ref{ass:types}, $\alpha^-$ and $\alpha^+$ are monotone when actions are finite. With infinite actions, however, it can fail. Furthermore, restricting attention to the perfect monotone equilibria need not restore a largest or a smallest element.

\paragraph{Finite actions.}
The type spaces may remain infinite.

\begin{theorem}\label{prop:general-finite}
Under Assumptions~\ref{ass:main} and~\ref{ass:types}, if every $A_i$ is finite, then the extremal perfect equilibria $\alpha^-$ and $\alpha^+$ are monotone.
\end{theorem}

Therefore, $\alpha^-$ and $\alpha^+$ are also the smallest and largest perfect monotone equilibria. As in Theorem~\ref{thm:main}, the bounds extend to nonmonotone and behavioral equilibria.

With finite actions, at almost every type, the equilibrium action is an exact best response along the supporting sequence for all sufficiently large indices. In the proof, we construct supporting profiles that tremble toward the largest (respectively, smallest) actions with vanishing probability. Against these profiles, the eventual largest (respectively, smallest) best responses define an increasing operator that preserves monotonicity in type. Its largest (respectively, smallest) fixed point in the lattice of monotone profiles is a perfect monotone equilibrium that bounds every pure perfect equilibrium, including nonmonotone ones. Theorem~\ref{thm:general-global} identifies it with the corresponding extremal perfect equilibrium.\footnote{The conditional tremble distributions used in these supporting profiles are type independent; see Section~\ref{app:finite-monotonicity}. }


\paragraph{Infinite actions.}
With infinitely many actions, the best responses along a supporting sequence can converge without ever choosing the limiting action. The following producer--retailer game demonstrates this issue. The producer's higher type has stochastically higher beliefs, but chooses a lower action in the largest perfect equilibrium.

\begin{example}\label{ex:regional-chain}
	A producer, player~1, supplies a retailer, player~2. The producer chooses either a standard product, \(a_1=0\), or a product specialized toward customers with high tastes. Specialization is represented by \(a_1=1+x\), where \(x\in[0,1]\) is the development intensity. The producer's action space is the compact chain \(A_1=\{0\}\cup[1,2]\). The retailer chooses an order \(b\in A_2=\{0,1,2\}\), and the residual state is customer taste \(\omega\in T_0=[0,1]\). Let \(\pi(x,\omega)=2(x+\omega-1)_+\), where \(c_+=\max\{c,0\}\) for real \(c\). The payoffs are
\begin{align*}
	u_1(a_1,b,\omega)&=
	\begin{cases}
		0,&a_1=0,\\
		\mathbf1_{\{b=2\}}\,\pi(x,\omega)-\mathbf1_{\{b=0\}}-x^2/2,&a_1=1+x,
	\end{cases}\\
	u_2(a_1,b,\omega)&=-(b-1)^2.
\end{align*}
	The retailer strictly prefers the standard order \(b=1\); a large order \(b=2\) or cancellation \(b=0\) is a mistake. Specialization earns \(\pi\) from a large order, but only when customer taste \(\omega\) exceeds \(1-x\). It costs \(x^2/2\) to develop and loses one if the order is cancelled. The standard product always yields zero. We call \(u_1\) and \(u_2\) the product specialization payoffs.

	The producer has two types, \(T_1=\{L,H\}\) with \(L<H\) and \(\lambda_1(L)=\lambda_1(H)=1/2\). The retailer's type space is \(T_2=[0,1]\), with reference probability \(\lambda_2=\frac12U[0,1]+\frac12\delta_1\), where \(U[0,1]\) is the uniform distribution on \([0,1]\). Both \(T_0\) and \(T_2\) carry their usual orders and Borel $\sigma$-algebras. The producer's two beliefs can be described using a draw \(s\sim U[0,1]\). At type \(L\), it believes that \((\omega,t_2)=(s,s)\). At type \(H\), it believes that \((\omega,t_2)=(s,1)\). The belief at \(L\) perfectly correlates customer taste with the retailer's type. At \(H\), the retailer's type is fixed. At every type, the retailer believes that \(t_1=L\) and \(\omega\sim U[0,1]\). For \(y,z\in\{0,1\}\), let \(\beta^{y,z}\) prescribe producer actions \(y\) at \(L\) and \(z\) at \(H\), and retailer action \(1\) at every type.
\end{example}

The retailer always prefers the standard order. The producer must nevertheless consider erroneous orders and cancellations, and its two types disagree about how those errors may be related to customer taste.

\begin{claim}\label{claim:regional-chain}
	The game in Example~\ref{ex:regional-chain} has the following properties.
	\begin{enumerate}[label=(\alph*)]
		\item It satisfies Assumptions~\ref{ass:main} and~\ref{ass:types}, and its pure Bayesian Nash equilibria are the four profiles \(\beta^{y,z}\), \(y,z\in\{0,1\}\), defined in Example~\ref{ex:regional-chain}.
		\item \(\PP^p=\{\beta^{0,0},\beta^{1,0}\}\), and \(\delta_{\beta^{0,0}}\st r\st\delta_{\beta^{1,0}}\) for every \(r\in\PP^b\). Thus \(\alpha^-=\beta^{0,0}\) and \(\alpha^+=\beta^{1,0}\).
		\item The largest perfect equilibrium \(\beta^{1,0}\) is not monotone. The smallest perfect equilibrium \(\beta^{0,0}\) is the unique perfect monotone equilibrium.
		\item The game obtained by reversing all action and type orders, including the order on residual states, satisfies Assumptions~\ref{ass:main} and~\ref{ass:types}. Its smallest perfect equilibrium is not monotone.
	\end{enumerate}
\end{claim}

Both producer types expect customer taste to be uniform on $[0,1]$. At $L$, the retailer's type equals customer taste; at $H$, it equals $1$. The higher type's belief is therefore obtained by raising the retailer's type while leaving customer taste unchanged. This verifies the stochastic comparison in Assumption~\ref{ass:types}(b). Private types do not enter ex post payoffs, so they affect the producer's choice only through beliefs.

At $H$, only the retailer's behavior at type $1$ matters. That type has positive reference probability, and its probability of an erroneous order must tend to zero along every supporting sequence. Once this probability is small enough, specialization is unprofitable, and the producer chooses the standard product. At $L$, the producer may instead expect frequent erroneous large orders on a shrinking interval $[1-h_n,1)$ of customer tastes. The mistakes vanish at every fixed retailer type, but for suitable supporting profiles the producer specializes with a small positive development intensity $x_n$. As $x_n\to0$, its actions converge to $1$, not to the standard product $0$.

A specialized product with zero development intensity still loses money upon cancellation. Indeed, action $1$ is weakly dominated by action $0$ and is never optimal against completely mixed retailer behavior.\footnote{For every $b,\omega$, $u_1(1,b,\omega)=-\mathbf1_{\{b=0\}}\le0=u_1(0,b,\omega)$, with strict inequality when $b=0$.} It is perfect because the best responses $1+x_n$ approach it. The retailer's supporting behavior is not monotone: large orders occur with high probability just below type $1$ and with low probability at type $1$. The construction and proof are in Section~\ref{app:sec3}.

\begin{remark}\label{rem:atomless-counterexample}
An extremal perfect equilibrium can also fail to be monotone when every reference type distribution is atomless. Retain the action spaces and payoffs of Example~\ref{ex:regional-chain}, but take \(T_1=[0,1]\), \(\lambda_1=U[0,1]\), \(T_2=[0,2]\), and \(\lambda_2=U[0,2]\). Let \(S,U\) be independent uniform random variables on \([0,1]\). At \(t_1<1/2\), the producer believes that \((\omega,t_2)=(S,S)\); at \(t_1\ge1/2\), it believes that \((\omega,t_2)=(S,1+U)\). At every type, the retailer believes that \(t_1\sim U[0,1/2]\) independently of \(\omega\sim U[0,1]\). These subjective beliefs satisfy Assumptions~\ref{ass:main} and~\ref{ass:types}. The largest perfect equilibrium is induced by the pure profile \(\widehat\alpha^+\) with \(\widehat\alpha_1^+(t_1)=\mathbf1_{\{t_1<1/2\}}\) and \(\widehat\alpha_2^+(t_2)=1\). It is not monotone. The smallest perfect equilibrium is induced by \(\widehat\alpha^-\), with producer action \(0\) and retailer action \(1\) at every type. This is also the unique perfect monotone equilibrium. The proof in Section~\ref{app:infinite-counterexamples} replaces the distinguished retailer type by a diffuse upper interval, while preserving the separation between its supporting trembles and those faced by low producer types.
\end{remark}

Although the largest perfect equilibrium in Example~\ref{ex:regional-chain} is not monotone, the game has a unique perfect monotone equilibrium. One could still use that equilibrium as both bounds within the subset of perfect monotone equilibria. In the next example, two perfect monotone equilibria prescribe incomparable production plans, and there is no perfect monotone equilibrium above both.

\begin{example}\label{ex:no-extremal}
A producer, player~1, supplies two products, \(x\) and \(y\), to a retailer, player~2. It chooses a production plan and a common development investment. The retailer can maintain or cancel the contract and place additional orders. Types, reference probabilities, and beliefs are as in Example~\ref{ex:regional-chain}.

The producer chooses a production plan \(l=(l_x,l_y)\) and investment \(v\in[0,1]\), where \(l_j\) is the scale of product \(j\). The available plans are \(C=\{O,X,Y,U,V\}\), where \(O=(0,0)\), \(X=(2,0)\), \(Y=(0,2)\), \(U=(2,2)\), and \(V=(3,3)\). Its action space is \(A_1=C\times[0,1]\). Under the coordinatewise order, \(X\) and \(Y\) are incomparable, with \(X\vee Y=U<V\).

The retailer chooses \(b=(b_x,b_y,z)\in D=\{(0,0,0),(0,0,1),(1,0,1),(0,1,1),(1,1,1)\}\). The action \((0,0,0)\) cancels the contract. Otherwise, \(z=1\), with \(b_j=1\) indicating an extra order for product \(j\). The retailer's payoff is \(u_2(l,v,b,t)=\mathbf1_{\{b=(0,0,1)\}}\), so the normal order \((0,0,1)\) is strictly dominant. Both action spaces carry the coordinatewise order and the sum of the coordinate distances.

An extra order earns \(\pi(v,\omega)=2(v+\omega-1)_+\) per unit, where \(c_+=\max\{c,0\}\), while cancellation loses one per unit. Development costs \(v^2/2\). Each active product also yields a fixed benefit \(\theta(t_1)\), independent of its scale, where \(\theta(L)=0\) and \(\theta(H)=1\). The producer's payoff is
\[
u_1(l,v,b,t)
=\sum_{j\in\{x,y\}}\bigl[l_j\{b_j\pi(v,\omega)-(1-z)\}
+\theta(t_1)\mathbf1_{\{l_j>0\}}\bigr]-\frac{v^2}{2}.
\]

For \(j=x,y\), write \(\zeta_j=b_j\pi(v,\omega)-(1-z)\). The table compares the production plans before subtracting the common development cost \(v^2/2\).
\begin{center}
\begin{tabular}{@{}ccl@{}}
\toprule
Production plan & \((l_x,l_y)\) & Payoff before development cost \\
\midrule
\(O\) & \((0,0)\) & \(0\) \\
\(X\) & \((2,0)\) & \(2\zeta_x+\theta(t_1)\) \\
\(Y\) & \((0,2)\) & \(2\zeta_y+\theta(t_1)\) \\
\(U\) & \((2,2)\) & \(2(\zeta_x+\zeta_y)+2\theta(t_1)\) \\
\(V\) & \((3,3)\) & \(3(\zeta_x+\zeta_y)+2\theta(t_1)\) \\
\bottomrule
\end{tabular}
\end{center}
\end{example}

\begin{claim}\label{claim:no-extremal}
The game in Example~\ref{ex:no-extremal} has the following properties.
\begin{enumerate}[label=(\alph*)]
\item It satisfies Assumptions~\ref{ass:main} and~\ref{ass:types}.
\item In every pure perfect equilibrium, the retailer uses the normal order and the producer's investment is zero at both types. The pure perfect equilibria are exactly the four profiles below, where the entries give the producer's production plan.
\[
\begin{array}{c|cc|c}
 & L&H&\text{Monotone}\\ \hline
\alpha^-&O&U&\text{Yes}\\
\alpha^x&X&U&\text{Yes}\\
\alpha^y&Y&U&\text{Yes}\\
\alpha^+&V&U&\text{No}
\end{array}
\]
The profiles \(\alpha^-\) and \(\alpha^+\) are the smallest and largest perfect equilibria. Their degenerate behavioral profiles bound every element of \(\PP^b\).
\item The perfect monotone equilibria are \(\alpha^-,\alpha^x,\alpha^y\). The profiles \(\alpha^x\) and \(\alpha^y\) have no common upper bound within this set, so it has no largest element.
\item The product of the game and its dual has four players and satisfies both assumptions. Its largest and smallest perfect equilibria are not monotone. Its nonempty set of perfect monotone equilibria has neither a largest nor a smallest element.
\end{enumerate}
\end{claim}

Any perfect monotone equilibrium above both $\alpha^x$ and $\alpha^y$ would prescribe at least $X\vee Y=U$ at $L$. Every perfect equilibrium prescribes $U$ at $H$. Monotonicity would therefore require $U$ at $L$ too, but that plan cannot be perfect for the low type.

To see this, fix completely mixed retailer behavior and an investment level. If the combined expected operating return on the two products is positive, type $L$ prefers $V$ to $U$. If it is nonpositive and investment is positive, the type prefers $(O,0)$. At zero investment, the positive risk of cancellation again makes $(O,0)$ strictly better. Thus $U$ is never a best response at $L$, whatever the investment level. Because the production plans are finite, any sequence of actions converging to a choice with plan $U$ must eventually use $U$. Unlike action $1$ in Example~\ref{ex:regional-chain}, the plan cannot be supported by nearby best responses. At $H$, the fixed benefit favors both products, and small cancellation risks favor $U$ over $V$.

Reversing the action and type orders gives a game with no smallest perfect monotone equilibrium. Taking its product with the original game gives the simultaneous failure in part~(d). Section~\ref{app:sec3} verifies the assumptions and constructs the supporting sequences.

\paragraph{Comparative statics.}
An increase in a payoff parameter can raise the return to a higher action regardless of how opponents behave. A change in beliefs is different: to infer the change in opponents' actions, we must also know how their behavior varies with type. This is why the finite-action restriction will be needed for one comparison but not the other.

Consider games with the same action spaces, type spaces, and reference probabilities. Write $\alpha^\pm(\theta)$ when only payoffs vary with a parameter $\theta$ in a partially ordered set, and $\alpha^\pm(p)$ when only the belief profile varies.

\begin{proposition}\label{prop:general-comparative}
\begin{enumerate}[label=(\roman*)]
\item If Assumption~\ref{ass:main} holds at every $\theta$ and each $u_i^\theta$ has increasing differences in $(a_i,\theta)$, then $\theta\le\theta'$ implies $\alpha^+(\theta)\le\alpha^+(\theta')$ and $\alpha^-(\theta)\le\alpha^-(\theta')$ almost everywhere.
\item If actions are finite, both games satisfy Assumptions~\ref{ass:main} and~\ref{ass:types}, and $p_i(\cdot\mid t_i)\st p_i'(\cdot\mid t_i)$ for every $i,t_i$, then $\alpha^+(p)\le\alpha^+(p')$ and $\alpha^-(p)\le\alpha^-(p')$ almost everywhere.
\item There are games satisfying Assumptions~\ref{ass:main} and~\ref{ass:types}, with one infinite action space, such that $p_i(\cdot\mid t_i)\st p_i'(\cdot\mid t_i)$ for every $i,t_i$, yet $\alpha^+(p')<\alpha^+(p)$.
\end{enumerate}
\end{proposition}

Part~(i) does not require finite actions or Assumption~\ref{ass:types}. At a higher parameter, the gain from a higher action is greater against any fixed opponents' behavior, including nonmonotone behavior. For part~(ii), under Assumption~\ref{ass:types}(a), higher beliefs raise that gain against opponents whose behavior is stochastically increasing in type. With finite actions, we can use the increasing supporting profiles constructed for Theorem~\ref{prop:general-finite}.

For the failure in part~(iii), give the producer in Example~\ref{ex:regional-chain} a single type and compare its belief at $L$ with its belief at $H$. Under the lower belief, it specializes in the largest perfect equilibrium; under the higher belief, it chooses the standard product. The largest equilibrium thus decreases as beliefs increase. 



\subsection{Comparison of the refinements}\label{sec:comparison}

An equilibrium under increasing perturbations is perfect under Definition~\ref{def:general}; see Lemma~\ref{lem:optimality}. Hence
\[
\PP_{\mathrm{inc}}^b\subseteq\PP^b,
\qquad
\PP_{\mathrm{inc}}^p\subseteq\PP^p.
\]
Under Assumptions~\ref{ass:main} and~\ref{ass:types}, we have
\[
\alpha^-\le\alpha_{\mathrm{inc}}^-\le\alpha_{\mathrm{inc}}^+\le\alpha^+
\quad\text{almost everywhere}.
\]
We aim to understand when restricting attention to increasing perturbations gives the same bounds.


In general, the bounds can be different. For example, in the product specialization game, only the standard-product equilibrium can be supported by increasing perturbed-game equilibria. In every perturbed-game equilibrium, the retailer intends to place the uniquely optimal standard order. Its probability of a large order is therefore at most its tremble probability at almost every type. Once that probability is small enough, specialization is unprofitable at both producer types. The frequent mistakes on shrinking intervals that supported $\beta^{1,0}$ are no longer possible. 

\begin{claim}\label{claim:regional-chain-inc}
In the game of Example~\ref{ex:regional-chain}, \(\PP_{\mathrm{inc}}^p=\{\beta^{0,0}\}\) and \(\PP_{\mathrm{inc}}^b=\{\delta_{\beta^{0,0}}\}\). Consequently, \(\alpha^-=\alpha_{\mathrm{inc}}^-=\alpha_{\mathrm{inc}}^+=\beta^{0,0}<\beta^{1,0}=\alpha^+\).
\end{claim}

With finite actions, both pairs of extrema are monotone. If players' type spaces are finite as well, the pairs coincide.

\begin{proposition}\label{prop:finite-type-endpoints}
Under Assumptions~\ref{ass:main} and~\ref{ass:types}, if every $A_i$ and every $T_i$ is finite, then $\alpha^-=\alpha_{\mathrm{inc}}^-$ and $\alpha^+=\alpha_{\mathrm{inc}}^+$.
\end{proposition}

The proof uses the type-independent supporting trembles constructed for Theorem~\ref{prop:general-finite}. At each type, the extremal equilibrium action is a best response once the trembles are sufficiently small. With finitely many players and types, a common positive threshold makes these conditions hold simultaneously outside the reference-null sets. The extremal profile is then an exact equilibrium of each sufficiently small perturbed game in the construction, and hence belongs to $\PP_{\mathrm{inc}}^p$. 


For a continuum of types, the distinction between eventual optimality at each type and simultaneous optimality across types can matter. Every fixed type may eventually prefer the prescribed action, although some other types still prefer to deviate at every index of the sequence. The following game gives different extrema under the two refinements with binary actions, independent types, and a common prior.

\begin{example}\label{ex:binary-separation}
Two firms choose whether to invest, \(a_i\in\{0,1\}\). Their types are independent and uniformly distributed on \([0,1]\), and there is no residual uncertainty. Firm~\(i\)'s payoff is \(u_i(a_i,a_j,t_i,t_j)=a_i(t_i^2+a_j-1)\). A firm pays a cost of one to obtain its private return \(t_i^2\), plus a benefit of one if the other firm also invests.
\end{example}

\begin{claim}\label{claim:binary-separation}
The game in Example~\ref{ex:binary-separation} has the following properties.
\begin{enumerate}[label=(\alph*)]
\item It satisfies Assumptions~\ref{ass:main} and~\ref{ass:types}. Up to equality almost everywhere, its only behavioral Bayesian Nash equilibria are \(\delta_{(0,0)}\) and \(\delta_{(1,1)}\).
\item \(\PP^p=\{(0,0),(1,1)\}\) and \(\PP^b=\{\delta_{(0,0)},\delta_{(1,1)}\}\).
\item \(\PP_{\mathrm{inc}}^p=\{(0,0)\}\) and \(\PP_{\mathrm{inc}}^b=\{\delta_{(0,0)}\}\). In particular, \(\alpha^-=\alpha_{\mathrm{inc}}^-=\alpha_{\mathrm{inc}}^+=(0,0)<(1,1)=\alpha^+\).
\end{enumerate}
\end{claim}

To support $(1,1)$ under perfection, let each firm refuse to invest with a type-independent probability $\eta_n\downarrow0$. A firm of type $t_i$ gains $t_i^2-\eta_n$ from investing. For every fixed $t_i>0$, the gain is positive once $n$ is large enough, as Definition~\ref{def:general} requires. At every $n$, however, types below $\sqrt{\eta_n}$ prefer not to invest. Their supporting behavior need not be a best response.

In a perturbed-game equilibrium, those low-return types must choose optimally. Their decision not to invest makes investment less attractive to the other firm, inducing further noninvestment. If $y_i$ is firm $i$'s equilibrium probability of intended noninvestment, the best response conditions imply
\begin{equation}\label{eq:binary-contagion}
y_1y_2\ge(1-\varepsilon_1)(1-\varepsilon_2).
\end{equation}
As tremble probabilities vanish, both $y_1$ and $y_2$ converge to one. Realized behavior has the same limit, so only $(0,0)$ is perfect under increasing perturbations.\footnote{We  prove the inequality and the equilibrium characterization in Section~\ref{app:sec3}. The exclusion of $(1,1)$ holds for arbitrary full-support conditional tremble distributions, provided tremble probabilities remain constant across types. Dropping stochastic monotonicity of those distributions would not recover the high-investment equilibrium. 
}

\FloatBarrier

\section{Interim complementarities}\label{sec:discussion}

A higher private return may come with worse news about other determinants of profitability. A firm with a more profitable project, for example, may also expect higher input costs. It can still have a greater incentive to invest if the direct effect of its return outweighs the adverse information. Such a game need not satisfy the belief condition in Assumption~\ref{ass:types}, even though the relevant payoff differences increase after taking expectations. In this section, we accommodate this observation by imposing the complementarity conditions on interim payoffs.

We retain the action and type spaces, reference probabilities, belief systems, payoff regularity, and compatibility condition of Section~\ref{sec:model}. We replace Assumptions~\ref{ass:main} and~\ref{ass:types} by the conditions below. For each player $i$, let $\mathcal F^{\mathrm{st}}_{-i}$ consist of opponents' behavioral profiles that are stochastically increasing in type: $h_j(\cdot\mid t_j)\st h_j(\cdot\mid t_j')$ whenever $t_j\le t_j'$, for every $j\ne i$.


\begin{assumption}\label{ass:interim}
For every player \(i\), the following conditions hold at every type for the payoffs and conditional beliefs.
\begin{enumerate}[label=(\roman*)]
\item For every \(h_{-i}\in\mathcal F^{\mathrm{st}}_{-i}\) and every \(t_i\), the function \(a_i\mapsto V_i(a_i,t_i;h_{-i})\) is supermodular.
\item For any \(a_i'\ge a_i\) and \(t_i\), let \(r_{-i},h_{-i}\) be opponents' behavioral profiles with at least one in \(\mathcal F^{\mathrm{st}}_{-i}\). If \(r_j(\cdot\mid t_j)\st h_j(\cdot\mid t_j)\) for every \(j\ne i\) and every \(t_j\), then \(V_i(a_i',t_i;h_{-i})-V_i(a_i,t_i;h_{-i})\ge V_i(a_i',t_i;r_{-i})-V_i(a_i,t_i;r_{-i})\).
\item For every \(h_{-i}\in\mathcal F^{\mathrm{st}}_{-i}\), the function \((a_i,t_i)\mapsto V_i(a_i,t_i;h_{-i})\) has increasing differences in \((a_i,t_i)\).
\end{enumerate}
\end{assumption}

In parts~(i) and~(iii), we require own-action supermodularity and increasing differences with own type only when opponents' behavior is stochastically increasing.
These are the profiles encountered along the monotone best response iterations. 
To bound equilibria outside the monotone class, we must also compare increasing opponents’ behavior with arbitrary behavioral profiles that lie above or below it. Part (ii) requires increasing differences for these comparisons. It imposes no restriction on comparisons between two nonmonotone profiles.

\begin{proposition}\label{prop:interim}
Under Assumption~\ref{ass:interim}, the results in Theorem~\ref{thm:main} and Propositions~\ref{prop:common} and~\ref{prop:closure} continue to hold.
\end{proposition}


We prove the proposition directly under the interim conditions in Section~\ref{app:sec5}. Under Assumptions~\ref{ass:main} and~\ref{ass:types}, Lemma~\ref{lem:interim} verifies all three conditions in Assumption~\ref{ass:interim}. The results of Section~\ref{sec:results} therefore follow as special cases. Along the iterations, parts~(i) and~(iii) give increasing extremal best responses. Part~(ii) allows us to order those iterations and to compare their limits with arbitrary behavioral equilibria.

The entry game in the introduction shows why part~(ii) cannot simply be omitted. It satisfies the remaining conditions, yet its three equilibria are pairwise incomparable and perfect under both refinements; see Claim~\ref{claim:entry}. Conditions on a player's response to her own information are not enough to order equilibria across different opponents' behavior.

For comparative statics, consider two games $\theta\le\theta'$ satisfying the hypotheses of Proposition~\ref{prop:interim}. They have the same action spaces, type spaces, reference probabilities, and permitted perturbations. Suppose that
\[
a_i\longmapsto V_i^{\theta'}(a_i,t_i;h_{-i})-V_i^\theta(a_i,t_i;h_{-i})
\]
is increasing for every player $i$, type $t_i$, and $h_{-i}\in\mathcal F^{\mathrm{st}}_{-i}$. In other words, against any fixed increasing opponents' behavior, the payoff gain from a higher own action is at least as large in the second game. Then both extremal equilibria under increasing perturbations increase almost everywhere from $\theta$ to $\theta'$. 
Unlike the primitive conditions of Section~\ref{sec:model}, the interim assumptions alone do not ensure this comparison merely because beliefs increase stochastically.

The following investment game satisfies the interim conditions even though higher types expect a less favorable common profitability component.

\begin{example}\label{ex:interim-beliefs}
Two firms choose whether to invest, $a_i\in\{0,1\}$. Their private types $t_1,t_2$ are independent and uniform on $[0,1]$. Under the common prior, a profitability component $\omega\in[-2,0]$ satisfies $\omega=-(t_1+t_2)$. It represents pressure on shared inputs: higher-return projects are associated with higher costs. Firm $i$'s payoff is
\[
u_i(a_i,a_j,t_i,t_j,\omega)=a_i(2t_i+a_j+\omega-\tfrac12).
\]
For an opponent's behavioral strategy $r_j$, let $m_j=\int_0^1r_j(\{1\}\mid t_j)\,\dd t_j$ be its ex ante investment probability.
\end{example}

A firm with a higher $t_i$ expects a lower $\omega$. But its direct return rises by twice the increase in type, whereas the conditional mean of $\omega$ falls by only that increase. Its interim gain from investing is therefore increasing in type. We record the resulting equilibrium bounds in the claim below.

\begin{claim}\label{claim:interim-beliefs}
The game in Example~\ref{ex:interim-beliefs} has the following properties.
\begin{enumerate}[label=(\alph*)]
\item It satisfies Assumption~\ref{ass:main} and Assumption~\ref{ass:types}(a). Moreover, $\mathbb E[\omega\mid t_i]=-t_i-1/2$, so Assumption~\ref{ass:types}(b) fails.
\item For every behavioral strategy $r_j$ of the other firm, firm $i$'s interim payoff is $V_i(a_i,t_i;r_j)=a_i(t_i+m_j-1)$.
\item It satisfies Assumption~\ref{ass:interim}.
\item The smallest and largest perfect equilibria under increasing perturbations are the constant profiles $(0,0)$ and $(1,1)$.
\end{enumerate}
\end{claim}

Thus, unfavorable information about a common cost need not prevent monotone extremal equilibria. What matters here is the firm's expected gain from investment after taking that information into account.


\section{Conclusion}\label{sec:conclusion}

In this paper, we study the extremal approach to perfect equilibrium under increasing perturbations. In monotone supermodular Bayesian games, the refinement has largest and smallest equilibria on every compact metric action lattice. Importantly, both extrema are pure and monotone. We further demonstrate that the results extend to the setting with interim complementarities. Under the broader notion of perfection, strategic complementarities still yield pure extremal perfect equilibria. In monotone supermodular games, these extrema are monotone in the setting with finite actions. With infinite actions, the monotonicity can fail, and the subset of perfect monotone equilibria can even lack extrema.  When actions and types are finite, the two refinements share the same extremal perfect equilibria. These are also the extremal perfect monotone equilibria. With continuously distributed types, their bounds can differ even in games with binary actions. 


\clearpage


\vspace{-0.2cm}

\appendix
\renewcommand{\thesection}{\Alph{section}}  
\setcounter{section}{0}  

\section*{Appendix A}
\addcontentsline{toc}{section}{Appendix A}  
\setcounter{section}{1}  
\label{appendixA}  

We collect the preliminary results in Section~\ref{app:lemmas}. The proofs for increasing perturbations follow in Section~\ref{app:sec4}, where one argument under the interim conditions covers the results of Sections~\ref{sec:results} and~\ref{sec:discussion}. That section also analyzes the market entry game of the Introduction together with the investment game of Section~\ref{sec:discussion}. The results on perfection, including the comparison of the two refinements in Section~\ref{sec:general-comparison}, are proved in Section~\ref{app:sec3}. Prescribed tremble distributions are the subject of Section~\ref{oa:fixed}. Throughout we rely on two elementary facts. Every compact metric space carries a Borel probability measure with full support, for example \(\sum_{k\ge1}2^{-k}\delta_{x_k}\) for a dense sequence \((x_k)\). Compactness also bounds each action metric \(d_i\), which lets us fix a common bound \(\bar d>0\).

\subsection{Technical preparations}\label{app:lemmas}

This subsection collects the order and measurability results used in the equilibrium proofs. Lemmas~\ref{lem:compact} to~\ref{lem:space} concern compact lattices, probability measures, and measurable strategies. The remaining ones, Lemmas~\ref{lem:interim} to~\ref{lem:glue}, apply these results to best responses and to limits under perturbations. We state standard results with their sources, while the adaptations and estimates needed below come with proofs. Readers may pass directly to the equilibrium proofs and consult these preparations when they are cited.

The following lemma collects basic properties of compact metric lattices. 
\citet[Lemma~9]{lawson1973} shows that monotone nets in compact subsets of topological lattices converge to their suprema or infima. This gives part~(c), while application to nets of finite joins or meets yields completeness in part~(b). Nonempty compact sublattices are also complete, hence have largest and smallest elements. Continuity of the lattice operations gives part~(a) and measurability of finite lattice operations. Part~(d) then follows by taking pointwise limits using part~(c). We state these standard facts without proof.

\begin{lemma}\label{lem:compact}
Let \(C\) be a compact metric lattice.
\begin{enumerate}[label=(\alph*)]
\item The order of \(C\) is closed, that is, \(\{(x,y)\in C\times C\mid x\le y\}\) is a closed subset of \(C\times C\).
\item The lattice \(C\) is complete. Every nonempty compact sublattice of \(C\) has a largest and a smallest element.
\item Every increasing sequence \((x_n)_{n\ge1}\) in \(C\) converges to \(\bigvee_{n\ge1}x_n\), and every decreasing sequence \((x_n)_{n\ge1}\) in \(C\) converges to \(\bigwedge_{n\ge1}x_n\).
\item Let \((S,\mathcal S)\) be a measurable space, and let \(f_1,f_2,\ldots\) be measurable maps from \(S\) to \(C\). Then the maps \(s\mapsto\bigvee_{n\ge1}f_n(s)\) and \(s\mapsto\bigwedge_{n\ge1}f_n(s)\) are measurable.
\end{enumerate}
\end{lemma}

The closed order in Lemma~\ref{lem:compact}(a) permits an application of the continuous utility theorem of \citet{levin}. It yields a continuous real function that increases weakly along the order and strictly along its strict part. For a partial order the strict part consists of the distinct comparable pairs, so the function increases strictly between any two distinct comparable points. Normalizing this bounded function to \([0,1]\) gives the following lemma.

\begin{lemma}\label{lem:isotone}
Let \(C\) be a compact metric lattice. There is a continuous function \(\chi\) from \(C\) to \([0,1]\) such that \(\chi(x)<\chi(y)\) whenever \(x\le y\) and \(x\ne y\).
\end{lemma}

Best responses in the proofs are maximizers of continuous functions on compact metric lattices. Part~(a) of the next lemma provides the largest and smallest maximizers of a supermodular objective, by \citet[Theorem~2.7.1]{topkis} together with Lemma~\ref{lem:compact}(b). Part~(b) orders the extremal maximizers of two objectives whose difference is increasing, which is how we compare best responses across opponents' strategies, types, and perturbations. Its standard form assumes both objectives to be supermodular, whereas we need only one, so we include the short proof.

\begin{lemma}\label{lem:choice}
Let \(C\) be a compact metric lattice, and let \(f\) and \(g\) be continuous real functions on \(C\).
\begin{enumerate}[label=(\alph*)]
\item If \(f\) is supermodular, then \(\argmax f\) is a nonempty compact sublattice of \(C\), and it has a largest and a smallest element.
\item Suppose that \(g-f\) is increasing. If \(g\) is supermodular, then every maximizer of \(f\) lies below the largest maximizer of \(g\). If \(f\) is supermodular, then every maximizer of \(g\) lies above the smallest maximizer of \(f\).
\end{enumerate}
\end{lemma}
\begin{proof}
\setlength{\parskip}{0pt}
(b) Let \(a\) maximize \(f\) and let \(b\) be the largest maximizer of the supermodular function \(g\). Then \(g(a\vee b)-g(b)\ge g(a)-g(a\wedge b)\ge f(a)-f(a\wedge b)\ge0\). Thus \(a\vee b\) maximizes \(g\), so \(a\le b\). Reverse the lattice order and interchange \(f,g\) to obtain the second assertion.
\end{proof}

Strategies are measurable maps of types, so the extremal maximizers of the previous lemma must be selected measurably. The distance to the maximizer set must also be jointly measurable in action and type, because the perfection criterion of Section~\ref{sec:general-comparison} as well as the estimates below are built from it. Part~(a) of the next lemma is the measurable maximum theorem of \citet[Theorem~18.19]{aliprantisborder} for a fixed constraint set, combined with their Theorem~18.5 and Lemma~4.51 for the distance function. Part~(b) applies the same theorem with \(\chi\) from Lemma~\ref{lem:isotone} as the objective over the maximizer correspondence, which selects its largest and smallest elements. We record these consequences without proof.

\begin{lemma}\label{lem:measurable}
Let \(C\) be a compact metric lattice with metric \(d\), and let \((S,\mathcal S)\) be a measurable space. Let \(v\) be a real function on \(C\times S\), continuous in its first argument. Suppose that \(v\) is measurable with respect to the product of the Borel $\sigma$-algebra of \(C\) and \(\mathcal S\), and let \(B(s)=\argmax_{x\in C}v(x,s)\).
\begin{enumerate}[label=(\alph*)]
\item The correspondence \(B\) is measurable with nonempty compact values, and the map \((x,s)\mapsto d(x,B(s))\) is measurable.
\item If \(B(s)\) is a sublattice of \(C\) for every \(s\in S\), then the maps that assign to \(s\) the largest and the smallest element of \(B(s)\) are measurable.
\end{enumerate}
\end{lemma}

The next lemma collects the properties of the stochastic order on action distributions that the comparisons of behavioral strategies require. Part~(a), the closedness of the order under weak convergence, rests on \citet[Theorem~2.6]{whitt1980}, who notes that a compact space with a closed partial order is normally ordered as his theorem requires. Parts~(b) to~(e) follow from the definition, with Fubini's theorem for the products in~(b).

\begin{lemma}\label{lem:stoch}
Let \(C,C_1,\ldots,C_m\) be compact metric lattices. Compare Borel probability measures on them in the stochastic order \(\st\) defined in Section~\ref{sec:model} by bounded Borel measurable increasing functions.
\begin{enumerate}[label=(\alph*)]
\item If \(\mu_n\st\nu_n\) for every \(n\), \(\mu_n\Rightarrow\mu\), and \(\nu_n\Rightarrow\nu\), then \(\mu\st\nu\).
\item If \(\mu_k\st\nu_k\) on \(C_k\) for \(k=1,\ldots,m\), then \(\mu_1\otimes\cdots\otimes\mu_m\st\nu_1\otimes\cdots\otimes\nu_m\) on the product lattice \(C_1\times\cdots\times C_m\).
\item If \(\mu\st\nu\) on \(C_1\) and \(\phi\) is a continuous increasing map from \(C_1\) to \(C\), then the image measures satisfy \(\mu\circ\phi^{-1}\st\nu\circ\phi^{-1}\). If \(X\) and \(Y\) are random elements of \(C\), defined on a common probability space, with \(X\le Y\) almost surely, then the distribution of \(X\) is stochastically dominated by the distribution of \(Y\).
\item Let \(\mu\) be a Borel probability measure on \(C\), and let \(b\in C\). Then \(\mu\st\delta_b\) if and only if \(\mu(\{x\in C\mid x\le b\})=1\). Likewise, \(\delta_b\st\mu\) if and only if \(\mu(\{x\in C\mid b\le x\})=1\). In particular, \(\delta_a\st\delta_b\) if and only if \(a\le b\).
\item If \(\mu_k\st\nu_k\) on \(C\) for \(k=1,\ldots,m\) and \(w_1,\ldots,w_m\ge0\) sum to one, then \(\sum_kw_k\mu_k\st\sum_kw_k\nu_k\).
\end{enumerate}
\end{lemma}

Compared through common upper and lower perturbations, the extremal equilibria of the perturbed games form directed families. The next lemma provides the smallest essential upper bound of such a family. It also provides a sequence from the family that approximates the bound in the integrated action metric. These sequences serve as supporting sequences for perfect equilibria under increasing perturbations. The bound rests on the lattice criterion of \citet[Lemma~A.3 and Remark~A.4]{nendel2020}, whose hypotheses we verify for measurable maps identified almost everywhere. The approximation uses monotone convergence in the compact lattice. We also record that the bound inherits increasing representatives from the family. This property makes the extremal perfect equilibria monotone.

\begin{lemma}\label{lem:directed}
	Let \((S,\mathcal S,\lambda)\) be a probability space. Let \(C\) be a compact metric lattice with metric \(d_C\), and let \(\Gamma\) be a nonempty family of measurable maps from \(S\) to \(C\). For such maps \(f,g\), we write \(f\le g\) almost everywhere if \(f(s)\le g(s)\) for \(\lambda\)-almost every \(s\). Suppose that \(\Gamma\) is upward directed, that is, for any \(f,g\in\Gamma\), there exists \(h\in\Gamma\) such that \(f\vee g\le h\) almost everywhere. Then the following statements hold.
	\begin{enumerate}[label=(\alph*)]
		\item There exists a measurable map \(H:S\to C\) such that \(f\le H\) almost everywhere for every \(f\in\Gamma\). If a measurable map \(K:S\to C\) satisfies the same inequalities, then \(H\le K\) almost everywhere. We call \(H\) the smallest essential upper bound of \(\Gamma\).\footnote{It is unique up to equality almost everywhere.}
		\item There exists a sequence \((f_n)_{n\ge1}\) in \(\Gamma\) such that \(\int_S d_C(f_n(s),H(s))\,\lambda(ds)\to0\).
		\item Suppose, in addition, that \(S\) is partially ordered. If every \(f\in\Gamma\) has a measurable increasing representative \(\widetilde f\), then \(H\) also has a measurable increasing representative \(\widetilde H\).\footnote{An increasing representative of \(f\) is a measurable map \(\widetilde f:S\to C\) that equals \(f\) almost everywhere and satisfies \(\widetilde f(s)\le\widetilde f(s')\) whenever \(s\le s'\).}
	\end{enumerate}
	For a downward directed family, the corresponding conclusions hold with a largest essential lower bound.
	
	The conclusions also extend to directed families of finite profiles \((f_i)_{i=1}^N\), ordered componentwise almost everywhere. Each \(f_i\) maps a probability space \((S_i,\mathcal S_i,\lambda_i)\) measurably into a compact metric lattice \(C_i\) with metric \(d_i\). The additional hypotheses in (c) are imposed on each component. In (b), the approximation becomes \(\sum_{i=1}^N\int_{S_i}d_i(f_{n,i}(s_i),H_i(s_i))\,\lambda_i(ds_i)\to0\).
\end{lemma}
\begin{proof}
\setlength{\parskip}{0pt}
(a) Let \(L\) be the lattice of measurable maps from \(S\) to \(C\),  identified almost everywhere. Lemma~\ref{lem:compact}(d) makes \(L\) Dedekind \(\sigma\)-complete, and the constant extreme maps bound it. With \(\chi\) from Lemma~\ref{lem:isotone}, define \(J(f)=\int_S\chi(f(s))\lambda(\dd s)\). If \(f\le g\) and \(f\ne g\) in \(L\), strict increase of \(\chi\) gives \(J(f)<J(g)\). Thus \(J\) is a strictly increasing map from \(L\) into the Dedekind super complete lattice \(\R\). By \citet[Lemma~A.3]{nendel2020}, \(L\) is Dedekind super complete. In particular, \(\Gamma\) and some countable subfamily \((g_n)\) have the same supremum in \(L\). As in \citet[Remark~A.4]{nendel2020}, directedness permits a sequence \((f_n)\) in \(\Gamma\) with \(f_1=g_1\) and \(f_{n-1}\vee g_n\le f_n\) almost everywhere. Since \(g_n\le f_n\) for every \(n\), the measurable map \(H=\bigvee_n f_n\) represents the supremum of \(\Gamma\) in \(L\), which is its smallest essential upper bound.
\par
(b) Outside one null set the sequence \((f_n)\) is increasing, so Lemma~\ref{lem:compact}(c) gives \(f_n\to H\). Bounded convergence then yields \(\int_S d_C(f_n(s),H(s))\lambda(\dd s)\to0\).
\par
	(c) Choose measurable increasing representatives \(\widetilde f_n\). Their pointwise join \(\widetilde H=\bigvee_n\widetilde f_n\) is measurable, increasing, and equal to \(H\) almost everywhere. Order reversal gives the lower bound assertions. For finite profiles, the same application of Nendel's criterion uses the product of the component lattices and \(J(f)=\sum_i\int_{S_i}\chi_i(f_i(s_i))\lambda_i(\dd s_i)\).
\end{proof}

The next lemma records the order, metric, and value structure of the strategy space. Part~(a) follows from Lemma~\ref{lem:directed}, applied to the finite joins or meets of a nonempty set of profiles, with part~(c) of that lemma keeping the bound in \(\mathcal Q_{\mathrm{inc}}\). Part~(b) is standard. Bounded action metrics make convergence in \(d_{\mathcal Q}\) equivalent to convergence in measure, for which completeness and almost everywhere convergent subsequences are classical. Part~(c) builds \(J\) from the functions of \citet{levin} in Lemma~\ref{lem:isotone}. Its uniform estimate serves the proof of Theorem~\ref{thm:general-global}, so we prove it.

\begin{lemma}\label{lem:space}
	Let \(\mathcal Q\) be the set of pure strategy profiles, identifying profiles that are equal almost everywhere. Order \(\mathcal Q\) by \(\le\), interpreted componentwise almost everywhere. Let \(\mathcal Q_{\mathrm{inc}}\subseteq\mathcal Q\) consist of profiles that admit measurable increasing representatives. Fix continuous functions \(\chi_i:A_i\to[0,1]\) as in Lemma~\ref{lem:isotone}. For \(s,z\in\mathcal Q\), define
	\begin{equation}\label{eq:strategy-functionals}
		d_{\mathcal Q}(s,z)=\sum_{i\in I}\int_{T_i}d_i\bigl(s_i(t_i),z_i(t_i)\bigr)\lambda_i(\dd t_i)
		\quad\text{and}\quad
		J(s)=\sum_{i\in I}\int_{T_i}\chi_i\bigl(s_i(t_i)\bigr)\lambda_i(\dd t_i).
	\end{equation}
	\begin{enumerate}[label=(\alph*)]
		\item The sets \(\mathcal Q\) and \(\mathcal Q_{\mathrm{inc}}\) are complete lattices. For every nonempty subset of \(\mathcal Q_{\mathrm{inc}}\), its join in \(\mathcal Q_{\mathrm{inc}}\) equals its join in \(\mathcal Q\). The same holds for meets.
		\item The function \(d_{\mathcal Q}\) is a metric on \(\mathcal Q\). The metric space \((\mathcal Q,d_{\mathcal Q})\) is complete. Every sequence that converges in \(d_{\mathcal Q}\) has a subsequence that converges almost everywhere to the same limit for every player.
		\item If \(s\le z\), then \(J(s)\le J(z)\). Equality holds only if \(s=z\). Moreover, for every \(\varepsilon>0\), there is \(\kappa>0\) such that \(d_{\mathcal Q}(s,z)<\varepsilon\) whenever \(s\le z\) and \(J(z)-J(s)<\kappa\).
	\end{enumerate}
\end{lemma}

\begin{proof}
\setlength{\parskip}{0pt}
(c) If \(s\le z\), then \(\chi_i(s_i)\le\chi_i(z_i)\) almost everywhere for every player, so \(J(s)\le J(z)\). Equality of the integrals forces \(\chi_i(s_i)=\chi_i(z_i)\) almost everywhere, hence \(s=z\) by Lemma~\ref{lem:isotone}. For the uniform estimate, fix \(\varepsilon>0\) and put \(\gamma=\varepsilon/(2N)\). For each \(i\), the set \(K_i=\{(a,b)\in A_i^2\mid a\le b,\ d_i(a,b)\ge\gamma\}\) is compact. If every \(K_i\) is empty, then \(d_{\mathcal Q}(s,z)\le N\gamma<\varepsilon\) for all \(s\le z\), and any \(\kappa>0\) works. Otherwise, let \(c>0\) be the minimum of \(\chi_i(b)-\chi_i(a)\) over the nonempty sets \(K_i\). Splitting each integral in \(d_{\mathcal Q}(s,z)\) according to whether \(d_i(s_i,z_i)<\gamma\) gives
	\[
	d_{\mathcal Q}(s,z)\le N\gamma+\frac{\bar d}{c}\bigl[J(z)-J(s)\bigr].
	\]
	Indeed, on the complementary event the integrand of \(J(z)-J(s)\) for player \(i\) is at least \(c\), while \(d_i\le\bar d\). Taking \(\kappa=c\varepsilon/(2\bar d)\) proves the assertion, with \(\kappa\) independent of \(s\) and \(z\).
\end{proof}

Interim payoffs inherit strategic complementarities from ex post payoffs through the integration argument of \citet{vzv}. The lemma below adapts this argument to behavioral strategies of the opponents, with Lemma~\ref{lem:stoch} supplying the stochastic comparisons. 

\begin{lemma}\label{lem:interim}
	Fix a player \(i\) together with two profiles \(g_{-i},h_{-i}\) of opponents' behavioral strategies.
	\begin{enumerate}[label=(\alph*)]
		\item The interim payoff \(V_i(a_i,t_i;g_{-i})\) is jointly measurable in \((a_i,t_i)\) and continuous in \(a_i\). At every \(t_i\), the set \(\BR_i(t_i;g_{-i})\) is nonempty and compact. The distance \(d_i\bigl(a_i,\BR_i(t_i;g_{-i})\bigr)\) is jointly measurable in \((a_i,t_i)\). Under Assumption~\ref{ass:main}, \(V_i\) is supermodular in \(a_i\), so \(\BR_i(t_i;g_{-i})\) is a compact sublattice of \(A_i\). Its largest element \(\overline{\BR}_i(t_i;g_{-i})\) and its smallest element \(\underline{\BR}_i(t_i;g_{-i})\) are measurable in \(t_i\).
		\item Under Assumption~\ref{ass:main}, if \(g_j(\cdot\mid t_j)\st h_j(\cdot\mid t_j)\) for every \(j\ne i\) at every \(t_j\), then
		\[
		V_i(b_i,t_i;g_{-i})-V_i(a_i,t_i;g_{-i})\le V_i(b_i,t_i;h_{-i})-V_i(a_i,t_i;h_{-i})
		\]
		for all \(a_i\le b_i\) in \(A_i\) at every \(t_i\). Consequently, \(\overline{\BR}_i(t_i;g_{-i})\le\overline{\BR}_i(t_i;h_{-i})\) and \(\underline{\BR}_i(t_i;g_{-i})\le\underline{\BR}_i(t_i;h_{-i})\) at every \(t_i\). The same conclusions hold at almost every \(t_i\) if the dominance holds only at almost every \(t_j\).
		\item If \(g_j(\cdot\mid t_j)=h_j(\cdot\mid t_j)\) for every \(j\ne i\) at almost every \(t_j\), then \(V_i(\cdot,t_i;g_{-i})=V_i(\cdot,t_i;h_{-i})\) at almost every \(t_i\).
		\item Let Assumptions~\ref{ass:main} and \ref{ass:types} hold. Suppose that \(g_j(\cdot\mid t_j)\st g_j(\cdot\mid t_j')\) whenever \(t_j\le t_j'\), for every \(j\ne i\). For \(a_i\le b_i\) in \(A_i\), define
		\[
		\Delta_i(t)=\int_{A_{-i}}\bigl[u_i(b_i,a_{-i},t)-u_i(a_i,a_{-i},t)\bigr]\bigotimes_{j\ne i}g_j(\dd a_j\mid t_j),\qquad t\in T.
		\]
		Then \(\Delta_i\) is bounded, measurable, and increasing on \(T\). Moreover, \(V_i(a_i,t_i;g_{-i})\) has increasing differences in \((a_i,t_i)\), so \(\overline{\BR}_i(t_i;g_{-i})\) and \(\underline{\BR}_i(t_i;g_{-i})\) are increasing in \(t_i\).
	\end{enumerate}
\end{lemma}

\begin{proof}
\setlength{\parskip}{0pt}
The argument adapts the integration methods of \citet[Propositions~8 and~11, Lemma~9, and Corollary~10]{vzv} to behavioral strategies. The opponents' pure actions are replaced by the product distributions \(\bigotimes_{j\ne i}g_j(\cdot\mid t_j)\). For~(a), integration against transition probabilities preserves joint measurability, while bounded convergence gives continuity in own action. Supermodularity, when imposed, is also preserved by integration. Lemmas~\ref{lem:choice} and~\ref{lem:measurable} give the assertions about maximizers and distance.
\par
	For~(b), Lemma~\ref{lem:stoch}(b) orders the product action distributions whenever their components are ordered. Increasing differences in \((a_i,a_{-i})\) then give the payoff comparison by integration, and Lemma~\ref{lem:choice}(b) orders the extremal maximizers. In~(d), increasing differences in \((a_i,t)\), together with stochastic monotonicity of the action distributions, make \(\Delta_i(t)\) increasing in the entire type profile. Its boundedness and measurability follow from bounded payoffs and measurable integration. For \(t_i\le t_i'\), first increasing the own type in \(\Delta_i\), then applying \(p_i(\cdot\mid t_i)\st p_i(\cdot\mid t_i')\), gives increasing differences of the interim payoff. Lemma~\ref{lem:choice}(b) yields monotonicity of both extremal best responses.
\par
	For the almost everywhere assertions in~(b) and~(c), choose measurable \(\lambda_j\)-null sets \(N_j\) outside which the respective hypotheses hold. By the compatibility condition,  the event that \(t_j\in N_j\) for some \(j\ne i\) has conditional probability zero at almost every \(t_i\). At each such own type, the conditional action integrals satisfy the required comparison or equality for conditionally almost every opponents' type profile. Integration gives the conclusions. The exceptional set of own types is independent of the own actions being compared.
\end{proof}

Before applying the iterations, we relate optimality in a perturbed game to best responses against the induced action distributions. The affine form of the trembles makes the relation explicit. This is the usual Selten perturbation calculation. We record the identities without a separate proof.

\begin{remark}\label{rem:perturbed}
 Let \(q\) be a perturbation, and let \(r\) be a behavioral strategy profile.
\begin{enumerate}[label=(\alph*)]
\item Let \(V_i^q(a_i,t_i;r_{-i})\) be player \(i\)'s interim payoff in \(G(q)\), defined by~\eqref{eq:V} with \(u_i^q\) in place of \(u_i\). For every player \(i\), type \(t_i\), and action \(a_i\),
\begin{equation}\label{eq:own}
V_i^q(a_i,t_i;r_{-i})=\varepsilon_i\int_{A_i}V_i(x_i,t_i;r_{-i}^{q_{-i}})\,\mu_i(\dd x_i\mid t_i)+(1-\varepsilon_i)V_i(a_i,t_i;r_{-i}^{q_{-i}}).
\end{equation}
Consequently, the set of maximizers of \(V_i^q(\cdot,t_i;r_{-i})\) is \(\BR_i(t_i;r_{-i}^{q_{-i}})\), and~\eqref{eq:perturbedoptimality} holds.
\item For every player \(i\), type \(t_i\), and bounded measurable function \(\phi\) on \(A_i\),
\begin{equation}\label{eq:small}
\left|\int_{A_i}\phi\dd r_i^{q_i}(\cdot\mid t_i)-\int_{A_i}\phi\dd r_i(\cdot\mid t_i)\right|\le2\varepsilon_i\|\phi\|_\infty.
\end{equation}
\item If a pure strategy \(\alpha_i\) is increasing in type, then \(\alpha_i^{q_i}\) is stochastically increasing in type.
\end{enumerate}
\end{remark}

Extremal equilibria of the perturbed games arise from the best response iterations of \citet{vives1990} and \citet{vzv}. Lemma~\ref{lem:iterate} below records their properties. Parts~(a), (b), and~(d) are those iterations in the present setting, so we only indicate the sources. Part~(c) compares the limits with arbitrary behavioral equilibria, which requires the null set adjustments, so we prove it in full.

\begin{lemma}\label{lem:iterate}
	Suppose that 	Assumption~\ref{ass:interim} holds. Let \(q\) be a perturbation.
	A pure strategy profile \(\alpha\) is increasing if
	\(\alpha_j(t_j)\le\alpha_j(t_j')\) whenever \(t_j\le t_j'\),
	for every player \(j\).
	\begin{enumerate}[label=(\alph*)]
		
		\item
		Let \(\alpha\) be increasing. For every player \(i\) and type
		\(t_i\), the set \(\BR_i(t_i;\alpha_{-i}^{q_{-i}})\) is a
		nonempty compact sublattice of \(A_i\). Its largest and smallest
		elements are measurable increasing functions of \(t_i\).
		Define the pure strategy profiles
		\(\overline{\BR}(\alpha^q)\) and
		\(\underline{\BR}(\alpha^q)\) by
		\(\bigl[\overline{\BR}(\alpha^q)\bigr]_i(t_i)
		=\overline{\BR}_i(t_i;\alpha_{-i}^{q_{-i}})\) and
		\(\bigl[\underline{\BR}(\alpha^q)\bigr]_i(t_i)
		=\underline{\BR}_i(t_i;\alpha_{-i}^{q_{-i}})\).
		If \(\beta\) is increasing with
		\(\alpha_j(t_j)\le\beta_j(t_j)\) for every \(j,t_j\), then
		\(\overline{\BR}(\alpha^q)\le\overline{\BR}(\beta^q)\) and
		\(\underline{\BR}(\alpha^q)\le\underline{\BR}(\beta^q)\)
		pointwise.
		
		\item
		Let \(\alpha^0\) be the constant profile of largest actions.
		Define
		\(\alpha^{n+1}=\overline{\BR}((\alpha^n)^q)\) for \(n\ge0\).
		Then \((\alpha^n)_{n\ge0}\) decreases pointwise to a measurable
		increasing profile \(\bar\alpha(q)\) satisfying
		\(\bar\alpha(q)=\overline{\BR}((\bar\alpha(q))^q)\) pointwise.
		Similarly, starting from the constant profile \(\beta^0\) of
		smallest actions, the sequence defined by
		\(\beta^{n+1}=\underline{\BR}((\beta^n)^q)\) increases pointwise
		to a measurable increasing profile \(\underline\alpha(q)\)
		satisfying
		\(\underline\alpha(q)
		=\underline{\BR}((\underline\alpha(q))^q)\) pointwise.
		
		\item
		Suppose, in addition, that the null compatibility condition holds.
		Then every \(r\in\EE(q)\) satisfies
		\(\delta_{\underline\alpha(q)}\st r\st\delta_{\bar\alpha(q)}\).
		Consequently, \(\bar\alpha(q)\) is the largest element of
		\(E^p(q)\), while \(\underline\alpha(q)\) is its smallest element.
		
		\item
		Consider a second game with the same action spaces, type spaces,
		and reference probabilities. Suppose that it satisfies
		Assumption~\ref{ass:interim}.
		Let \(V_i'\) denote its interim payoffs.
		For a perturbation \(q'\), let \(\bar\alpha'(q')\) and
		\(\underline\alpha'(q')\) be the profiles obtained in~(b)
		for this game.
		Suppose that
		\(a_i\mapsto
		V_i'(a_i,t_i;\beta_{-i}^{q_{-i}'})
		-V_i(a_i,t_i;\alpha_{-i}^{q_{-i}})\)
		is increasing for every \(i,t_i\) whenever \(\alpha,\beta\)
		are increasing profiles with
		\(\alpha_j(t_j)\le\beta_j(t_j)\) for every \(j,t_j\).
		Then \(\bar\alpha(q)\le\bar\alpha'(q')\) and
		\(\underline\alpha(q)\le\underline\alpha'(q')\) pointwise.
		
	\end{enumerate}
\end{lemma}

\begin{proof}
\setlength{\parskip}{0pt}
(a) Remark~\ref{rem:perturbed}(c) makes \(\alpha_{-i}^{q_{-i}}\) stochastically increasing in type. Lemma~\ref{lem:interim}(a), Assumption~\ref{ass:interim}(i), and Lemmas~\ref{lem:choice} and~\ref{lem:measurable} give a nonempty compact sublattice of maximizers with measurable extrema. Assumption~\ref{ass:interim}(iii) and Lemma~\ref{lem:choice}(b) make these extrema increasing in type. If increasing profiles satisfy \(\alpha\le\beta\) pointwise, Lemma~\ref{lem:stoch}(d),(e) orders their perturbed distributions at every type. Both induced opponents' profiles belong to \(\mathcal F^{\mathrm{st}}_{-i}\) by Remark~\ref{rem:perturbed}(c). Assumption~\ref{ass:interim}(ii), followed by Lemma~\ref{lem:choice}(b), therefore orders both extremal best responses.
\par(b) The iteration follows the Cournot tâtonnement of \citet[Theorem~6.1]{vives1990}, adapted to Bayesian games by \citet[Lemma~6]{vzv}. By~(a), the profiles \(\alpha^n\) are measurable and increasing in type, while decreasing pointwise in \(n\). Lemma~\ref{lem:compact} gives a measurable limit \(\bar\alpha(q)\), which is increasing by closedness of the order. The perturbed action distributions converge weakly at every opponents' type profile. Finite product convergence \citep[Theorem~2.8(ii)]{billingsley}, continuity of payoffs in actions, and bounded convergence allow passage to the limit in the best response inequalities. Thus the limit is a best response to its induced opponents' profile. Together with~(a), this gives
	\(\bar\alpha(q)\le\overline{\BR}((\bar\alpha(q))^q)\le\alpha^{n+1}\).
	Letting \(n\to\infty\) proves the fixed point identity. Starting from the smallest actions, the lower iteration increases to \(\underline\alpha(q)\) and gives
	\(\beta^{n+1}\le\underline{\BR}((\underline\alpha(q))^q)\le\underline\alpha(q)\).
	Taking limits yields its fixed point identity.
\par(c) The fixed point identities in~(b), together with Remark~\ref{rem:perturbed}(a), place both limits in \(E^p(q)\). Fix \(r\in\EE(q)\). We prove inductively that \(r_i(\cdot\mid t_i)\st\delta_{\alpha_i^n(t_i)}\) almost everywhere for every player. The case \(n=0\) follows from maximality of \(\alpha^0\). Suppose the claim holds for \(n\). Choose measurable null sets \(N_j\) outside which it holds, and set \(\widetilde r_j=r_j\) outside \(N_j\), with \(\widetilde r_j=\delta_{\underline a_j}\) on \(N_j\). Lemma~\ref{lem:stoch}(e) gives
	\(\widetilde r_j^{q_j}(\cdot\mid t_j)\st(\alpha_j^n)^{q_j}(\cdot\mid t_j)\)
	at every type. Since \(\alpha^n\) is increasing, \((\alpha_{-i}^n)^{q_{-i}}\in\mathcal F^{\mathrm{st}}_{-i}\) by Remark~\ref{rem:perturbed}(c). Assumption~\ref{ass:interim}(ii) applies to its comparison with \(\widetilde r_{-i}^{q_{-i}}\). Lemma~\ref{lem:interim}(c) then replaces the latter by \(r_{-i}^{q_{-i}}\) at almost every \(t_i\), so
	\(V_i(\cdot,t_i;(\alpha_{-i}^n)^{q_{-i}})-V_i(\cdot,t_i;r_{-i}^{q_{-i}})\)
	is increasing for almost every \(t_i\).
\par
	By~(a), the first objective is supermodular with largest maximizer \(\alpha_i^{n+1}(t_i)\). Lemma~\ref{lem:choice}(b) places every maximizer of the second objective below it. Since~\eqref{eq:perturbedoptimality} concentrates \(r_i(\cdot\mid t_i)\) on these maximizers, Lemma~\ref{lem:stoch}(d) completes the induction. Outside a common null set for each player, all comparisons hold simultaneously. Lemma~\ref{lem:stoch}(a) then yields \(r\st\delta_{\bar\alpha(q)}\).
\par
	For the lower bound, inductively suppose that \(\delta_{\beta_j^n(t_j)}\st r_j(\cdot\mid t_j)\) almost everywhere. Replacing \(r_j\) by \(\delta_{\overline a_j}\) on the exceptional null sets gives \((\beta_j^n)^{q_j}\st\widetilde r_j^{q_j}\) at every type. Here \((\beta_{-i}^n)^{q_{-i}}\in\mathcal F^{\mathrm{st}}_{-i}\), so Assumption~\ref{ass:interim}(ii) and Lemma~\ref{lem:interim}(c) make \(V_i(\cdot,t_i;r_{-i}^{q_{-i}})-V_i(\cdot,t_i;(\beta_{-i}^n)^{q_{-i}})\) increasing almost everywhere. The second assertion of Lemma~\ref{lem:choice}(b) places every maximizer of the first objective above \(\beta_i^{n+1}(t_i)\), the smallest maximizer of the supermodular second objective. Equation~\eqref{eq:perturbedoptimality} and Lemma~\ref{lem:stoch}(d) complete the induction, whose initial case follows from minimality of \(\beta^0\). Passing to the limit outside a common null set gives \(\delta_{\underline\alpha(q)}\st r\). Since pure equilibria are degenerate behavioral equilibria, these bounds also establish the extrema of \(E^p(q)\).
\par(d) The comparison follows the induction used in \citet[Proposition~16]{vzv}. The assumed payoff comparison and Lemma~\ref{lem:choice}(b) order both extremal best responses across the two games whenever the increasing input profiles are ordered pointwise. Their upper and lower iterations start from the same respective extreme action profiles. Induction orders the iterates, hence their pointwise limits.
\end{proof}

The next lemma uses the weak convergence and Prohorov metric criteria in \citet{billingsley}. Part~(a) is the standard criterion for convergence to a point mass. The proof treats the distance to moving best response sets in parts~(b) and~(c).

\begin{lemma}\label{lem:criterion}
 Let \((g^n)_{n\ge1}\) be a sequence of completely mixed behavioral strategy profiles.
\begin{enumerate}[label=(\alph*)]
\item For every player \(i\), every sequence \((\mu_n)_{n\ge1}\) in \(\mathcal M(A_i)\), and every \(a\in A_i\), we have \(\mu_n\Rightarrow\delta_a\) if and only if \(\int_{A_i}d_i(x,a)\,\mu_n(\dd x)\to0\).
\item Fix a player \(i\), a type \(t_i\), and \(\nu\in\mathcal M(A_i)\). Then \(\rho_i(\nu,\mathcal B_i(t_i;g_{-i}^n))\to0\) if and only if there are \(\nu^n\in\mathcal B_i(t_i;g_{-i}^n)\) with \(\nu^n\Rightarrow\nu\). For \(\nu=\delta_a\), both conditions are equivalent to \(d_i(a,\BR_i(t_i;g_{-i}^n))\to0\). Consequently, \((g^n)_{n\ge1}\) satisfies~\eqref{eq:general-behavioral-criterion} for \(r=\delta_\alpha\) if and only if it satisfies~\eqref{eq:general-criterion} for \(\alpha\).
\item If a behavioral strategy profile \(r\) and \((g^n)_{n\ge1}\) satisfy~\eqref{eq:general-behavioral-criterion}, then, for every player \(i\) and almost every \(t_i\),
\[
\int_{A_i}d_i\bigl(a_i,\BR_i(t_i;g_{-i}^n)\bigr)\,g_i^n(\dd a_i\mid t_i)\longrightarrow0.
\]
\end{enumerate}
\end{lemma}
\begin{proof}
\setlength{\parskip}{0pt}
\par
(b) Put \(C^n=\BR_i(t_i;g_{-i}^n)\). The first equivalence follows by choosing a distribution within \(1/n\) of the infimum defining the distance, since \(\rho_i\) metrizes weak convergence. If \(\nu^n(C^n)=1\) and \(\nu^n\Rightarrow\delta_a\), then \(d_i(a,C^n)\le\int d_i(x,a)\nu^n(\dd x)\to0\) by~(a). Conversely, closest points \(b^n\in C^n\) give \(\delta_{b^n}\Rightarrow\delta_a\). This also establishes the equivalence of the two perfection criteria for pure profiles.
\par
 (c) Fix \(i,t_i\) where both convergence requirements hold. Put \(C^n=\BR_i(t_i;g_{-i}^n)\). The triangle inequality gives \(\rho_i(g_i^n(\cdot\mid t_i),\mathcal B_i(t_i;g_{-i}^n))\to0\). For any \(\varepsilon>0\), eventually there is \(\nu^n\) supported on \(C^n\) with \(\rho_i(g_i^n(\cdot\mid t_i),\nu^n)<\varepsilon\). The definition of the Prohorov metric gives \(g_i^n((C^n)^\varepsilon\mid t_i)\ge1-\varepsilon\), so the integral in~(c) is at most \((1+\bar d)\varepsilon\).
\end{proof}

Limits of supporting sequences are Bayesian Nash equilibria. This is the usual continuity argument for perfection, see \citet{simonStinchcombe1995} and, for Bayesian games, \cite{bfv16} and \cite{cn21}. Lemma~\ref{lem:optimality} verifies it under null compatibility, which handles the exceptional type sets of the opponents. Its second part shows that supporting sequences under increasing perturbations also support in Definition~\ref{def:general}, so every equilibrium selected by increasing perturbations is perfect.

\begin{lemma}\label{lem:optimality}
	\begin{enumerate}[label=(\alph*)]
		\item Every \(r\in\PP^b\) is a Bayesian Nash equilibrium.
		\item Let \(r\in\PP_{\mathrm{inc}}^b\), and let \(q^n\) and \(r^n\in\EE(q^n)\) be as in Definition~\ref{def:perfect}. Then \(r_i^n(\cdot\mid t_i)\Rightarrow r_i(\cdot\mid t_i)\) for every player \(i\) and almost every \(t_i\). Moreover, the completely mixed profiles \((r^n)^{q^n}\) support \(r\) in Definition~\ref{def:general}. Consequently, \(\PP_{\mathrm{inc}}^b\subseteq\PP^b\) and \(\PP_{\mathrm{inc}}^p\subseteq\PP^p\). In particular, every element of \(\PP_{\mathrm{inc}}^b\) is a Bayesian Nash equilibrium.
	\end{enumerate}
\end{lemma}

\begin{proof}
\setlength{\parskip}{0pt}
(a) Let \(g^n\) support \(r\) in Definition~\ref{def:general}. By null compatibility, at almost every own type \(t_i\) the opponents' distributions converge weakly for \(p_i(\cdot\mid t_i)\)-almost every \(t_{-i}\). At such a type where the distance condition also holds, Lemma~\ref{lem:criterion}(b) supplies \(\nu_i^n\in\mathcal B_i(t_i;g_{-i}^n)\) with \(\nu_i^n\Rightarrow r_i(\cdot\mid t_i)\). The product measures \(\nu_i^n\otimes\bigotimes_{j\ne i}g_j^n(\cdot\mid t_j)\) then converge weakly at conditionally almost every opponents' type profile by \citet[Theorem~2.8(ii)]{billingsley}. Continuity of payoffs in actions gives convergence of the action integrals, and bounded convergence over types passes the best response inequalities of \(\nu_i^n\) to the limit. Hence
	\[
	\int_{A_i}V_i(b_i,t_i;r_{-i})\,r_i(\dd b_i\mid t_i)\ge V_i(a_i,t_i;r_{-i})\qquad\text{for every }a_i\in A_i,
	\]
	with an exceptional set of own types that does not depend on \(a_i\). Since the interim payoff is continuous in own action, \(r_i(\cdot\mid t_i)\) is concentrated on its maximizers.
\par
	(b) Put \(g^n=(r^n)^{q^n}\). Full support of the trembles makes \(g^n\) completely mixed, while~\eqref{eq:small} gives \(r_i^n(\cdot\mid t_i)\) the same weak limit as \(g_i^n(\cdot\mid t_i)\). Outside a common null set, Remark~\ref{rem:perturbed}(a) gives \(r_i^n(\cdot\mid t_i)\in\mathcal B_i(t_i;g_{-i}^n)\) for every \(n\). Therefore
	\[
	\rho_i\bigl(r_i(\cdot\mid t_i),\mathcal B_i(t_i;g_{-i}^n)\bigr)\le\rho_i\bigl(r_i(\cdot\mid t_i),r_i^n(\cdot\mid t_i)\bigr)\longrightarrow0,
	\]
	so \(g^n\) supports \(r\) in Definition~\ref{def:general}. The inclusions follow, and part~(a) gives the equilibrium property.
\end{proof}

The construction below supplies the comparison step in the proof of Theorem~\ref{thm:general-global}. It uses the fixed point theorem of \citet{tarski} and the best response comparisons of Lemma~\ref{lem:interim} to obtain a pure best response \(z\) to a completely mixed profile \(r\). The profile \(z\) dominates \(s\) and all best responses to \(h\), while~\eqref{eq:gluing-estimate} controls \(R(z,r)\) by \(R(s,g)+e(h)\) and the mixing weight.

\begin{lemma}\label{lem:glue}
	Suppose that Assumption~\ref{ass:main} holds. For a pure strategy profile \(s\) and a behavioral strategy profile \(g\), let
	\begin{equation}\label{eq:discrepancy-functionals}
		\begin{aligned}
			R(s,g)&=\sum_{i\in I}\int_{T_i}\int_{A_i}d_i\bigl(a_i,s_i(t_i)\bigr)\,g_i(\dd a_i\mid t_i)\,\lambda_i(\dd t_i),\\
			e(g)&=\sum_{i\in I}\int_{T_i}\int_{A_i}d_i\bigl(a_i,\BR_i(t_i;g_{-i})\bigr)\,g_i(\dd a_i\mid t_i)\,\lambda_i(\dd t_i).
		\end{aligned}
	\end{equation}
	Let \(\bar d>0\) be the common bound on the action metrics. For \(\varepsilon\ge0\), define
	\[
	w(\varepsilon)=\sup\bigl\{d_i(a\vee x\vee y,\,a)\bigm|\, i\in I,\ a,b,c,x,y\in A_i,\ b\le a,\ c\le a,\ d_i(x,b)+d_i(y,c)\le\varepsilon\bigr\},
	\]
	and let \(\Phi(\tau)=Nw(\sqrt{\tau})+\bar d\sqrt{\tau}\) for \(\tau\ge0\).
	\begin{enumerate}[label=(\alph*)]
		\item The function \(\Phi\) is nondecreasing with \(\Phi(0)=0\), and \(\Phi(\tau)\to0\) as \(\tau\downarrow0\).
		\item Let \(g\) and \(h\) be completely mixed behavioral strategy profiles. Let \(s\) be a pure strategy profile with \(s_i(t_i)\le\overline{\BR}_i(t_i;g_{-i})\) for every player \(i\) and almost every \(t_i\). For every \(\eta\in(0,1)\), there exist a pure strategy profile \(z\) and a completely mixed behavioral strategy profile \(r\) such that, for every player \(i\) and almost every \(t_i\),
		\[
		z_i(t_i)\in\BR_i(t_i;r_{-i}),\qquad s_i(t_i)\le z_i(t_i),\qquad
		b\le z_i(t_i)\ \text{for every }b\in\BR_i(t_i;h_{-i}),
		\]
		and
		\begin{equation}\label{eq:gluing-estimate}
			R(z,r)\le\Phi\bigl(R(s,g)+e(h)\bigr)+N\bar d\eta.
		\end{equation}
	\end{enumerate}
\end{lemma}

\begin{proof}
\setlength{\parskip}{0pt}
(a) Both integrands in~\eqref{eq:discrepancy-functionals} are jointly measurable by continuity of the action metrics and Lemma~\ref{lem:interim}(a), so \(R\) and \(e\) are well defined. The defining sets make \(w\) nondecreasing with \(0\le w\le\bar d\). Since \(a\vee b\vee c=a\) whenever \(b,c\le a\), we have \(w(0)=0\). Uniform continuity of the join on each compact \(A_i^3\), together with finiteness of \(I\), gives \(w(\varepsilon)\to0\) as \(\varepsilon\downarrow0\). The assertions about \(\Phi\) follow.
\par(b)
We construct completely mixed profiles using joins, obtain
\(z\) by a fixed point argument, then establish the estimate.
\par Step 1.
For each \(i,t_i\), let \(X_i,Y_i\) be independent random
actions with distributions
\(g_i(\cdot\mid t_i),h_i(\cdot\mid t_i)\).
Let \(\xi_i(\cdot\mid t_i)\) be the distribution of
\(X_i\vee Y_i\). For a pure profile \(v\), define
\(r_i^v(\cdot\mid t_i)\) as the mixture assigning weight
\(1-\eta\) to the distribution of
\(v_i(t_i)\vee X_i\vee Y_i\) and weight \(\eta\) to
\(\xi_i(\cdot\mid t_i)\).
Measurability of product distributions and continuity of
the join make these behavioral strategies.
\par
Each \(\xi_i(\cdot\mid t_i)\) has full support.
For any open \(U\ni a\), continuity at \((a,a)\) gives an
open \(O\ni a\) with \(O\vee O\subseteq U\). Hence
\(\xi_i(U\mid t_i)\ge
g_i(O\mid t_i)h_i(O\mid t_i)>0\).
Since \(\eta>0\), every \(r_i^v\) is completely mixed.
Lemma~\ref{lem:stoch}(c),(e) gives
\(g_i(\cdot\mid t_i)\st\xi_i(\cdot\mid t_i)
\st r_i^v(\cdot\mid t_i)\) and
\(h_i(\cdot\mid t_i)\st\xi_i(\cdot\mid t_i)\).
It also gives
\(r_i^v(\cdot\mid t_i)\st r_i^{v'}(\cdot\mid t_i)\)
whenever \(v_i(t_i)\le v_i'(t_i)\).
\par Step 2.
On the complete lattice \(\mathcal Q\) of
Lemma~\ref{lem:space}, define
\(\Psi(v)_i=\overline{\BR}_i(\cdot;r_{-i}^v)\).
Lemma~\ref{lem:interim}(a),(c) makes \(\Psi\) well defined
on equivalence classes, while part~(b) makes it increasing.
The fixed point theorem of \citet{tarski} yields a fixed
point. Choose a measurable representative \(z\) and put
\(r=r^z\). Then
\(z_i(t_i)=\overline{\BR}_i(t_i;r_{-i})\)
almost everywhere.
\par
Since \(r\) dominates \(g\) and \(h\),
Lemma~\ref{lem:interim}(b) gives
\(s_i(t_i)\le\overline{\BR}_i(t_i;g_{-i})\le z_i(t_i)\)
and
\(\overline{\BR}_i(t_i;h_{-i})\le z_i(t_i)\)
almost everywhere. This proves the order assertions.
\par Step 3.
Fix \(\varepsilon>0\), a player \(i\), and a type \(t_i\)
in the full measure set where the preceding comparisons
hold. Write \(C_i=\BR_i(t_i;h_{-i})\), and put
\(\Delta_i=d_i(X_i,s_i(t_i))+d_i(Y_i,C_i)\),
\(D_i=d_i(z_i(t_i)\vee X_i\vee Y_i,z_i(t_i))\), and
\(m_i(t_i)=\mathbb E[\Delta_i]\).
All expectations are at the fixed type \(t_i\).
\par
On \(\{\Delta_i\le\varepsilon\}\), choose \(c_i\in C_i\)
attaining \(d_i(Y_i,C_i)\). Since
\(s_i(t_i),c_i\le z_i(t_i)\), the definition of \(w\)
gives \(D_i\le w(\varepsilon)\).
On the complementary event, \(D_i\le\bar d\).
Markov's inequality therefore yields
\[
\mathbb E[D_i]
\le w(\varepsilon)+\bar d\,\mathbb P(\Delta_i>\varepsilon)
\le w(\varepsilon)+\tfrac{\bar d}{\varepsilon}m_i(t_i).
\]
\par
The mixture defining \(r\) consequently satisfies
\[
\int_{A_i}d_i(a_i,z_i(t_i))\,r_i(\dd a_i\mid t_i)
\le(1-\eta)\mathbb E[D_i]+\eta\bar d
\le w(\varepsilon)+\tfrac{\bar d}{\varepsilon}m_i(t_i)
+\eta\bar d.
\]
\par
By the distributions of \(X_i,Y_i\),
\(\sum_i\int_{T_i}m_i(t_i)\lambda_i(\dd t_i)
=R(s,g)+e(h)\).
Integrating over types and summing over players gives
\[
R(z,r)\le
Nw(\varepsilon)
+\tfrac{\bar d}{\varepsilon}\bigl[R(s,g)+e(h)\bigr]
+N\bar d\eta.
\]
\par
Let \(\tau=R(s,g)+e(h)\).
If \(\tau>0\), take \(\varepsilon=\sqrt{\tau}\) to obtain
\eqref{eq:gluing-estimate}.
If \(\tau=0\), let \(\varepsilon\downarrow0\).
The profiles \(z,r\) do not depend on \(\varepsilon\), so
the same estimate follows from \(w(\varepsilon)\to0\).
\end{proof}

The arguments below first construct upper equilibrium bounds. Lower bounds will then follow by reversing the relevant orders. We record the consequences of this familiar order duality for the assumptions and refinements used here.

\begin{remark}\label{rem:dual}
Consider the game obtained from a given Bayesian game by reversing the order on every action space, while retaining the metrics, type spaces, beliefs, reference probabilities, and payoff functions.
\begin{enumerate}[label=(\alph*)]
\item Each action space with the reversed order is a compact metric lattice. The new game satisfies Assumption~\ref{ass:main} whenever the original game does. The two games have the same sets \(\PP^b\) and \(\PP^p\). The comparison \(r\st r'\) holds in the new game if and only if \(r'\st r\) holds in the original game.
\item Suppose that the orders on every type space \(T_i\) and on \(T_0\) are also reversed. Then the new game satisfies Assumption~\ref{ass:types} whenever the original game does. A pure strategy is monotone in the new game if and only if it is monotone in the original game. Likewise, a behavioral strategy is stochastically increasing in type in the new game if and only if it is so in the original game.
\item Suppose that the payoff functions depend on a parameter \(\theta\) in a partially ordered set and have increasing differences in \((a_i,\theta)\). This property is preserved when the orders on every action space and on the parameter set are reversed.
\end{enumerate}
\end{remark}

\subsection{Proofs for Sections~\ref{sec:introduction}, \ref{sec:results}, and~\ref{sec:discussion}}\label{app:sec4}\label{app:sec5}
We first analyze the market entry game of the Introduction. We then prove Proposition~\ref{prop:interim} under the interim conditions. Lemma~\ref{lem:interim} then gives Theorem~\ref{thm:main},  and Propositions~\ref{prop:common} and~\ref{prop:closure} under the primitive conditions stated in Section~\ref{sec:model}.

\subsubsection{The market entry game}\label{app:entry-continuous}

We formalize the market entry game of the Introduction. The private types \(t_1\) and \(t_2\) are independent and uniformly distributed on \(T_i=[0,1]\), which carries the usual order. The space of residual states is a singleton. The common prior determines the reference probabilities and interim beliefs. Thus \(\lambda_i\) is uniform on \([0,1]\), while \(p_i(\cdot\mid t_i)=\lambda_j\) for every \(t_i\), where \(j\ne i\). We identify each action with the constant strategy that prescribes it at every type.
 Let \(f(t_i)=\mathbf 1_{\{t_i>1/2\}}\), and define \(s^A=(1,0)\), \(s^B=(0,1)\), and \(s^C=(f,f)\). 
The following claim collects the properties of this game that we use in the main text.

\begin{claim}\label{claim:entry}
The market entry game has the following properties.
\begin{enumerate}[label=(\alph*)]
\item It satisfies every condition in Assumptions~\ref{ass:main} and~\ref{ass:types} except the increasing differences in \((a_i,a_{-i})\) required by Assumption~\ref{ass:main}. Payoffs have strictly increasing differences in \((a_i,t_i)\), and the game satisfies the hypotheses of the existence theorems of \citet{athey}, \citet{mcadams}, and \citet{reny}. Conditions~(i) and~(iii) of Assumption~\ref{ass:interim} hold even against all behavioral strategies, whereas condition~(ii) fails.
\item Up to equality almost everywhere, the pure Bayesian Nash equilibria are \(s^A\), \(s^B\), and \(s^C\). The behavioral Bayesian Nash equilibria are \(\delta_{s^A}\), \(\delta_{s^B}\), and \(\delta_{s^C}\). All three pure equilibria are monotone.
\item All three equilibria are perfect under both refinements, and
\[
\PP_{\mathrm{inc}}^p=\PP^p=\{s^A,s^B,s^C\},\PP_{\mathrm{inc}}^b=\PP^b=\{\delta_{s^A},\delta_{s^B},\delta_{s^C}\}.
\]
In particular, \(s^A\), \(s^B\), and \(s^C\) are perfect monotone equilibria in the sense of \citet{heSunXu2026}.
\item The profiles \(s^A\), \(s^B\), and \(s^C\) are pairwise incomparable. Consequently, the set of monotone pure Bayesian Nash equilibria has neither a largest nor a smallest element. The same holds for \(\PP_{\mathrm{inc}}^p\), \(\PP_{\mathrm{inc}}^b\), \(\PP^p\), and \(\PP^b\).
\end{enumerate}
\end{claim}

\begin{proof}
	
\noindent(a) Fix a behavioral strategy \(r_j\) of the other firm, and let \(m_j=\int_0^1 r_j(\{1\}\mid t_j)\,\dd t_j\) be its ex ante entry probability. Independence of types gives
	\begin{equation}\label{eq:entry-gain}
		V_i(1,t_i;r_j)-V_i(0,t_i;r_j)=1+t_i-3m_j.
	\end{equation}
	This gain is strictly increasing in own type and decreasing in the opponent's entry probability. Supermodularity in own action is automatic on the binary action chain. Boundedness, measurability, continuity in actions, constant beliefs, and null compatibility follow directly from the primitives. Thus every condition in Assumptions~\ref{ass:main} and~\ref{ass:types} holds except strategic increasing differences. Equation~\eqref{eq:entry-gain} also verifies the interim single crossing conditions used by \citet{athey}, \citet{mcadams}, and \citet{reny}. Types are independent and atomless, with one dimension. Actions are finite, so the remaining regularity conditions in their existence theorems hold. The same calculation establishes conditions~(i) and~(iii) of Assumption~\ref{ass:interim} against every behavioral strategy. Condition~(ii) fails already between the constant opponent strategies \(\delta_0\) and \(\delta_1\), both of which are stochastically increasing in type. The gain also verifies the corresponding incentive conditions of \citet{heSunXu2026}.

\par\addvspace{\baselineskip}\noindent(b) Consider any behavioral Bayesian Nash equilibrium \(r\). Write \(\theta_i(t_i)=r_i(\{1\}\mid t_i)\), so that \(m_i=\int_0^1\theta_i(t_i)\,\dd t_i\). By~\eqref{eq:entry-gain}, optimality requires \(\theta_i(t_i)=0\) when \(t_i<3m_j-1\) and \(\theta_i(t_i)=1\) when \(t_i>3m_j-1\), for almost every \(t_i\). A nondegenerate mixture can be optimal only at the indifferent type \(t_i=3m_j-1\), if this type lies in \([0,1]\). Since the type distribution is atomless, this singleton has probability zero. Hence every behavioral equilibrium is pure almost everywhere.
	
	Integrating these best responses over types gives \(m_1=\beta(m_2)\) and \(m_2=\beta(m_1)\), where \(\beta(m)=\min\{1,\max\{0,2-3m\}\}\). If \(m_2\le1/3\), then \(m_1=1\), which implies \(m_2=0\). If \(m_2\ge2/3\), then \(m_1=0\), which implies \(m_2=1\). The same argument applies with the players interchanged. Otherwise, both probabilities lie in \((1/3,2/3)\), so the equations become \(m_1=2-3m_2\) and \(m_2=2-3m_1\). Their unique solution is \(m_1=m_2=1/2\).
	
	In the last case, the gain from entry is \(t_i-1/2\). Thus optimality requires \(\theta_i(t_i)=f(t_i)\) almost everywhere, rather than merely requiring its average to equal \(1/2\). Mixing at \(t_i=1/2\) does not change the strategy up to equality almost everywhere. The associated equilibrium profiles are therefore precisely \(s^A,s^B,s^C\), with behavioral representations \(\delta_{s^A},\delta_{s^B},\delta_{s^C}\). Each pure profile is increasing in type.

\par\addvspace{\baselineskip}\noindent(c) Give both firms a common tremble probability \(0<\varepsilon<2/3\), with \(\mu_i(\cdot\mid t_i)=\tfrac12\delta_0+\tfrac12\delta_1\) at every type. These distributions have full support and are stochastically increasing in type because they are constant. A strategy with average entry probability \(m_i\) induces actual entry probability \((1-\varepsilon)m_i+\varepsilon/2\). By Remark~\ref{rem:perturbed}, optimality in the perturbed game can be checked using the original interim payoff against the opponent's induced strategy.
	
	Under \(s^A\), the actual entry probabilities are \(1-\varepsilon/2\) for firm~1 and \(\varepsilon/2\) for firm~2. Equation~\eqref{eq:entry-gain} gives firm~1's gain from entry as \(1+t_1-3\varepsilon/2>0\) at every type, and firm~2's gain as \(t_2-2+3\varepsilon/2<0\) at every type. Hence \(s^A\) remains an equilibrium of the perturbed game. By symmetry, the same holds for \(s^B\).
	
	Under \(s^C\), both actual entry probabilities remain \((1-\varepsilon)/2+\varepsilon/2=1/2\). The gain from entry is therefore \(t_i-1/2\), so \(s^C\) also remains an equilibrium. Letting \(\varepsilon\downarrow0\), the induced behavioral strategies converge to the corresponding pure profiles at every type. Thus all three equilibria are perfect under increasing perturbations.
	
	Lemma~\ref{lem:optimality} implies that perfection under increasing perturbations entails perfection, while every perfect equilibrium is a Bayesian Nash equilibrium. Part~(b) therefore excludes any additional equilibrium under either refinement, giving the asserted equalities. The same supporting sequences satisfy the exact best response requirement in \citet[Definition~5]{heSunXu2026}, so all three profiles are perfect monotone equilibria in their sense.

\par\addvspace{\baselineskip}\noindent(d) The profiles \(s^A\) and \(s^B\) prescribe opposite strict rankings for the two firms. Relative to \(s^C\), the profile \(s^A\) prescribes a higher action for firm~1 at types below \(1/2\), but a lower action for firm~2 at types above \(1/2\). The comparison between \(s^B\) and \(s^C\) is symmetric. These type sets have positive probability, so the three profiles are pairwise incomparable. Their degenerate behavioral representations are likewise incomparable in stochastic order. Since they exhaust the respective equilibrium sets, none of these sets has a largest or smallest element.
\end{proof}

\subsubsection{Existence under interim complementarities}\label{app:increasing-existence}

The proof proceeds in four steps. Step 1 constructs common upper and lower perturbations. Step 2 obtains monotone pure perfect equilibria as limits of extremal equilibria of perturbed games. Step 3 shows that these limits bound every behavioral perfect equilibrium under increasing perturbations. Step 4 establishes closedness under convergence in measure.

\begin{proof}[Proof of Proposition~\ref{prop:interim}]
	We maintain Assumption~\ref{ass:interim}. Lemma~\ref{lem:iterate} supplies the increasing extremal pure equilibria \(\bar\alpha(q)\) and \(\underline\alpha(q)\) of each perturbed game, together with their bounds on all behavioral equilibria. 

\par\addvspace{\baselineskip}\noindent Step 1.
	Fix a player and suppress the player index. For \(k=1,2\), write \(q^k=(\varepsilon_k,\mu_k)\). Set \(\underline\varepsilon=\min\{\varepsilon_1,\varepsilon_2\}\) and \(\overline\varepsilon=\max\{\varepsilon_1,\varepsilon_2\}\). At each type, draw independent actions from \(\mu_1\) and \(\mu_2\). Let \(\mu^\vee\) be the distribution of their join, with \(\mu^\wedge\) defined analogously using their meet. Products of conditional distributions, followed by the continuous lattice operations, give measurable conditional distributions. Both have full support. Indeed, for every neighborhood \(U\) of an action \(a\), continuity provides a neighborhood \(W\) of \(a\) satisfying \(W\vee W\subseteq U\) and \(W\wedge W\subseteq U\). Each original distribution assigns positive probability to \(W\), so both new distributions assign positive probability to \(U\). Lemma~\ref{lem:stoch}(b),(c) shows that they are stochastically increasing in type.

	For the upper perturbation, define \(\mu^+=(\underline\varepsilon/ \overline{\varepsilon})\mu^\vee+(1-\underline\varepsilon/\overline{\varepsilon})\delta_{\overline a}\) and \(q^+=(\overline{\varepsilon},\mu^+)\). Conditional on a tremble, the coefficient \(\underline\varepsilon/\overline{\varepsilon}\) is the weight on the join distribution. The remaining weight is placed on the largest action. Similarly, for the lower perturbation, define \(\mu^-=(\underline\varepsilon/\overline{\varepsilon})\mu^\wedge+(1-\underline\varepsilon/\overline{\varepsilon})\delta_{\underline a}\) and \(q^-=(\overline{\varepsilon},\mu^-)\). 
	All mixture weights are constant across types. Each conditional tremble distribution is therefore stochastically increasing, with full support inherited from its join or meet component. Thus both perturbations are admissible. They are type independent whenever the original tremble distributions are type independent.\footnote{Thus the construction remains valid when \(\mu_i^n(\cdot\mid t_i)=\mu_i^n\) at every \(n\), with \(\mu_i^n\) allowed to vary along the supporting sequence. Section~\ref{oa:fixed} explains why its conclusions also hold under this restriction. Prescribing a single profile \(\mu\) for the entire sequence is a further restriction.}

	For any action \(a\), decompose \(a^{q^k}\) into the four nonnegative components \((1-\bar\varepsilon)\delta_a\), \(\underline\varepsilon\mu_k\), \((\varepsilon_k-\underline\varepsilon)\mu_k\), and \((\bar\varepsilon-\varepsilon_k)\delta_a\). The last two components have total mass \(\bar\varepsilon-\underline\varepsilon\). To obtain an upper bound, replace \(\mu_k\) in the second component by \(\mu^\vee\) and place the last two components at \(\overline a\). Since \(\mu_k\st\mu^\vee\), the resulting distribution is \(a^{q^+}\), so \(a^{q^k}(\cdot\mid t)\st a^{q^+}(\cdot\mid t)\). To obtain a lower bound, replace \(\mu_k\) in the second component by \(\mu^\wedge\) and place the last two components at \(\underline a\). Since \(\mu^\wedge\st\mu_k\), this gives \(a^{q^-}(\cdot\mid t)\st a^{q^k}(\cdot\mid t)\). Both comparisons hold for \(k=1,2\) at every type. Applying these constructions to each player yields \(\|q^+\|=\|q^-\|=\max\{\|q^1\|,\|q^2\|\}\).

	For increasing profiles satisfying \(\alpha\le\beta\) at every type, the same comparisons give
	\(\alpha_j^{q_j^k}\st\beta_j^{q_j^+}\) and
	\(\alpha_j^{q_j^-}\st\beta_j^{q_j^k}\) at every type.
	All induced opponents' profiles in these comparisons belong to \(\mathcal F^{\mathrm{st}}_{-i}\) by Remark~\ref{rem:perturbed}(c), since \(q^k,q^+,q^-\) are admissible.
	Assumption~\ref{ass:interim}(ii) and Lemma~\ref{lem:iterate}(d) therefore imply
	\(\bar\alpha(q^+)\ge\bar\alpha(q^1)\vee\bar\alpha(q^2)\) and
	\(\underline\alpha(q^-)\le\underline\alpha(q^1)\wedge\underline\alpha(q^2)\).
	This proves the conclusion of Proposition~\ref{prop:common}.

\par\addvspace{\baselineskip}\noindent Step 2.
	For \(n\ge2\), let
	\(\Gamma_n^+=\{\bar\alpha(q)\mid\|q\|<1/n\}\) and
	\(\Gamma_n^-=\{\underline\alpha(q)\mid\|q\|<1/n\}\).
	These families are nonempty. Step 1 makes \(\Gamma_n^+\) upward directed and \(\Gamma_n^-\) downward directed. Lemma~\ref{lem:directed} supplies increasing measurable representatives \(H_n\) of the smallest essential upper bounds of \(\Gamma_n^+\), together with increasing measurable representatives \(L_n\) of the largest essential lower bounds of \(\Gamma_n^-\).
	
	The inclusion \(\Gamma_{n+1}^+\subseteq\Gamma_n^+\) implies \(H_{n+1}\le H_n\) almost everywhere. Similarly, \(\Gamma_{n+1}^-\subseteq\Gamma_n^-\) implies \(L_n\le L_{n+1}\) almost everywhere. Define
	\begin{equation}\label{eq:increasing-extrema-construction}
	\alpha_{\mathrm{inc}}^+=\bigwedge_{n\ge2}H_n,\alpha_{\mathrm{inc}}^-=\bigvee_{n\ge2}L_n
	\end{equation}
	at every type. Both profiles are measurable and increasing. Outside one null set, Lemma~\ref{lem:compact} gives \(H_n\to\alpha_{\mathrm{inc}}^+\) and \(L_n\to\alpha_{\mathrm{inc}}^-\).
	By Lemma~\ref{lem:directed}, choose
	\(\bar\alpha(q^n)\in\Gamma_n^+\) satisfying
	\(d_{\mathcal Q}(\bar\alpha(q^n),H_n)<1/n\), where \(d_{\mathcal Q}\) is defined in~\eqref{eq:strategy-functionals}.
	Bounded convergence gives
	\(d_{\mathcal Q}(H_n,\alpha_{\mathrm{inc}}^+)\to0\).
	The triangle inequality therefore yields
	\(d_{\mathcal Q}(\bar\alpha(q^n),\alpha_{\mathrm{inc}}^+)\to0\).
	Passing to a subsequence, these equilibria converge almost everywhere to \(\alpha_{\mathrm{inc}}^+\).
	
	Since \(\|q^n\|\to0\), equation~\eqref{eq:small} gives
	\((\bar\alpha_i(q^n))^{q_i^n}(\cdot\mid t_i)
	\Rightarrow\delta_{\alpha_{\mathrm{inc},i}^+(t_i)}\)
	for every player \(i\) and almost every type \(t_i\).
	Hence \(\alpha_{\mathrm{inc}}^+\in\PP_{\mathrm{inc}}^p\), supported by largest equilibria of perturbed games. Applying the same argument to \(\Gamma_n^-\) gives
	\(\alpha_{\mathrm{inc}}^-\in\PP_{\mathrm{inc}}^p\), supported by smallest equilibria.

\par\addvspace{\baselineskip}\noindent Step 3.
	Take any \(r\in\PP_{\mathrm{inc}}^b\), supported by equilibria
	\(r^k\in\EE(\widehat q^k)\) with \(\|\widehat q^k\|\to0\).
	Lemma~\ref{lem:iterate}(c) bounds these arbitrary behavioral equilibria by comparisons with the monotone upper and lower iterations.
	Lemma~\ref{lem:optimality}(b) gives
	\(r_i^k(\cdot\mid t_i)\Rightarrow r_i(\cdot\mid t_i)\)
	almost everywhere.
	Fix \(n\). For all sufficiently large \(k\), we have
	\(\|\widehat q^k\|<1/n\).
	The equilibrium bounds in Lemma~\ref{lem:iterate}(c), together with the definitions of \(L_n,H_n\), imply
	\(\delta_{L_{n,i}(t_i)}\st r_i^k(\cdot\mid t_i)
	\st\delta_{H_{n,i}(t_i)}\)
	almost everywhere.
	
	There are only countably many pairs \((n,k)\), so these comparisons hold outside one common null set. Letting \(k\to\infty\), closedness of stochastic order gives
	\(\delta_{L_{n,i}(t_i)}\st r_i(\cdot\mid t_i)
	\st\delta_{H_{n,i}(t_i)}\).
	Letting \(n\to\infty\) then gives \(\delta_{\alpha_{\mathrm{inc}}^-}\st r\st\delta_{\alpha_{\mathrm{inc}}^+}\).
		Step 2 shows that both bounding profiles belong to \(\PP_{\mathrm{inc}}^p\). They are therefore the smallest and largest behavioral perfect equilibria under increasing perturbations, proving the conclusion of Theorem~\ref{thm:main}. In particular, they provide monotone pure lower and upper bounds within the refinement for any two elements of \(\PP_{\mathrm{inc}}^b\). By Lemma~\ref{lem:stoch}(d), this equilibrium set is a singleton exactly when its extrema coincide almost everywhere. 

\par\addvspace{\baselineskip}\noindent Step 4.
	Suppose that \(\alpha^k\in\PP_{\mathrm{inc}}^p\) converges in measure to \(\alpha\).
	Use the functional \(R\) defined in~\eqref{eq:discrepancy-functionals}. It is well defined for measurable profiles with bounded action metrics, including under the present assumptions.
	Since the action metrics are bounded, choose increasing indices \(k_n\) such that
	\(d_{\mathcal Q}(\alpha^{k_n},\alpha)<2^{-n}\).
	Definition~\ref{def:perfect}, Lemma~\ref{lem:criterion}(a), and bounded convergence allow us to select
	\(r^n\in\EE(q^n)\) satisfying
	\(\|q^n\|<1/n\) and
	\(R(\alpha^{k_n},(r^n)^{q^n})<2^{-n}\).
	
	The triangle inequality gives
	\(R(\alpha,(r^n)^{q^n})<2^{1-n}\).
	Consequently, \(\sum_nR(\alpha,(r^n)^{q^n})<\infty\).
	Tonelli's theorem implies
	\(\int_{A_i}d_i(a_i,\alpha_i(t_i))\,
	(r_i^n)^{q_i^n}(\dd a_i\mid t_i)\to0\)
	for every player \(i\) and almost every type \(t_i\).
	By Lemma~\ref{lem:criterion}(a), this is equivalent to
	\((r_i^n)^{q_i^n}(\cdot\mid t_i)
	\Rightarrow\delta_{\alpha_i(t_i)}\).
	Thus \(\alpha\in\PP_{\mathrm{inc}}^p\), proving the conclusion of Proposition~\ref{prop:closure}.
\end{proof}

The proof of Proposition~\ref{prop:interim} applies when conditional tremble distributions are required to be type independent. This class contains arbitrarily small perturbations and is closed under the common upper and lower constructions. The comparison argument below also applies. The explicit supports for perfection under increasing perturbations in Proposition~\ref{prop:finite-type-endpoints} and the examples use type independent tremble distributions. The remaining example conclusions follow from the same nonemptiness and exclusion arguments.

\subsubsection{Comparative statics and applications}\label{app:increasing-comparison}
The proof proceeds in three steps. Step 1 compares extremal equilibria under each fixed perturbation. Step 2 uses the essential bounds from the proof of Proposition~\ref{prop:interim} to compare extremal perfect equilibria. Step 3 verifies the comparison condition for Corollary~\ref{cor:comparative}.

\begin{proof}[Interim comparative statics and Corollary~\ref{cor:comparative}]
\noindent Step 1.
Let \(\theta\le\theta'\) index two games satisfying the hypotheses and the additional comparison condition in Section~\ref{sec:discussion}. Fix a common admissible perturbation \(q\) and increasing profiles \(\alpha,\beta\) with \(\alpha\le\beta\) at every type. Remark~\ref{rem:perturbed}(c) places both \(\alpha_{-i}^{q_{-i}}\) and \(\beta_{-i}^{q_{-i}}\) in \(\mathcal F^{\mathrm{st}}_{-i}\). Lemma~\ref{lem:stoch} gives \(\alpha_j^{q_j}\st\beta_j^{q_j}\) at every type. Decompose the difference in interim payoffs as
\begin{align*}
V_i^{\theta'}(\cdot,t_i;\beta_{-i}^{q_{-i}})-V_i^\theta(\cdot,t_i;\alpha_{-i}^{q_{-i}})
&=\bigl[V_i^{\theta'}(\cdot,t_i;\beta_{-i}^{q_{-i}})-V_i^{\theta'}(\cdot,t_i;\alpha_{-i}^{q_{-i}})\bigr]
\\&\quad+\bigl[V_i^{\theta'}(\cdot,t_i;\alpha_{-i}^{q_{-i}})-V_i^\theta(\cdot,t_i;\alpha_{-i}^{q_{-i}})\bigr],
\end{align*}
where the first bracket is increasing in own action by Assumption~\ref{ass:interim}(ii). The second bracket is increasing in own action by the additional comparison condition. Lemma~\ref{lem:iterate}(d) therefore gives \(\bar\alpha(q;\theta)\le\bar\alpha(q;\theta')\) and \(\underline\alpha(q;\theta)\le\underline\alpha(q;\theta')\) pointwise, where these profiles are the largest and smallest equilibria of the corresponding perturbed games.

\par\addvspace{\baselineskip}\noindent Step 2.
For \(n\ge2\) and \(\vartheta\in\{\theta,\theta'\}\), define \(\Gamma_n^+(\vartheta)=\{\bar\alpha(q;\vartheta)\mid\|q\|<1/n\}\) and \(\Gamma_n^-(\vartheta)=\{\underline\alpha(q;\vartheta)\mid\|q\|<1/n\}\). Apply Step 2 of the proof of Proposition~\ref{prop:interim} in Section~\ref{app:increasing-existence} separately to each game. In that construction, \(H_n(\vartheta)\) represents the smallest essential upper bound of \(\Gamma_n^+(\vartheta)\), while \(L_n(\vartheta)\) represents the largest essential lower bound of \(\Gamma_n^-(\vartheta)\). These increasing measurable representatives are supplied by Lemma~\ref{lem:directed}. The essential bounds are taken componentwise with respect to the common reference probabilities \(\lambda_i\).

For every admissible \(q\) with \(\|q\|<1/n\), Step 1 gives \(\bar\alpha(q;\theta)\le\bar\alpha(q;\theta')\le H_n(\theta')\) almost everywhere. Thus \(H_n(\theta')\) is an essential upper bound of \(\Gamma_n^+(\theta)\). Since \(H_n(\theta)\) is its smallest essential upper bound, \(H_n(\theta)\le H_n(\theta')\) almost everywhere. Similarly, \(L_n(\theta)\le\underline\alpha(q;\theta)\le\underline\alpha(q;\theta')\) almost everywhere for every such \(q\). Hence \(L_n(\theta)\) is an essential lower bound of \(\Gamma_n^-(\theta')\), which implies \(L_n(\theta)\le L_n(\theta')\) almost everywhere.

The construction in~\eqref{eq:increasing-extrema-construction} applies separately to each game. Since there are only countably many \(n\), the comparisons of the essential bounds hold outside one common null set for each player. Taking the corresponding meets and joins therefore yields \(\alpha_{\mathrm{inc}}^+(\theta)\le\alpha_{\mathrm{inc}}^+(\theta')\) and \(\alpha_{\mathrm{inc}}^-(\theta)\le\alpha_{\mathrm{inc}}^-(\theta')\) almost everywhere.

\par\addvspace{\baselineskip}\noindent Step 3.
Under the hypotheses of Corollary~\ref{cor:comparative}, Lemma~\ref{lem:interim} verifies Assumption~\ref{ass:interim} in both games. It remains to check the additional comparison condition.

\par\addvspace{\baselineskip}\noindent(i) Fix \(h_{-i}\in\mathcal F^{\mathrm{st}}_{-i}\), \(t_i\), and \(a_i\le b_i\). Let \(\Delta_i(t_i,t_{-i})\) be the expected payoff gain from choosing \(b_i\) instead of \(a_i\), integrating opponents' actions against \(h_{-i}\) as in Lemma~\ref{lem:interim}(d). That lemma shows that \(\Delta_i(t_i,\cdot)\) is bounded, measurable, and increasing on \(T_{-i}\). The stochastic order \(p_i(\cdot\mid t_i)\st p_i'(\cdot\mid t_i)\) therefore gives \(\int\Delta_i(t_i,t_{-i})\,p_i(\dd t_{-i}\mid t_i)\le\int\Delta_i(t_i,t_{-i})\,p_i'(\dd t_{-i}\mid t_i)\). This says that the interim payoff gain from \(b_i\) rather than \(a_i\) is weakly larger under \(p_i'\), which is precisely the additional comparison condition.

\par\addvspace{\baselineskip}\noindent(ii) Increasing differences of \(u_i^\theta\) in \((a_i,\theta)\) mean that the payoff gain from \(b_i\) rather than \(a_i\) is weakly larger at \(\theta'\) than at \(\theta\), for each fixed opponents' action profile and type profile. Integrating this inequality against the same opponents' strategies and conditional beliefs verifies the additional comparison condition. Applying Steps 1 and 2 proves both parts of the corollary.
\end{proof}

\begin{proof}[Proof of Claim~\ref{claim:interim-beliefs}]

\noindent(a) At each type \(t_i\), the opponent's type \(t_j\) is uniformly distributed on \([0,1]\), with \(\omega=-t_i-t_j\). These measurable conditional distributions come from the stated common prior, whose type marginals are the reference probabilities. The common prior observation gives null compatibility. On the binary action chains, payoff regularity and supermodularity in own action are immediate, and the gain \(2t_i+a_j+\omega-1/2\) increases in \(a_j,t_i,\omega\) throughout the product space. Thus Assumption~\ref{ass:main} and Assumption~\ref{ass:types}(a) hold. Since the increasing function \(\omega\) has conditional expectation \(-t_i-1/2\), Assumption~\ref{ass:types}(b) fails. This proves~(a).

\par\addvspace{\baselineskip}\noindent(b) For every behavioral strategy \(r_j\), integration gives
\(V_i(a_i,t_i;r_j) =a_i\int_0^1\bigl(t_i+r_j(\{1\}\mid t_j)-t_j-\tfrac12\bigr)\dd t_j =a_i(t_i+m_j-1)\), 
proving~(b).

\par\addvspace{\baselineskip}\noindent(c) The investment gain increases in own type and in the opponent's strategy under pointwise stochastic dominance, for all behavioral strategies. This implies condition~(ii) whenever at least one opponent strategy is stochastically increasing. Together with the binary action chain, these observations verify all three conditions of Assumption~\ref{ass:interim}, proving~(c).

\par\addvspace{\baselineskip}\noindent(d) 
For \(m\in[0,1]\), define \(\alpha_i^m(t_i)=\mathbf1_{\{t_i>1-m\}}\) and \(\alpha^m=(\alpha_1^m,\alpha_2^m)\). Consider any behavioral Bayesian Nash equilibrium \(r\), and let \(m_i=\int_0^1r_i(\{1\}\mid t_i)\,\dd t_i\). By part~(b), the gain from investment is \(t_i+m_j-1\). Thus player \(i\) invests with probability one at almost every type above \(1-m_j\), and with probability zero at almost every type below \(1-m_j\). Mixing is possible only at the single cutoff type, which has measure zero. It follows that \(r_i=\delta_{\alpha_i^{m_j}}\) almost everywhere and \(m_i=\int_0^1\mathbf1_{\{t_i>1-m_j\}}\,\dd t_i=m_j\). Hence \(m_1=m_2=m\) for some \(m\in[0,1]\), with \(r=\delta_{\alpha^m}\) almost everywhere. Conversely, each profile \(\delta_{\alpha^m}\) is an equilibrium, since each firm's cutoff strategy is a best response to the other firm's investment probability \(m\). This characterizes all behavioral Bayesian Nash equilibria.

Fix \(m\in[0,1]\). For \(n\ge1\), set \(m_n=(nm+1)/(n+2)\) and \(\varepsilon_n=1/(n+2)\). Then \(0<m_n<1\), \(m_n\to m\), and \(\varepsilon_n\to0\). Give both firms tremble probability \(\varepsilon_n\) and conditional tremble distribution \(\mu_i^n(\cdot\mid t_i)=(1-m_n)\delta_0+m_n\delta_1\) at every type. These distributions have full support. They are stochastically increasing in type because they are constant across types. Let \(\varepsilon^n=(\varepsilon_n,\varepsilon_n)\) and
\(\mu^n=(\mu_1^n,\mu_2^n)\).
These choices define the perturbation
\(q^n=(\varepsilon^n,\mu^n)\).
For each firm \(i\), define the behavioral strategy
\(r_i^n(\cdot\mid t_i)=\delta_{\alpha_i^{m_n}(t_i)}\).
Thus \(r_i^n\) assigns probability one to the action prescribed
by the cutoff strategy \(\alpha_i^{m_n}\).
Write \(r^n=(r_1^n,r_2^n)\).

Write \(g^n=(r^n)^{q^n}\) for the induced behavioral profile. Its investment probability at type \(t_i\) is \(g_i^n(\{1\}\mid t_i)=(1-\varepsilon_n)\alpha_i^{m_n}(t_i)+\varepsilon_n m_n\). Since \(\int_0^1\alpha_i^{m_n}(t_i)\,\dd t_i=m_n\), we obtain \(\int_0^1g_i^n(\{1\}\mid t_i)\,\dd t_i=(1-\varepsilon_n)m_n+\varepsilon_n m_n=m_n\). Thus the opponent's ex ante investment probability remains \(m_n\) after perturbation.

By part~(b) and Remark~\ref{rem:perturbed}, the gain from choosing action \(1\) rather than \(0\) in \(G(q^n)\) is \(V_i^{q^n}(1,t_i;r_j^n)-V_i^{q^n}(0,t_i;r_j^n)=(1-\varepsilon_n)(t_i+m_n-1)\). Its sign is the sign of \(t_i+m_n-1\). Therefore \(\alpha_i^{m_n}(t_i)\) is optimal at every type. At the cutoff \(t_i=1-m_n\), both actions are optimal, so the prescribed action \(0\) is also a best response. It follows that \(r^n\in\EE(q^n)\).

We have \(\|q^n\|=\varepsilon_n\to0\). For every \(t_i\ne1-m\), the convergence \(m_n\to m\) implies \(\alpha_i^{m_n}(t_i)=\alpha_i^m(t_i)\) for all sufficiently large \(n\). The formula for \(g_i^n(\{1\}\mid t_i)\) then gives \(g_i^n(\{1\}\mid t_i)\to\alpha_i^m(t_i)\). On the binary action space, this is equivalent to \(g_i^n(\cdot\mid t_i)\Rightarrow\delta_{\alpha_i^m(t_i)}\). The exceptional cutoff type has measure zero. Definition~\ref{def:perfect} therefore gives \(\delta_{\alpha^m}\in\PP_{\mathrm{inc}}^b\).

Conversely, Lemma~\ref{lem:optimality} implies that every element of \(\PP_{\mathrm{inc}}^b\) is a Bayesian Nash equilibrium. Thus, we have \(\PP_{\mathrm{inc}}^b=\{\delta_{\alpha^m}\mid m\in[0,1]\}\). Since \(\alpha^m\) is increasing in \(m\), this set has smallest element \(\delta_{\alpha^0}\) and largest element \(\delta_{\alpha^1}\). We have \(\alpha^0=(0,0)\), while \(\alpha^1=(1,1)\) almost everywhere because the type \(0\) has measure zero. This proves~(d).
\end{proof}

\subsection{Proofs for Section~\ref{sec:general-comparison}}\label{app:sec3}
\subsubsection{Existence of extremal perfect equilibria}\label{app:global-existence}
The proof proceeds in four steps. Step 1 selects pure best response profiles close to their completely mixed supports. Step 2 proves convergence to a pure perfect equilibrium. Step 3 shows that this equilibrium bounds every behavioral perfect equilibrium from above. Step 4 obtains the smallest equilibrium by reversing the action orders.

\begin{proof}[Proof of Theorem~\ref{thm:general-global}]
\noindent Step 1.
Let \(\mathcal Q\) be as in Lemma~\ref{lem:space}, with \(d_{\mathcal Q}\) and \(J\) defined by~\eqref{eq:strategy-functionals}. Let \(R\) and \(e\) be defined by~\eqref{eq:discrepancy-functionals}, with \(\Phi\) as in Lemma~\ref{lem:glue}. For \(\tau>0\), let \(\mathcal Z_\tau\) be the set of pure strategy profiles \(s\) for which there is a completely mixed behavioral strategy profile \(g\) such that
\[
s_i(t_i)\in\BR_i(t_i;g_{-i}),R(s,g)\le\tau
\]
for every player \(i\) and almost every type \(t_i\).
The set \(\mathcal Z_\tau\) is nonempty for every \(\tau>0\). Indeed, fix distributions \(\nu_i\) with full support on \(A_i\). Choose \(\varepsilon\in(0,1)\) with \(N\bar d\varepsilon\le\tau\). For \(s\in\mathcal Q\), define the completely mixed behavioral strategy profile \(\sigma(s)\) by \(\sigma_i(s)(\cdot\mid t_i)=(1-\varepsilon)\delta_{s_i(t_i)}+\varepsilon\nu_i\). If \(s\le s'\), then \(\sigma_j(s)(\cdot\mid t_j)\st\sigma_j(s')(\cdot\mid t_j)\) for every \(j\) and almost every \(t_j\) by Lemma~\ref{lem:stoch}. Parts~(a), (b), and~(c) of Lemma~\ref{lem:interim} therefore show that the map that assigns to \(s\) the profile of largest best responses \(\overline{\BR}_i(\cdot;\sigma_{-i}(s))\) is a well defined increasing map on \(\mathcal Q\). Since \(\mathcal Q\) is a complete lattice by Lemma~\ref{lem:space}, the fixed point theorem of \citet{tarski} gives a fixed point \(s\). By the definitions of \(R\) and \(\sigma(s)\),
\[
\begin{aligned}
	R(s,\sigma(s))
	&=(1-\varepsilon)\sum_{i\in I}\int_{T_i}
	d_i\bigl(s_i(t_i),s_i(t_i)\bigr)\,\lambda_i(\dd t_i)
	+\varepsilon\sum_{i\in I}\int_{T_i}\int_{A_i}
	d_i\bigl(a_i,s_i(t_i)\bigr)\,
	\nu_i(\dd a_i)\lambda_i(\dd t_i)\\
	&=\varepsilon\sum_{i\in I}\int_{T_i}\int_{A_i}
	d_i\bigl(a_i,s_i(t_i)\bigr)\,
	\nu_i(\dd a_i)\lambda_i(\dd t_i) \le\varepsilon\sum_{i\in I}\bar d\,\lambda_i(T_i)
	=N\bar d\,\varepsilon\le\tau.
\end{aligned}
\]
The Dirac component contributes zero because
\(d_i(s_i(t_i),s_i(t_i))=0\).
The bound uses \(d_i\le\bar d\), together with
\(\nu_i(A_i)=\lambda_i(T_i)=1\).
Thus \(s\in\mathcal Z_\tau\).

Let \(M_\tau=\sup_{s\in\mathcal Z_\tau}J(s)\). If \(0<\tau\le\tau'\), then \(\mathcal Z_\tau\subseteq\mathcal Z_{\tau'}\), so \(M_\tau\le M_{\tau'}\). Since \(0\le J\le N\), the limit \(M=\lim_{\tau\downarrow0}M_\tau=\inf_{\tau>0}M_\tau\) exists. For every \(n\ge1\), choose \(s^n\in\mathcal Z_{2^{-n}}\) and a completely mixed behavioral strategy profile \(g^n\) such that
\begin{equation}\label{eq:global-approximation}
J(s^n)\ge M_{2^{-n}}-\tfrac1n,R(s^n,g^n)\le2^{-n},s_i^n(t_i)\in\BR_i(t_i;g_{-i}^n)
\end{equation}
for every player \(i\) and almost every type \(t_i\).
Then \(J(s^n)\to M\).

\par\addvspace{\baselineskip}\noindent Step 2.
Apply Lemma~\ref{lem:glue}(b) with \(s=s^n\), \(g=g^n\), \(h=g^m\), and \(\eta=1/(n+m)\) for \(n, m \in \mathbb{N}\). 
For every player \(i\), the best response property in~\eqref{eq:global-approximation} and the definition of distance to a set give
\(d_i(a_i,\BR_i(t_i;g_{-i}^m))
\le d_i(a_i,s_i^m(t_i))\)
for every \(a_i\in A_i\) and almost every \(t_i\).
Integrating this inequality and using~\eqref{eq:discrepancy-functionals} gives
\[
e(g^m)\le R(s^m,g^m)\le2^{-m}.
\]
The exceptional null sets do not affect these integrals.
The last inequality follows from~\eqref{eq:global-approximation}. The lemma yields a pure strategy profile \(z^{nm}\) with \(z_i^{nm}(t_i)\ge s_i^n(t_i)\vee s_i^m(t_i)\) for every player \(i\) and almost every \(t_i\), and a completely mixed behavioral strategy profile \(r^{nm}\). They satisfy \(z_i^{nm}(t_i)\in\BR_i(t_i;r_{-i}^{nm})\) almost everywhere. By~\eqref{eq:gluing-estimate}, \(R(z^{nm},r^{nm})\le\tau_{nm}=\Phi(2^{-n}+2^{-m})+N\bar d/(n+m)\), so \(z^{nm}\in\mathcal Z_{\tau_{nm}}\). For \(n,m\ge k\), monotonicity of \(\Phi\) gives
\[
0<\tau_{nm}\le\Phi(2^{1-k})+\frac{N\bar d}{2k}\longrightarrow0.
\]
Monotonicity of \(M_\tau\) therefore gives \(M_{\tau_{nm}}\to M\) uniformly as \(\min\{n,m\}\to\infty\). Since \(J(s^n)\to M\) and
\(\max\{J(s^n),J(s^m)\}\le J(z^{nm})\le M_{\tau_{nm}}\),
both \(J(z^{nm})-J(s^n)\) and \(J(z^{nm})-J(s^m)\) tend to zero in this same sense. For any \(\gamma>0\), Lemma~\ref{lem:space}(c) supplies a \(\kappa>0\) independent of the ordered profiles such that each metric distance is less than \(\gamma/2\) whenever the corresponding \(J\)-gap is less than \(\kappa\). Both gaps satisfy this bound for all sufficiently large \(n,m\). The triangle inequality then gives \(d_{\mathcal Q}(s^n,s^m)<\gamma\), proving that \((s^n)_{n\ge1}\) is Cauchy. By Lemma~\ref{lem:space}(b), it converges to a pure strategy profile \(\alpha^+\), and a subsequence converges to \(\alpha^+\) almost everywhere for every player. We pass to this subsequence and relabel it, which preserves \(J(s^n)\to M\) and the discrepancy bound in~\eqref{eq:global-approximation}.

The discrepancy bound in~\eqref{eq:global-approximation} gives \(\sum_n R(s^n,g^n)<\infty\), so Tonelli's theorem implies that
\(\int_{A_i}d_i(a_i,s_i^n(t_i))\,g_i^n(\dd a_i\mid t_i)\to0\)
for every player \(i\) and almost every type \(t_i\).
Together with \(s_i^n(t_i)\to\alpha_i^+(t_i)\), the triangle inequality then yields
\(\int_{A_i}d_i(a_i,\alpha_i^+(t_i))\,g_i^n(\dd a_i\mid t_i)\to0\)
at almost every type.
Hence \(g_i^n(\cdot\mid t_i)\Rightarrow\delta_{\alpha_i^+(t_i)}\)
by Lemma~\ref{lem:criterion}(a).
Moreover, the best response property in~\eqref{eq:global-approximation} gives
\(d_i(\alpha_i^+(t_i),\BR_i(t_i;g_{-i}^n))
\le d_i(\alpha_i^+(t_i),s_i^n(t_i))\to0\)
for every player \(i\) and almost every type \(t_i\).
These two convergence properties imply \(\alpha^+\in\PP^p\)
by Lemma~\ref{lem:criterion}(b).

\par\addvspace{\baselineskip}\noindent Step 3.
Fix \(r\in\PP^b\) and a sequence \((h^n)_{n\ge1}\) of completely mixed behavioral strategy profiles that supports \(r\). By Lemma~\ref{lem:criterion}(c) and bounded convergence, \(e(h^n)\to0\). Apply Lemma~\ref{lem:glue}(b) with \(s=s^n\), \(g=g^n\), \(h=h^n\), and \(\eta=1/(n+1)\). It yields a pure strategy profile \(z^n\) and a completely mixed behavioral strategy profile \(\widetilde r^n\). For every \(i\) and almost every \(t_i\), these profiles satisfy \(z_i^n(t_i)\in\BR_i(t_i;\widetilde r_{-i}^n)\) and \(z_i^n(t_i)\ge s_i^n(t_i)\). At each such type, every element of \(\BR_i(t_i;h_{-i}^n)\) lies below \(z_i^n(t_i)\). By~\eqref{eq:gluing-estimate}, \(R(z^n,\widetilde r^n)\le\tau_n=\Phi\bigl(R(s^n,g^n)+e(h^n)\bigr)+N\bar d/(n+1)\), where \(\tau_n\to0\) by Lemma~\ref{lem:glue}(a). Since \(\tau_n>0\), it follows that \(z^n\in\mathcal Z_{\tau_n}\), and hence \(J(s^n)\le J(z^n)\le M_{\tau_n}\to M\). Lemma~\ref{lem:space}(c) gives \(d_{\mathcal Q}(z^n,s^n)\to0\), and hence \(d_{\mathcal Q}(z^n,\alpha^+)\to0\). Passing to a further subsequence, we may assume that \(z_i^n(t_i)\to\alpha_i^+(t_i)\) for every player \(i\) and almost every \(t_i\). 

Fix a player \(i\) and a type \(t_i\) at which \(z_i^n(t_i)\to\alpha_i^+(t_i)\), the distance condition in~\eqref{eq:general-behavioral-criterion} holds, and every element of \(\BR_i(t_i;h_{-i}^n)\) lies below \(z_i^n(t_i)\) for every \(n\). By Lemma~\ref{lem:criterion}(b), there are \(\nu^n\in\mathcal B_i(t_i;h_{-i}^n)\) with \(\nu^n\Rightarrow r_i(\cdot\mid t_i)\). Each \(\nu^n\) is concentrated on \(\BR_i(t_i;h_{-i}^n)\), which lies below \(z_i^n(t_i)\), so \(\nu^n\st\delta_{z_i^n(t_i)}\) by Lemma~\ref{lem:stoch}(d). Since \(\delta_{z_i^n(t_i)}\Rightarrow\delta_{\alpha_i^+(t_i)}\), Lemma~\ref{lem:stoch}(a) gives \(r_i(\cdot\mid t_i)\st\delta_{\alpha_i^+(t_i)}\). The set of such types has full measure, so this comparison holds for every player \(i\) and almost every \(t_i\). Since \(r\in\PP^b\) is arbitrary and \(\delta_{\alpha^+}\in\PP^b\), the profile \(\delta_{\alpha^+}\) is the largest element of \(\PP^b\), and Lemma~\ref{lem:stoch}(d) shows that \(\alpha^+\) is the largest element of \(\PP^p\).

\par\addvspace{\baselineskip}\noindent Step 4.
By Remark~\ref{rem:dual}, the game obtained by reversing the order on every action space satisfies Assumption~\ref{ass:main}, has the same sets \(\PP^b\) and \(\PP^p\), and reverses the order on behavioral strategy profiles. Applying the preceding argument to that game yields \(\alpha^-\in\PP^p\) such that \(\delta_{\alpha_i^-(t_i)}\st r_i(\cdot\mid t_i)\) for every \(r\in\PP^b\), every player \(i\), and almost every \(t_i\). Hence \(\delta_{\alpha^-}\) is the smallest element of \(\PP^b\), and \(\alpha^-\) is the smallest element of \(\PP^p\).
\end{proof}

\subsubsection{Monotonicity with finite actions}\label{app:finite-monotonicity}
The proof of Theorem~\ref{prop:general-finite} proceeds in three steps. Step 1 uses stabilization of best responses to construct an increasing operator that preserves monotonicity in type. Step 2 shows that every pure perfect equilibrium lies below its image under this operator. Step 3 proves that the largest fixed point on monotone profiles is the largest pure perfect equilibrium, then obtains the smallest equilibrium by reversing the orders.

\begin{proof}[Proof of Theorem~\ref{prop:general-finite}]
\noindent Step 1.
Let \(N=|I|\). Let \(\mathcal Q\) and \(\mathcal Q_{\mathrm{inc}}\) be the spaces of pure profiles and monotone pure profiles in Lemma~\ref{lem:space}. Fix a probability distribution \(\nu_i\) with full support on each \(A_i\). For \(\alpha\in\mathcal Q\) and \(0<\varepsilon<1/3\), define the completely mixed behavioral strategy profile \(\sigma^\varepsilon(\alpha)\) by
\begin{equation}\label{eq:general-canonical}
\sigma_i^\varepsilon(\alpha)(\cdot\mid t_i)
=(1-\varepsilon-\varepsilon^{N+1})\delta_{\alpha_i(t_i)}
+\varepsilon\delta_{\overline a_i}+\varepsilon^{N+1}\nu_i.
\end{equation}
Let \(\Psi_i^\varepsilon(\alpha)(t_i)=\overline{\BR}_i(t_i;\sigma_{-i}^\varepsilon(\alpha))\). Lemmas~\ref{lem:stoch} and~\ref{lem:interim} imply that \(\Psi^\varepsilon\) is a well defined increasing map on \(\mathcal Q\). If \(\alpha\in\mathcal Q_{\mathrm{inc}}\), choose increasing representatives of its components. Each \(\sigma_j^\varepsilon(\alpha)\) is then stochastically increasing in type, so Lemma~\ref{lem:interim}(d) makes \(\Psi_i^\varepsilon(\alpha)\) increasing at every type. Thus \(\Psi^\varepsilon\) maps \(\mathcal Q_{\mathrm{inc}}\) into itself.

Fix \(\alpha,i,t_i\). For \(a,b\in A_i\), let
\(P_{a,b}(\varepsilon)
=V_i(a,t_i;\sigma_{-i}^{\varepsilon}(\alpha))
-V_i(b,t_i;\sigma_{-i}^{\varepsilon}(\alpha))\).
The product expansion shows that \(P_{a,b}\) is a polynomial.
If it is identically zero, the two actions yield the same payoff
for every \(\varepsilon\). Otherwise, let \(c_k\varepsilon^k\)
be its lowest degree nonzero term. Then
\(P_{a,b}(\varepsilon)/\varepsilon^k\to c_k\ne0\)
as \(\varepsilon\downarrow0\). Since \(\varepsilon^k>0\),
the payoff difference has the same sign as \(c_k\) for all
sufficiently small positive \(\varepsilon\).
The threshold for this sign constancy may depend on the pair \(a,b\).
Since there are finitely many such pairs, taking the minimum of
their positive thresholds gives
\(\bar\varepsilon_{\alpha,i,t_i}\in(0,1/3)\), after shrinking it
if necessary, such that every pairwise payoff difference has
a constant sign on \((0,\bar\varepsilon_{\alpha,i,t_i})\).
An action \(a\) is a best response precisely when
\(P_{a,b}(\varepsilon)\ge0\) for every \(b\in A_i\).
All these comparisons remain unchanged on this interval, so the entire best response set is constant there. The threshold may depend on \(\alpha,i,t_i\).
Define \(\Psi_i(\alpha)(t_i)\) as the largest element of this eventual best response set. If \(N=1\), there are no opponents, so the best response set is independent of \(\varepsilon\) from the outset.

Choose a fixed sequence \(\varepsilon_n\downarrow0\) in \((0,1/3)\). At every type, \(\Psi_i^{\varepsilon_n}(\alpha)(t_i)\) eventually equals \(\Psi_i(\alpha)(t_i)\). This representation makes \(\Psi(\alpha)\) measurable. For any two representatives of the same profile, take the union of the null sets on which the corresponding values of \(\Psi^{\varepsilon_n}\) differ. Outside this null set their limits agree, so \(\Psi\) is well defined on \(\mathcal Q\). The same argument applied to \(\alpha\le\beta\) shows that \(\Psi(\alpha)\le\Psi(\beta)\). For increasing representatives of \(\alpha\), the maps \(\Psi^{\varepsilon_n}(\alpha)\) are increasing at every type. Their pointwise limit is increasing because the action order is closed. Thus \(\Psi\) also maps \(\mathcal Q_{\mathrm{inc}}\) into itself.

\par\addvspace{\baselineskip}\noindent Step 2.
We show that \(\alpha\le\Psi(\alpha)\) for every \(\alpha\in\PP^p\), including profiles that are not monotone. Let \((g^n)_{n\ge1}\) support such an \(\alpha\). By Lemma~\ref{lem:optimality}, \(\alpha\) is a Bayesian Nash equilibrium. Fix \(i\) and a type \(t_i\) outside the null sets in this equilibrium property and~\eqref{eq:general-criterion}. Write \(a=\alpha_i(t_i)\). Since \(A_i\) is finite, \(d_i(a,\BR_i(t_i;g_{-i}^n))\to0\) implies \(a\in\BR_i(t_i;g_{-i}^n)\) for all sufficiently large \(n\).\footnote{The minimum distance from \(a\) to any other action is positive when such actions exist. With infinite actions, best responses can instead converge to \(a\) without ever containing it, as in Example~\ref{ex:regional-chain}. This is the finite action step needed in the zero gap argument below.}

For a fixed lower action \(c\le a\), define the payoff difference \(\Delta^{(a,c)}(a_{-i},t_{-i})=u_i(a,a_{-i},t_i,t_{-i})-u_i(c,a_{-i},t_i,t_{-i})\). Define \(\Delta_0^{(a,c)}=\int_{T_{-i}}\Delta^{(a,c)}(\alpha_{-i}(t_{-i}),t_{-i})\,p_i(\dd t_{-i}\mid t_i)\), where \(\alpha_{-i}(t_{-i})=(\alpha_j(t_j))_{j\ne i}\). Also define \(\overline\Delta^{(a,c)}=\int_{T_{-i}}\Delta^{(a,c)}(\overline a_{-i},t_{-i})\,p_i(\dd t_{-i}\mid t_i)\). Increasing differences make \(\Delta^{(a,c)}\) increasing in \(a_{-i}\). The equilibrium property gives \(\Delta_0^{(a,c)}\ge0\), hence \(\overline\Delta^{(a,c)} \ge\Delta_0^{(a,c)} \ge0\). Let \(\bar u\) bound \(|u_i|\), and abbreviate \(V^\varepsilon(x)=V_i(x,t_i;\sigma_{-i}^\varepsilon(\alpha))\).

Suppose first that \(\overline\Delta^{(a,c)}>0\). Expand \(V^\varepsilon(a)-V^\varepsilon(c)\) according to the three mixture components in~\eqref{eq:general-canonical}. A term using only the components \(\delta_{\alpha_j(t_j)}\) and \(\delta_{\overline a_j}\) has an action profile weakly above \(\alpha_{-i}(t_{-i})\) at every \(t_{-i}\). Its contribution is therefore at least its mixture weight times \(\Delta_0^{(a,c)}\), which is nonnegative. In particular, the term selecting the component \(\varepsilon\delta_{\overline a_j}\) for every opponent contributes \(\varepsilon^{N-1}\overline\Delta^{(a,c)}\). The total weight of terms selecting at least one \(\nu_j\) component is at most \((N-1)\varepsilon^{N+1}\). Since \(|\Delta^{(a,c)}|\le2\bar u\), it follows that
\[
V^\varepsilon(a)-V^\varepsilon(c)
\ge \varepsilon^{N-1}\bigl[\overline\Delta^{(a,c)}-2\bar u(N-1)\varepsilon^2\bigr]>0
\]
for all sufficiently small \(\varepsilon>0\). Thus the possible losses from the full support components vanish faster than the gain from the top action components.

Suppose instead that \(\overline\Delta^{(a,c)}=0\). Choose \(n\) such that \(a\in\BR_i(t_i;g_{-i}^n)\). The expected value of \(\Delta^{(a,c)}\) against \(g_{-i}^n\) is nonnegative by this exact best response property. Because \(\Delta^{(a,c)}\) is increasing in opponents' actions, we have
\[
\int_{T_{-i}}\int_{A_{-i}}
\bigl[\Delta^{(a,c)}(\overline a_{-i},t_{-i})-\Delta^{(a,c)}(a_{-i},t_{-i})\bigr]
\bigotimes_{j\ne i}g_j^n(\dd a_j\mid t_j)\,
p_i(\dd t_{-i}\mid t_i)=0.
\]
The integrand is nonnegative. At almost every \(t_{-i}\) with respect to \(p_i(\cdot\mid t_i)\), its average over actions is therefore zero. At each such \(t_{-i}\), this nonnegative difference is continuous in \(a_{-i}\), and the product of the completely mixed action distributions has full support. If the difference were positive at any action profile, continuity would make it positive on an open neighborhood with positive probability, contradicting its zero integral. Thus \(\Delta^{(a,c)}(a_{-i},t_{-i})=\Delta^{(a,c)}(\overline a_{-i},t_{-i})\) for every \(a_{-i}\). This implication uses continuity and full support; finiteness was needed above to obtain the exact best response \(a\). Hence \(\Delta^{(a,c)}\) is constant in opponents' actions at almost every opponents' type profile. Its interim expectation against any behavioral strategy profile equals \(\overline\Delta^{(a,c)}=0\). In particular, \(V^\varepsilon(a)=V^\varepsilon(c)\).

In either case, \(V^\varepsilon(a)\ge V^\varepsilon(c)\) for all sufficiently small \(\varepsilon>0\). There are finitely many \(c\le a\), so a common positive threshold makes these inequalities hold simultaneously.\footnote{The threshold for each comparison can depend on \(c\). Finite actions allow us to take a positive minimum over these thresholds. With infinitely many lower actions, the infimum of these thresholds may be zero. The threshold here may still depend on the player and type.} For \(\varepsilon\) below this threshold, let \(b=\Psi_i^\varepsilon(\alpha)(t_i)\). Since \(a\wedge b\le a\), supermodularity yields
\[
V^\varepsilon(a\vee b)
\ge V^\varepsilon(a)+V^\varepsilon(b)-V^\varepsilon(a\wedge b)
\ge V^\varepsilon(b).
\]
Thus \(a\vee b\) is a best response. Since \(b\) is the largest best response, \(a\le a\vee b\le b\). Letting \(\varepsilon\downarrow0\) along the sequence from Step 1 gives \(a\le\Psi_i(\alpha)(t_i)\). This proves \(\alpha\le\Psi(\alpha)\).

\par\addvspace{\baselineskip}\noindent Step 3.
By Lemma~\ref{lem:space}(a), \(\mathcal Q_{\mathrm{inc}}\) is a complete lattice. The fixed point theorem of \citet{tarski} gives a largest fixed point \(\alpha^+\) of \(\Psi\) on \(\mathcal Q_{\mathrm{inc}}\). It also implies that \(\alpha^+\) lies above every \(\beta\in\mathcal Q_{\mathrm{inc}}\) satisfying \(\beta\le\Psi(\beta)\).

First, \(\alpha^+\) is perfect. Choose increasing representatives of its components. The profiles \(\sigma^{\varepsilon_n}(\alpha^+)\) are completely mixed and stochastically increasing in type. They converge weakly to \(\delta_{\alpha^+}\) at every type. The fixed point identity \(\Psi_i(\alpha^+)(t_i)=\alpha_i^+(t_i)\) and stabilization in Step 1 give
\begin{equation}\label{eq:finite-eventual-optimality}
\alpha_i^+(t_i)\in\BR_i(t_i;\sigma_{-i}^\varepsilon(\alpha^+))
\end{equation}
for every player \(i\), almost every type \(t_i\), and all sufficiently small \(\varepsilon>0\). The threshold may depend on \(t_i\), which suffices for~\eqref{eq:general-criterion} along \((\varepsilon_n)_{n\ge1}\). Lemma~\ref{lem:criterion}(b) therefore gives \(\alpha^+\in\PP^p\).

Next, fix any \(\alpha\in\PP^p\). To compare this possibly nonmonotone profile with \(\alpha^+\), let \(\widehat\alpha\) be the meet of all \(\beta\in\mathcal Q_{\mathrm{inc}}\) satisfying \(\beta\ge\alpha\). This family is nonempty because it contains the constant profile of largest actions. By Lemma~\ref{lem:space}(a), its meet in \(\mathcal Q_{\mathrm{inc}}\) agrees with its meet in \(\mathcal Q\). Thus \(\widehat\alpha\ge\alpha\), so \(\widehat\alpha\) is the least monotone profile above \(\alpha\). Monotonicity of \(\Psi\) and Step 2 give \(\Psi(\widehat\alpha)\ge\Psi(\alpha)\ge\alpha\). Since \(\Psi(\widehat\alpha)\in\mathcal Q_{\mathrm{inc}}\), it belongs to the family defining \(\widehat\alpha\). Therefore \(\widehat\alpha\le\Psi(\widehat\alpha)\). The fixed point comparison gives \(\widehat\alpha\le\alpha^+\), hence \(\alpha\le\alpha^+\). We have proved that \(\alpha^+\) is the largest element of \(\PP^p\). It therefore coincides with the largest pure perfect equilibrium in Theorem~\ref{thm:general-global}.

Finally, reverse all action and type orders. Remark~\ref{rem:dual} shows that the hypotheses remain valid. The perfect equilibria are unchanged, while their order is reversed. Simultaneous reversal of the action and type orders also leaves monotone strategies unchanged. Applying the preceding argument to this game makes the smallest element \(\alpha^-\) of \(\PP^p\) monotone. The corresponding supporting profiles are given by~\eqref{eq:general-canonical} with \(\underline a_i\) in place of \(\overline a_i\). They are stochastically increasing in type, with the same eventual best response property. By Theorem~\ref{thm:general-global}, \(\delta_{\alpha^-}\) and \(\delta_{\alpha^+}\) are the extremal perfect equilibria in \(\PP^b\). Since both pure profiles are monotone, they are also the extremal perfect monotone equilibria.
\end{proof}

\begin{proof}[Proof of Proposition~\ref{prop:finite-type-endpoints}]
	It suffices to show that \(\alpha^+\) and \(\alpha^-\) belong to
	\(\PP_{\mathrm{inc}}^p\). Indeed,
	\(\PP_{\mathrm{inc}}^b\subseteq\PP^b\) by
	Lemma~\ref{lem:optimality}, so the largest and smallest elements
	of \(\PP^b\) remain the extrema of this subset whenever they belong to it.
	
	Let \(N=|I|\), and use the full support distributions \(\nu_i\)
	chosen in Step 1 of the proof of Theorem~\ref{prop:general-finite}.
	For each player \(i\), let \(T_i^*\) be a full measure set on which
	the fixed point identity \(\Psi_i(\alpha^+)(t_i)=\alpha_i^+(t_i)\) holds.
	At each \(t_i\in T_i^*\), stabilization in Step 1 of that proof supplies
	a threshold \(\bar\varepsilon_{\alpha^+,i,t_i}>0\) below which
	\eqref{eq:finite-eventual-optimality} holds, with
	\(\sigma^\varepsilon\) defined by~\eqref{eq:general-canonical}.
	There are finitely many players and each \(T_i^*\) is finite.
	The minimum of these finitely many positive thresholds is therefore
	positive and makes the best response conditions hold simultaneously
	at every \(t_i\in T_i^*\).
	
	For \(0<\varepsilon<1/3\) below this threshold, put
	\(\eta_\varepsilon=\varepsilon+\varepsilon^{N+1}\).
	Define
	\(\mu_i^{\varepsilon,+}
	=(\varepsilon/\eta_\varepsilon)\delta_{\overline a_i}
	+(\varepsilon^{N+1}/\eta_\varepsilon)\nu_i\)
	and \(q_i^{\varepsilon,+}=(\eta_\varepsilon,\mu_i^{\varepsilon,+})\).
	Each tremble distribution has full support and is independent of
	type, hence is stochastically increasing in type.
	Moreover,
	\((\alpha^+)^{q^{\varepsilon,+}}
	=\sigma^\varepsilon(\alpha^+)\).
	Remark~\ref{rem:perturbed} therefore implies that \(\alpha^+\)
	is an equilibrium of \(G(q^{\varepsilon,+})\).
	As \(\varepsilon\downarrow0\), we have \(\eta_\varepsilon\to0\).
	Thus the constant sequence of equilibrium profiles \(\alpha^+\),
	together with these increasing perturbations, gives
	\(\alpha^+\in\PP_{\mathrm{inc}}^p\).
	
	For the smallest equilibrium, define
	\(\mu_i^{\varepsilon,-}
	=(\varepsilon/\eta_\varepsilon)\delta_{\underline a_i}
	+(\varepsilon^{N+1}/\eta_\varepsilon)\nu_i\)
	and \(q_i^{\varepsilon,-}=(\eta_\varepsilon,\mu_i^{\varepsilon,-})\).
	The order reversal in Step 3 of the proof of
	Theorem~\ref{prop:general-finite} gives the eventual best response
	property for \(\alpha^-\), with \(\underline a_i\) replacing
	\(\overline a_i\) in~\eqref{eq:general-canonical}.
	Finiteness of the type spaces again gives a common positive threshold.
	Since these tremble distributions also have full support and are
	independent of type, Remark~\ref{rem:perturbed} makes \(\alpha^-\)
	an equilibrium of \(G(q^{\varepsilon,-})\) for every sufficiently
	small \(\varepsilon>0\).
	Letting \(\varepsilon\downarrow0\) yields
	\(\alpha^-\in\PP_{\mathrm{inc}}^p\).
\end{proof}

\subsubsection{Counterexamples and comparisons}\label{app:infinite-counterexamples}\label{app:general-comparative}

	The following claim collects calculations specific to Example~\ref{ex:regional-chain}.

\begin{claim}\label{claim:specialization}
		Consider the product specialization payoffs \(u_1\) and \(u_2\) of Example~\ref{ex:regional-chain} as functions of \((a_1,b,\omega)\in A_1\times A_2\times[0,1]\).
		\begin{enumerate}[label=(\alph*)]
			\item Both functions are bounded, jointly measurable, continuous in actions, and supermodular in own action. The function \(u_1\) has increasing differences in \((a_1,b)\) and in \((a_1,\omega)\), and \(u_2\) depends neither on \(a_1\) nor on \(\omega\).
			\item Suppose that \(\omega\) is uniformly distributed on \([0,1]\), and let \(0<\bar x\le1\). For every \(x\in[0,1]\),
			\begin{equation}\label{eq:specialization-moments}
			\mathbb E[\pi(x,\omega)]=x^2,\mathbb E\bigl[\pi(x,\omega)\mathbf1_{\{\omega\ge1-\bar x\}}\bigr]=
			\begin{cases}
				x^2,&x\le\bar x,\\
				2x\bar x-\bar x^2,&x\ge\bar x.
			\end{cases}
			\end{equation}
			\item Suppose that \(\omega\) is uniformly distributed on \([0,1]\). The retailer's order \(b\) is random, with \(\Pr(b=2\mid\omega)\le\varepsilon<1/2\) for almost every \(\omega\). Suppose also that \(\Pr(b=0)>0\). Then the standard product \(a_1=0\) is the unique maximizer of the producer's expected payoff \(\mathbb E[u_1(a_1,b,\omega)]\) over \(a_1\in A_1\).
		\end{enumerate}
	\end{claim}
	\begin{proof}
\noindent(a) The function \(\pi\) is continuous on \([0,1]^2\) and satisfies \(0\le\pi(x,\omega)\le2\). The payoff formulas therefore imply boundedness and joint measurability. Continuity in actions follows from continuity of \(\pi\) on the specialization interval, since \(0\) is isolated in \(A_1=\{0\}\cup[1,2]\) and \(A_2\) is finite. On a chain, \(a\vee a'\) and \(a\wedge a'\) are simply the larger and smaller of \(a,a'\). Hence \(f(a\vee a')+f(a\wedge a')=f(a)+f(a')\) for every real function \(f\), so both payoffs are supermodular in own action.
		
		To verify increasing differences for \(u_1\), first compare two specialization actions \(1+x<1+x'\). Their payoff difference is
		\(\mathbf1_{\{b=2\}}\bigl[\pi(x',\omega)-\pi(x,\omega)\bigr]-(x'^2-x^2)/2\).
		For \(0\le x<x'\le1\),
		\[
		\pi(x',\omega)-\pi(x,\omega)=
		\begin{cases}
			0,&\omega\le1-x',\\
			2(x'+\omega-1),&1-x'<\omega<1-x,\\
			2(x'-x),&\omega\ge1-x.
		\end{cases}
		\]
		This difference is nonnegative and nondecreasing in \(\omega\). The payoff difference between the two specialization actions is therefore nondecreasing in both \(b\) and \(\omega\), since the cost difference is constant in these variables.
		
		For a comparison between \(0\) and \(1+x\), the payoff difference is
		\(\mathbf1_{\{b=2\}}\pi(x,\omega)-\mathbf1_{\{b=0\}}-x^2/2\).
		As \(b\) increases from \(0\) to \(1\) to \(2\), this difference takes the values
		\(-1-x^2/2\), \(-x^2/2\), and \(\pi(x,\omega)-x^2/2\), respectively. These values are nondecreasing because \(\pi\ge0\). For each fixed \(b\), the difference is also nondecreasing in \(\omega\). These comparisons cover all ordered pairs of producer actions, proving increasing differences in \((a_1,b)\) and \((a_1,\omega)\). Finally, \(u_2=-(b-1)^2\) depends neither on \(a_1\) nor on \(\omega\).

\par\addvspace{\baselineskip}\noindent(b) Since \(\omega\) is uniformly distributed on \([0,1]\) and \(\pi(x,\omega)=0\) whenever \(\omega\le1-x\),
		\[
		\mathbb E[\pi(x,\omega)]
		=\int_{1-x}^{1}2(x+\omega-1)\,\dd\omega
		=\bigl[(x+\omega-1)^2\bigr]_{1-x}^{1}
		=x^2.
		\]
		For the second expectation, the integrand can be nonzero only when both \(\omega\ge1-x\) and \(\omega\ge1-\bar x\). Thus
	$
		\mathbb E\bigl[\pi(x,\omega)\mathbf1_{\{\omega\ge1-\bar x\}}\bigr]
		=\int_{\max\{1-x,1-\bar x\}}^{1}
		2(x+\omega-1)\,\dd\omega.
	$
		If \(x\le\bar x\), the lower endpoint is \(1-x\), so the integral equals \(x^2\). If \(x\ge\bar x\), the lower endpoint is \(1-\bar x\), giving
		\(x^2-(x-\bar x)^2=2x\bar x-\bar x^2\).
		The two expressions agree at \(x=\bar x\).

\par\addvspace{\baselineskip}\noindent(c) Fix \(x\in[0,1]\). Since \(\pi(x,\omega)\) is nonnegative and depends only on \(\omega\) for fixed \(x\), iterated expectation gives
		\[
		\begin{aligned}
			\mathbb E\bigl[\mathbf1_{\{b=2\}}\pi(x,\omega)\bigr]
			&=\mathbb E\bigl[
			\pi(x,\omega)\mathbb E[\mathbf1_{\{b=2\}}\mid\omega]
			\bigr]=\mathbb E\bigl[
			\pi(x,\omega)\Pr(b=2\mid\omega)
			\bigr] \le\varepsilon\,\mathbb E[\pi(x,\omega)]
			=\varepsilon x^2.
		\end{aligned}
		\]
		The inequality uses the conditional probability bound, while the last equality follows from~\eqref{eq:specialization-moments}. This estimate allows dependence between \(b\) and \(\omega\). Consequently,
		\[
		\begin{aligned}
			\mathbb E[u_1(1+x,b,\omega)]
			&=\mathbb E\bigl[\mathbf1_{\{b=2\}}\pi(x,\omega)\bigr]
			-\Pr(b=0)-\tfrac{x^2}{2} \le\left(\varepsilon-\tfrac12\right)x^2-\Pr(b=0).
		\end{aligned}
		\]
		For \(x>0\), the first term on the right is strictly negative because \(\varepsilon<1/2\). For \(x=0\), the right side equals \(-\Pr(b=0)<0\). Thus every specialization action \(1+x\) yields strictly negative expected payoff. Since \(u_1(0,b,\omega)=0\) for every \(b,\omega\), the standard product \(a_1=0\) is the unique maximizer.
	\end{proof}

\begin{proof}[Proof of Claim~\ref{claim:regional-chain}]
\noindent(a) The product specialization payoffs do not depend on private types. By Claim~\ref{claim:specialization}, they satisfy Assumption~\ref{ass:main}. They also have increasing differences in \((a_1,\omega)\). Thus Assumption~\ref{ass:types}(a) holds, with equality in all comparisons involving private types.

We couple the producer's beliefs by a common uniform draw \(s\in[0,1]\). The pair \((\omega,t_2)\) equals \((s,s)\) at \(L\) and \((s,1)\) at \(H\). This coupling is ordered, so the producer's beliefs are stochastically increasing. The retailer's beliefs are constant. A null set of \(\lambda_2\) is a Lebesgue null subset of \([0,1]\) and excludes \(1\), so both producer types assign it probability zero. Since both producer types have positive reference probability, the only null set of \(\lambda_1\) is empty. Null compatibility follows, and the game satisfies Assumptions~\ref{ass:main} and~\ref{ass:types}.

The retailer's unique strictly dominant action is \(1\). Against this action, the producer obtains \(-x^2/2\) from \(1+x\) and zero from \(0\), so its best response set is \(\{0,1\}\) at each type. For \(y,z\in\{0,1\}\), let \(\beta^{y,z}\) prescribe producer actions \(y\) at \(L\) and \(z\) at \(H\), and retailer action \(1\) at every type. Hence the pure Bayesian Nash equilibria are precisely the four profiles \(\beta^{y,z}\), proving~(a).

\par\addvspace{\baselineskip}\noindent(b) We first exclude perfect equilibria that prescribe action \(1\)
at \(H\). 
By Lemma~\ref{lem:optimality}, every pure perfect equilibrium is
one of the profiles \(\beta^{y,z}\).
Let \((g^n)_{n\ge1}\) support such a profile.
Since \(\lambda_2(\{1\})=1/2>0\), convergence of the supporting
sequence holds at retailer type \(1\).
In particular,
\(e_n=g_2^n(\{2\}\mid1)\to0\).
Complete mixing also gives
\(c_n=g_2^n(\{0\}\mid1)>0\).

Under \(p_1(\cdot\mid H)\), \(t_2=1\) almost surely and
\(\omega\) is uniform on \([0,1]\).
Thus the retailer's conditional order distribution given
\(\omega\) is \(g_2^n(\cdot\mid1)\).
Using~\eqref{eq:specialization-moments}, we obtain
\begin{equation}\label{eq:chain-high-payoff}
V_1(1+x,H;g_2^n)
=\left(e_n-\tfrac12\right)x^2-c_n.
\end{equation}
For all sufficiently large \(n\), we have \(e_n<1/2\), so this payoff is strictly negative for every \(x\in[0,1]\).
Since action \(0\) yields zero,
\(\BR_1(H;g_2^n)=\{0\}\) for all such \(n\).
If \(z=1\), then
\(d_1(1,\BR_1(H;g_2^n))=d_1(1,0)>0\),
contrary to~\eqref{eq:general-criterion}.
Hence \(z=0\), proving
\(\PP^p\subseteq\{\beta^{0,0},\beta^{1,0}\}\).

We construct a supporting sequence for \(\beta^{1,0}\).
Let \(h_n=1/(n+1)\),
\(\varepsilon_n=h_n^2/8\), and \(I_n=[1-h_n,1)\).
Define the retailer's strategy by the following probabilities.
\begin{center}
	\begin{tabular}{lccc}
		\toprule
		Retailer type & \(b=0\) & \(b=1\) & \(b=2\) \\
		\midrule
		\(t_2\in I_n\)
		& \(\varepsilon_n\)
		& \(\varepsilon_n\)
		& \(1-2\varepsilon_n\) \\
		\(t_2\notin I_n\)
		& \(\varepsilon_n\)
		& \(1-2\varepsilon_n\)
		& \(\varepsilon_n\) \\
		\bottomrule
	\end{tabular}
\end{center}
Fix a probability distribution \(\nu\) with full support on \(A_1\),
and let
\(g_1^n(\cdot\mid t_1)
=(1-\varepsilon_n)\delta_{\beta_1^{1,0}(t_1)}
+\varepsilon_n\nu\).
Both players' strategies are completely mixed.
For each fixed \(t_2<1\), we eventually have \(h_n<1-t_2\), so \(t_2\notin I_n\).
The type \(t_2=1\) never belongs to \(I_n\).
Thus \(g_2^n(\cdot\mid t_2)\Rightarrow\delta_1\) at every type.
The producer's strategies also converge to the prescribed actions, so \(g^n\) converges to \(\delta_{\beta^{1,0}}\) at every type.

Under \(p_1(\cdot\mid L)\), \(t_2=\omega\) almost surely and
\(\omega\) is uniform on \([0,1]\).
The conditional probability of a large order is therefore
\(g_2^n(\{2\}\mid\omega)
=\varepsilon_n+(1-3\varepsilon_n)\mathbf1_{I_n}(\omega)\).
Define
\(A_n(x)=\int_{1-h_n}^{1}\pi(x,\omega)\,\dd\omega\).
By~\eqref{eq:specialization-moments}, the expected payoff from
\(1+x\) is
$
V_1(1+x,L;g_2^n)
=\varepsilon_nx^2+(1-3\varepsilon_n)A_n(x)
-\varepsilon_n-\tfrac{x^2}{2}.
$
Equation~\eqref{eq:specialization-moments} also gives \(A_n(h_n)=h_n^2\).
Consequently,
\[
\begin{aligned}
	V_1(1+h_n,L;g_2^n)
	&=\left(\tfrac12-2\varepsilon_n\right)h_n^2-\varepsilon_n
	=\left(\tfrac38-2\varepsilon_n\right)h_n^2>0.
\end{aligned}
\]
Action \(0\) yields zero, whereas action \(1\) yields
\(V_1(1,L;g_2^n)=-\varepsilon_n<0\).
Every best response at \(L\) therefore has the form \(1+x_n\)
with \(x_n>0\).
To show that these best responses approach action \(1\),
note that \(0\le A_n(x)\le2h_n\), since \(0\le\pi(x,\omega) \le2\).
For every \(x\in[0,1]\),
\[
\begin{aligned}
	V_1(1+x,L;g_2^n)
	&=\varepsilon_n(x^2-1)+(1-3\varepsilon_n)A_n(x)
	-\tfrac{x^2}{2}
	\le A_n(x)-\tfrac{x^2}{2}
	\le2h_n-\tfrac{x^2}{2}.
\end{aligned}
\]
Since \(V_1(1+x_n,L;g_2^n)\ge V_1(1+h_n,L;g_2^n)>0\), we obtain
\(0<x_n<2\sqrt{h_n}\to0\).
Hence
\(d_1(1,\BR_1(L;g_2^n))
\le d_1(1,1+x_n)\to0\).

At type \(H\), the producer assigns probability one to retailer
type \(1\), which lies outside \(I_n\).
Equation~\eqref{eq:chain-high-payoff}, with \(e_n=c_n=\varepsilon_n\),
therefore makes the payoff from every \(1+x\) strictly negative.
Thus action \(0\) is the unique best response at \(H\).
The retailer's unique best response remains \(1\) at every type.
Together with convergence of \(g^n\), these observations verify
\eqref{eq:general-criterion}.
Lemma~\ref{lem:criterion}(b) yields
\(\beta^{1,0}\in\PP^p\).

To support \(\beta^{0,0}\), let the retailer use the probabilities
\((\varepsilon_n,1-2\varepsilon_n,\varepsilon_n)\)
at every type.
Let the producer use
\(g_1^n(\cdot\mid t_1)
=(1-\varepsilon_n)\delta_0+\varepsilon_n\nu\)
at both types.
Because the retailer's order distribution is constant in type,
\eqref{eq:chain-high-payoff}, with \(e_n=c_n=\varepsilon_n\), applies
at either producer type and makes every specialization payoff strictly negative.
Thus action \(0\) is the producer's unique best response,
while \(1\) remains the retailer's unique best response.
These completely mixed profiles converge to
\(\delta_{\beta^{0,0}}\), proving
\(\beta^{0,0}\in\PP^p\).

We have established
\(\PP^p=\{\beta^{0,0},\beta^{1,0}\}\).
Since \(\beta^{0,0}\le\beta^{1,0}\),
Theorem~\ref{thm:general-global} identifies
\(\delta_{\beta^{0,0}}\) and \(\delta_{\beta^{1,0}}\)
as the smallest and largest elements of \(\PP^b\).
In particular,
\(\delta_{\beta^{0,0}}\st r\st\delta_{\beta^{1,0}}\)
for every \(r\in\PP^b\).

\par\addvspace{\baselineskip}\noindent(c) The producer's strategy in \(\beta^{1,0}\) decreases from \(1\) at \(L\) to \(0\) at \(H\), so \(\beta^{1,0}\) is not monotone. The profile \(\beta^{0,0}\) is monotone and is therefore the unique perfect monotone equilibrium, proving Claim~\ref{claim:regional-chain}(c).

\par\addvspace{\baselineskip}\noindent(d) Reverse all action and type orders, including the order on residual states. By Remark~\ref{rem:dual}, the resulting game satisfies Assumptions~\ref{ass:main} and~\ref{ass:types}. It has the same perfect equilibria and monotone strategies, with the order on profiles reversed. Its smallest perfect equilibrium, \(\beta^{1,0}\), is therefore not monotone, which proves Claim~\ref{claim:regional-chain}(d).

\end{proof}

\begin{proof}[Proof of Remark~\ref{rem:atomless-counterexample}]
The action and payoff conditions follow from Claim~\ref{claim:specialization}(a), since payoffs do not depend on private types. On a common realization of \((S,U)\), the producer's low and high type beliefs satisfy \((S,S)\le(S,1+U)\). The retailer's beliefs are constant. The relevant opponent type marginals are \(U[0,1]\) and \(U[1,2]\) for the producer, and \(U[0,1/2]\) for the retailer. They are absolutely continuous with respect to the corresponding reference probabilities. This proves stochastic monotonicity and null compatibility, hence both assumptions.

Let \(r\in\PP^b\), supported by \((g^n)\). By Lemma~\ref{lem:optimality}, \(r\) is a Bayesian Nash equilibrium. Thus the retailer chooses \(1\) almost everywhere, while the producer's action distribution is supported on \(\{0,1\}\) almost everywhere. Consequently, \(g_2^n(\cdot\mid t_2)\Rightarrow\delta_1\) for \(\lambda_2\)-almost every \(t_2\). Put
\[
E_n=\int_1^2g_2^n(\{2\}\mid t_2)\,\dd t_2,C_n=\int_1^2g_2^n(\{0\}\mid t_2)\,\dd t_2.
\]
Bounded convergence gives \(E_n\to0\), while complete mixing gives \(C_n>0\). For \(t_1\ge1/2\), independence of \(\omega\) and \(t_2\), together with~\eqref{eq:specialization-moments}, yields \(V_1(1+x,t_1;g_2^n)=(E_n-\tfrac12)x^2-C_n\). This is strictly negative for all \(x\in[0,1]\) eventually, so the best response set is \(\{0\}\). Definition~\ref{def:general} therefore implies \(r_1(\cdot\mid t_1)=\delta_0\) almost everywhere on \([1/2,1]\). It follows that \(r\st\delta_{\widehat\alpha^+}\).

We next support \(\widehat\alpha^+\). Let \(h_n=1/(n+1)\), \(\varepsilon_n=h_n^2/8\), and \(I_n=[1-h_n,1)\). Let \(g_2^n\) assign probabilities \((\varepsilon_n,\varepsilon_n,1-2\varepsilon_n)\) to orders \((0,1,2)\) on \(I_n\), and \((\varepsilon_n,1-2\varepsilon_n,\varepsilon_n)\) elsewhere in \([0,2]\). Fix a full support probability \(\nu\) on \(A_1\), and set \(g_1^n(\cdot\mid t_1)=(1-\varepsilon_n)\delta_{\widehat\alpha_1^+(t_1)}+\varepsilon_n\nu\). These profiles are completely mixed and converge to \(\delta_{\widehat\alpha^+}\) at every type.

For \(t_1<1/2\), put \(A_n(x)=\int_{1-h_n}^1\pi(x,\omega)\,\dd\omega\). The low type payoff is \(\varepsilon_nx^2+(1-3\varepsilon_n)A_n(x)-\varepsilon_n-x^2/2\). By~\eqref{eq:specialization-moments}, \(A_n(h_n)=h_n^2\), while \(0\le A_n(x)\le2h_n\). Hence
\[
V_1(1+h_n,t_1;g_2^n)=(\tfrac38-2\varepsilon_n)h_n^2>0,V_1(1+x,t_1;g_2^n)\le2h_n-\tfrac{x^2}{2}.
\]
Every low type best response is therefore \(1+x_n\), with \(0<x_n<2\sqrt{h_n}\to0\). In particular, \(d_1(1,\BR_1(t_1;g_2^n))\to0\) for low types. High types face the constant order probabilities outside \(I_n\), since their retailer types lie in \([1,2]\). Their specialization payoff is \((\varepsilon_n-\tfrac12)x^2-\varepsilon_n<0\), so their unique best response is \(0\). The retailer's unique best response remains \(1\). Thus~\eqref{eq:general-criterion} holds, proving \(\widehat\alpha^+\in\PP^p\) and establishing the largest element of \(\PP^b\).

To support \(\widehat\alpha^-\), use retailer probabilities \((\varepsilon_n,1-2\varepsilon_n,\varepsilon_n)\) at every type and producer strategies \((1-\varepsilon_n)\delta_0+\varepsilon_n\nu\). Every specialization payoff is then \((\varepsilon_n-\tfrac12)x^2-\varepsilon_n<0\) at either producer belief. Hence \(\widehat\alpha^-\) is perfect. Since \(0\) is the smallest producer action and every perfect equilibrium prescribes retailer action \(1\), its degenerate profile is the smallest element of \(\PP^b\). Finally, any increasing representative of a pure perfect strategy must be \(0\) almost everywhere on the lower half because it is \(0\) almost everywhere on the upper half. Thus \(\widehat\alpha^-\) is the unique perfect monotone equilibrium, whereas \(\widehat\alpha^+\) is not monotone.
\end{proof}

\begin{proof}[Proof of Claim~\ref{claim:no-extremal}]

\noindent(a) 
The finite sets \(C\) and \(D\) are closed under coordinatewise
maxima and minima. Hence \(C\times[0,1]\) and \(D\), equipped
with the specified metrics, are compact metric lattices.
Payoffs are bounded and jointly measurable.
They are continuous in actions because \(C,D\) are finite
and \(\pi\) is continuous.
The type spaces, reference probabilities, and beliefs coincide
with those in Example~\ref{ex:regional-chain}.
The verification in part~(a) of the proof of
Claim~\ref{claim:regional-chain} therefore establishes
the null compatibility condition and Assumption~\ref{ass:types}(b).

As established in part~(a) of the proof of
Claim~\ref{claim:specialization}, \(\pi\) is nonnegative,
increasing in each argument, and has increasing differences.
All sums below are over \(j\in\{x,y\}\).
For the producer's supermodularity, take any two actions
\((l,v)\) and \((l',v')\), with \(v\le v'\) without loss
of generality. Since coordinatewise maxima and minima
preserve both the sum of production scales and the number
of positive coordinates, the cancellation losses and
type-dependent benefits cancel in the supermodular difference.
The investment costs also cancel. Thus
\[
\begin{aligned}
	&u_1(l\vee l',v',b,t)+u_1(l\wedge l',v,b,t) -u_1(l,v,b,t)-u_1(l',v',b,t)\\
	&=
	\sum_j b_j\Big[
	\bigl(\max\{l_j,l'_j\}-l'_j\bigr)\pi(v',\omega)
	+\bigl(\min\{l_j,l'_j\}-l_j\bigr)\pi(v,\omega)
	\Big]\\
	&=
	\sum_j b_j(l_j-l'_j)_+
	\bigl[\pi(v',\omega)-\pi(v,\omega)\bigr]\ge0.
\end{aligned}
\]

To verify increasing differences, now suppose also that
\(l'\ge l\).
In the payoff difference
\(u_1(l',v',b,t)-u_1(l,v,b,t)\), the coefficient of \(b_j\) is
\[
l'_j\pi(v',\omega)-l_j\pi(v,\omega)
=(l'_j-l_j)\pi(v',\omega)
+l_j\bigl[\pi(v',\omega)-\pi(v,\omega)\bigr]\ge0.
\]
The coefficient of \(z\) is \(\sum_j(l'_j-l_j)\ge0\).
Hence this payoff difference is increasing in the retailer's
action.
The same decomposition shows that the coefficient of each
\(b_j\) is increasing in \(\omega\), using the increasing
differences of \(\pi\).
Since \(b_j\ge0\), the payoff difference is increasing
in \(\omega\).
Its coefficient of \(\theta(t_1)\) is
\(\sum_j(\mathbf1_{\{l'_j>0\}}-\mathbf1_{\{l_j>0\}})\ge0\), so it is increasing in \(t_1\).
It is independent of \(t_2\).
This proves increasing differences in the entire type vector.

The retailer's payoff is independent of the producer's action
and all types, so its increasing-differences conditions hold
with equality.
Its only incomparable pair of actions is
\((1,0,1)\) and \((0,1,1)\).
Their meet is \((0,0,1)\), while their join is \((1,1,1)\).
The corresponding supermodular inequality is \(0+0\le1+0\).
Comparable pairs satisfy the inequality with equality.
Thus both players satisfy the required payoff order conditions,
which completes the proof of~(a).

 \par\addvspace{\baselineskip}\noindent(b)
The retailer's normal order \((0,0,1)\) is strictly dominant.
Against this action, the producer's payoff is
\(\theta(t_1)(\mathbf1_{\{l_x>0\}}+\mathbf1_{\{l_y>0\}})-v^2/2\).
Thus every pure Bayesian Nash equilibrium has zero investment
at both types, an arbitrary production plan at \(L\), and
either \(U\) or \(V\) at \(H\).
By Lemma~\ref{lem:optimality}, these restrictions also apply
to every pure perfect equilibrium.

Fix a completely mixed retailer strategy \(g_2\).
Write \(e_j=g_2(\{b_j=1\}\mid1)\) for \(j=x,y\), and
\(e_0=g_2(\{z=0\}\mid1)>0\).
Under \(p_1(\cdot\mid H)\), \(t_2=1\) almost surely and
\(\omega\) is uniform on \([0,1]\).
Equation~\eqref{eq:specialization-moments} therefore gives
\begin{equation}\label{eq:portfolio-high-payoff}
V_1((l,v),H;g_2)
=\bigl(l_xe_x+l_ye_y-\tfrac12\bigr)v^2
-(l_x+l_y)e_0
+\mathbf1_{\{l_x>0\}}+\mathbf1_{\{l_y>0\}}.
\end{equation}
Along any supporting sequence, \(e_x,e_y,e_0\to0\),
because retailer type \(1\) has positive reference probability.
Once \(e_x+e_y<1/6\), the coefficient of \(v^2\) is strictly
negative for every \(l\in C\), so every best response has \(v=0\).
If also \(e_0<1/2\), plan \(U\) uniquely maximizes the payoff.
Indeed, its payoff \(2-4e_0\) strictly exceeds
\(0\), \(1-2e_0\), \(1-2e_0\), and \(2-6e_0\),
the respective payoffs of \(O,X,Y,V\).
Hence every pure perfect equilibrium prescribes \((U,0)\)
at \(H\).

We next exclude plan \(U\) at \(L\).
Under \(p_1(\cdot\mid L)\), \(t_2=\omega\) almost surely and
\(\omega\) is uniform on \([0,1]\).
Since \(\theta(L)=0\) and \(\pi(0,\omega)=0\),
complete mixing gives
\(V_1((U,0),L;g_2)
=-4\int_0^1g_2(\{z=0\}\mid\omega)\,\dd\omega<0\),
whereas \((O,0)\) yields zero.
For fixed investment \(v\), changing the production plan
from \(U=(2,2)\) to \(V=(3,3)\) multiplies the operating
return by \(3/2\) without changing the investment cost.
Hence
\[
V_1((V,v),L;g_2)-V_1((U,v),L;g_2)
=\tfrac12V_1((U,v),L;g_2)+\tfrac{v^2}{4}.
\]
If \(V_1((U,v),L;g_2)<0\), then \((O,0)\) is strictly better.
Otherwise, the strict negativity at \(v=0\) implies \(v>0\), so the displayed difference is strictly positive.
Thus \((V,v)\) is strictly better.
No best response at type \(L\) therefore uses plan \(U\).

Moreover, every other production plan has coordinate distance
at least \(2\) from \(U\).
Consequently,
\(d_1((U,0),\BR_1(L;g_2))\ge2\)
for every completely mixed \(g_2\).
Best responses along a supporting sequence therefore cannot
converge to \((U,0)\).
Since \(\lambda_1(\{L\})>0\), this contradicts
\eqref{eq:general-criterion} for any pure perfect equilibrium
prescribing \((U,0)\) at \(L\).

Let \(\alpha^x\) be given by
\(\alpha_1^x(L)=(X,0)\), \(\alpha_1^x(H)=(U,0)\), and
\(\alpha_2^x(t_2)=(0,0,1)\) at every retailer type.
To construct a supporting sequence, set
\(h_n=1/(16n)\), \(\eta_n=\tfrac52h_n^2\), and
\(\varepsilon_n=h_n^4\).
Write \(d_n(t_2)=(\mathbf1_{\{1-h_n\le t_2<1\}},0,1)\).
Let \(\nu_2\) be uniform on \(D\), and let \(\nu_1\)
have full support on \(C\times[0,1]\).
Define
\[
g_2^n(\cdot\mid t_2)=(1-\varepsilon_n)\bigl[\eta_n\delta_{(0,0,0)}+(1-\eta_n)\delta_{d_n(t_2)}\bigr]+\varepsilon_n\nu_2,g_1^n(\cdot\mid t_1)=(1-\varepsilon_n)\delta_{\alpha_1^x(t_1)}+\varepsilon_n\nu_1.
\]
These profiles are completely mixed.
Every fixed retailer type eventually lies outside
\([1-h_n,1)\), including type \(1\), which never belongs
to this interval.
Thus \(g^n\) converges to \(\delta_{\alpha^x}\) at every type.

Put \(A_n(v)=\int_{1-h_n}^1\pi(v,\omega)\,\dd\omega\).
Equation~\eqref{eq:specialization-moments}, with \(x=v\) and
\(\bar x=h_n\), determines \(A_n\).
Before mixing with \(\nu_2\), the low-type payoff is
\((1-\eta_n)l_xA_n(v)-\eta_n(l_x+l_y)-v^2/2\).
Since \(A_n(v)\le2h_n\), \(\eta_n=\tfrac52h_n^2\), and
\(\varepsilon_n=h_n^4\), there is a constant \(K>0\),
independent of \(n,l,v\), such that
\[
|
V_1((l,v),L;g_2^n)-l_xA_n(v)
+\tfrac{v^2}{2}+\tfrac52(l_x+l_y)h_n^2
|
\le K h_n^3
\]
for every \((l,v)\in C\times[0,1]\).
After division by \(h_n^2\), the error is bounded by
\(K h_n\to0\), uniformly over all actions.
For \(l_x\in\{0,2,3\}\), the maximum of
\(l_xA_n(v)-v^2/2\) is \((2l_x^2-l_x)h_n^2\),
attained at \(v=2l_xh_n\).
Consequently, the maximal payoffs for \(O,X,Y,U,V\),
divided by \(h_n^2\), converge respectively to
\(0,1,-5,-4,0\).
Thus every best response at \(L\) uses plan \(X\)
for all sufficiently large \(n\).

The payoffs converge uniformly to \(-v^2/2\).
Since \((O,0)\) yields zero, best-response investments
must converge to zero.
Hence there are best responses \((X,v_n)\) with \(v_n\to0\), so \(d_1((X,0),\BR_1(L;g_2^n))\le v_n\to0\).
Equation~\eqref{eq:portfolio-high-payoff} gives the unique best response
\((U,0)\) at \(H\) for all sufficiently large \(n\).
The retailer's unique best response remains \((0,0,1)\).
These facts verify \eqref{eq:general-criterion}, so Lemma~\ref{lem:criterion}(b) gives \(\alpha^x\in\PP^p\).
Interchanging the products gives \(\alpha^y\in\PP^p\),
where \(\alpha^y\) differs from \(\alpha^x\) only by
prescribing \((Y,0)\) at \(L\).

Theorem~\ref{thm:general-global} supplies the smallest and
largest pure perfect equilibria.
The smallest lies below both \(\alpha^x\) and \(\alpha^y\),
forcing plan \(X\wedge Y=O\) at \(L\).
The largest lies above both, requiring plan \(U\) or \(V\)
at \(L\).
Since \(U\) has been excluded, its plan must be \(V\).
Both extrema have zero investment, prescribe \(U\) at \(H\),
and prescribe the retailer's normal order.
They are therefore \(\alpha^-\) and \(\alpha^+\),
respectively.
Together with the restrictions already established,
this gives
\(\PP^p=\{\alpha^-,\alpha^x,\alpha^y,\alpha^+\}\).
Theorem~\ref{thm:general-global} further implies
\(\delta_{\alpha^-}\st r\st\delta_{\alpha^+}\)
for every \(r\in\PP^b\), proving~(b).

\par\addvspace{\baselineskip}\noindent(c) Exactly the first three profiles are monotone. A monotone perfect upper bound of \(\alpha^x\) and \(\alpha^y\) would require \(l(L)\ge X\vee Y=U\). Since \(l(H)=U\), monotonicity would force \(l(L)=U\), which perfection excludes. This proves~(c).

\par\addvspace{\baselineskip}\noindent(d) Reverse all action and type orders as in Remark~\ref{rem:dual}, and take the product of the original game with its dual using four distinct players. Use the product residual state space. Extend each player's beliefs by an independent fixed distribution on the other component with the prescribed reference marginals. The payoff, stochastic monotonicity, and null compatibility conditions are preserved. Each player's payoff and interim best response set depend only on opponents in the same factor. Her action space and its metric are unchanged, so her Prohorov distance to that best response set is also unchanged. A completely mixed supporting sequence for the product therefore projects to supporting sequences for both factors. Conversely, combine the \(n\)th profiles from any two supporting sequences. Complete mixing, weak convergence, and the distance condition in Definition~\ref{def:general} then hold separately for each player, giving a supporting sequence for the product. Thus its perfect equilibrium set is the Cartesian product of those of the factors. Since each player's type space and order are unchanged, the same is true of their monotone pure subsets. The original monotone subset has no largest element, and its dual has no smallest element, so their Cartesian product has neither extremum. The largest perfect equilibrium has a nonmonotone original component, while the smallest has a nonmonotone dual component. This proves~(d).
\end{proof}

\begin{proof}[Proof of Proposition~\ref{prop:general-comparative}]

\noindent(i) Use the pairs \((s^n,g^n)\) satisfying~\eqref{eq:global-approximation} and their counterparts \((s'^n,g'^n)\) at \(\theta\le\theta'\). Use the same functions \(\chi_i\), so \(d_{\mathcal Q},J,R,\Phi\) are common to both games, and primes distinguish the objects that depend on the game. Increasing differences and Lemma~\ref{lem:choice}(b) give \(s_i^n(t_i)\le\overline{\BR}_i'(t_i;g_{-i}^n)\) almost everywhere. Lemma~\ref{lem:glue}(b), applied in the second game with \(s=s^n\), \(g=g^n\), \(h=g'^n\), and \(\eta=1/(n+1)\), yields, by~\eqref{eq:gluing-estimate},
\(s^n\vee s'^n\le z^n\in\mathcal Z'_{\tau_n}, \tau_n=\Phi\bigl(R(s^n,g^n)+e'(g'^n)\bigr)+\tfrac{N\bar d}{n+1}\longrightarrow0\), 
because \(e'(g'^n)\le R(s'^n,g'^n)\to0\). Therefore \(J(s'^n)\le J(z^n)\le M'_{\tau_n}\to M'\). Since \(J(s'^n)\to M'\), Lemma~\ref{lem:space}(c) gives \(d_{\mathcal Q}(z^n,s'^n)\to0\). Passing to an almost everywhere convergent subsequence in \(s^n\le z^n\) yields \(\alpha^+(\theta)\le\alpha^+(\theta')\). Reversing the action and parameter orders as in Remark~\ref{rem:dual} gives the comparison of the smallest equilibria.

\par\addvspace{\baselineskip}\noindent(ii) Use the operators \(\Psi^\varepsilon,\Psi\) from the proof of Theorem~\ref{prop:general-finite}, with the same full support measures and vanishing sequence in both games. At every increasing profile \(\alpha\), the behavioral strategies \(\sigma_j^\varepsilon(\alpha)\) are stochastically increasing in type. Lemma~\ref{lem:interim}(d) and \(p_i(\cdot\mid t_i)\st p_i'(\cdot\mid t_i)\) therefore make
\(V_i^{p'}(\cdot,t_i;\sigma_{-i}^\varepsilon(\alpha))- V_i^p(\cdot,t_i;\sigma_{-i}^\varepsilon(\alpha))\) 
increasing in own action. Lemma~\ref{lem:choice}(b) gives \(\Psi^\varepsilon(\alpha)\le\Psi'^\varepsilon(\alpha)\), and hence \(\Psi\le\Psi'\) on \(\mathcal Q_{\mathrm{inc}}\). The proof of Theorem~\ref{prop:general-finite} identifies \(\alpha^+(p)\) and \(\alpha^+(p')\) with the largest fixed points of these operators on \(\mathcal Q_{\mathrm{inc}}\). Since \(\alpha^+(p)\le\Psi'(\alpha^+(p))\), Tarski's fixed point theorem gives \(\alpha^+(p)\le\alpha^+(p')\). Reversing all action and type orders gives the comparison of the smallest equilibria.

\par\addvspace{\baselineskip}\noindent(iii) Give the producer in Example~\ref{ex:regional-chain} a singleton type. Under \(p\), it believes that \((\omega,t_2)=(s,s)\), and under \(p'\), it believes that \((\omega,t_2)=(s,1)\), where \(s\sim U[0,1]\). The retailer assigns probability one to the producer's unique type. The coupling and null compatibility verification in Claim~\ref{claim:regional-chain} apply unchanged. The proof of that claim at each producer belief gives largest perfect actions \(1\) under \(p\) and \(0\) under \(p'\), while the retailer chooses \(1\) in both games. Hence \(\alpha^+(p')<\alpha^+(p)\).
\end{proof}

\begin{proof}[Proof of Claim~\ref{claim:regional-chain-inc}]
For any perturbation with \(\varepsilon_2<1/2\), Remark~\ref{rem:perturbed} preserves the retailer's unique best response \(1\). Its induced probabilities of orders \(0\) and \(2\) are therefore positive and at most \(\varepsilon_2\) almost everywhere. Null compatibility and Claim~\ref{claim:specialization}(c) then make \(0\) the producer's unique best response at both types. Thus the perturbed game has the unique behavioral equilibrium \(\delta_{\beta^{0,0}}\). Any vanishing sequence of such perturbations supports this profile, including sequences with type independent tremble distributions. The claimed singleton sets follow, and Claim~\ref{claim:regional-chain}(b) gives the comparison of extrema.
\end{proof}

\begin{proof}[Proof of Claim~\ref{claim:binary-separation}]

\noindent(a) On binary action chains, supermodularity in own action and continuity in actions are automatic. The gain \(t_i^2+a_j-1\) increases in \(t_i\) and \(a_j\). Independent uniform types give constant beliefs and null compatibility, so both maintained assumptions hold.

Against noninvestment probability \(x_j\), the unique best response outside one null type is \(\mathbf1_{\{t_i>\sqrt{x_j}\}}\), with noninvestment probability \(\sqrt{x_j}\). Hence every behavioral equilibrium is pure almost everywhere, and its noninvestment probabilities satisfy \(y_1^2=y_2\) and \(y_2^2=y_1\). The only solutions are \((0,0)\) and \((1,1)\), proving~(a).

\par\addvspace{\baselineskip}\noindent(b) To support the profile \((1,1)\) under perfection, let both players use the type independent distributions \(\varepsilon_n\delta_0+(1-\varepsilon_n)\delta_1\), where \(0<\varepsilon_n\downarrow0\). The investment gain \(t_i^2-\varepsilon_n\) is eventually positive at every \(t_i>0\), so~\eqref{eq:general-criterion} holds. Part~(c) below supplies \((0,0)\) under increasing perturbations. Its inclusion in \(\PP^b\), together with~(a), proves~(b).

\par\addvspace{\baselineskip}\noindent(c) For any perturbation \(q=(\varepsilon,\mu)\) and \(r\in\EE(q)\), let \(y_i=\int_0^1r_i(\{0\}\mid t_i)\dd t_i\) and \(z_i=\int_0^1\mu_i(\{0\}\mid t_i)\dd t_i\in(0,1)\). The same cutoff calculation, using Remark~\ref{rem:perturbed}, gives \(y_i^2=(1-\varepsilon_j)y_j+\varepsilon_jz_j\ge(1-\varepsilon_j)y_j\) for \(i\ne j\). Full support makes both \(y_i\) positive. Multiplication and division by \(y_1y_2\) yield~\eqref{eq:binary-contagion}.
Thus \(y_1,y_2\to1\) along every sequence with vanishing trembles. By~\eqref{eq:small}, the equilibrium profiles and their induced profiles have the same conditional weak limits. Bounded convergence therefore makes any such limit equal to \(\delta_{(0,0)}\) almost everywhere. This exclusion requires neither type independence nor stochastic monotonicity of the tremble distributions.

By Theorem~\ref{thm:main}, \(\PP_{\mathrm{inc}}^b\) is nonempty, so it equals \(\{\delta_{(0,0)}\}\), proving~(c).
\end{proof}

\subsection{Prescribed tremble distributions}\label{oa:fixed}
In Definition~\ref{def:perfect}, the conditional tremble distributions may vary along the supporting sequence, even when they are required to be type independent. Here we impose the additional restriction that one profile \(\mu\) is fixed for the entire sequence, so that only the tremble probabilities vary. The distributions may still depend on type, and the tremble probabilities may converge to zero at different rates across players. We show in Example~\ref{ex:private-value-errors} that this restriction can tighten equilibrium bounds and in Proposition~\ref{prop:failure} that it can eliminate largest and smallest elements.

Fix a profile \(\mu=(\mu_i)_{i\in I}\) of conditional tremble distributions satisfying the requirements of Section~\ref{sec:definition}. Let \(\PP_\mu^b\) be the set of behavioral strategy profiles that satisfy Definition~\ref{def:perfect} with perturbations of the form \(q^n=(\varepsilon^n,\mu)\), and let \(\PP_\mu^p=\{\alpha\mid\delta_\alpha\in\PP_\mu^b\}\). Since all other requirements of Definition~\ref{def:perfect} remain in force,
\begin{equation}\label{eq:inclusions}
\PP_\mu^p\subseteq\PP_{\mathrm{inc}}^p,\PP_\mu^b\subseteq\PP_{\mathrm{inc}}^b.
\end{equation}
All conclusions of this subsection also hold if Definition~\ref{def:perfect} in Section~\ref{sec:results} is restricted to type independent distributions \(\mu_i^n\) that may vary with \(n\). In that version, the prescribed distributions considered below are also type independent. The notation \(\PP_\mu^b\) retains the additional requirement \(\mu_i^n=\mu_i\) for every \(n\).

Prescribing the conditional tremble distributions leaves the prediction unchanged when all sufficiently small perturbed games share one equilibrium strategy profile, as in Example~\ref{ex:regional-chain}. The next example uses the interim investment game already analyzed in Claim~\ref{claim:interim-beliefs}. It shows how prescribed distributions affect the unique equilibrium of each perturbed game and its possible limits.

\begin{example}\label{ex:private-value-errors}
Two firms choose whether to invest, \(a_i\in\{0,1\}\). Their types \(t_1\) and \(t_2\) are independent, each uniformly distributed on \([0,1]\). Firm \(i\)'s payoff is \(u_i(a_i,a_j,t_i,t_j)=a_i(t_i+a_j-1)\). Investment yields the private return \(t_i\), costs one, and yields a complementary benefit of one if the other firm also invests. For \(m\in[0,1]\), let \(\alpha_i^m(t_i)=\mathbf1_{\{t_i>1-m\}}\) and \(\alpha^m=(\alpha_1^m,\alpha_2^m)\). For a profile \(\mu\) of conditional tremble distributions that satisfies the requirements of Section~\ref{sec:definition}, let \(\bar\mu_i=\int_0^1\mu_i(\{1\}\mid t_i)\dd t_i\in(0,1)\).
\end{example}

The following claim summarizes the findings, and we prove it at the end of this section.

\begin{claim}\label{claim:private-value-errors}
The game in Example~\ref{ex:private-value-errors} has the following properties.
\begin{enumerate}[label=(\alph*)]
\item It satisfies Assumptions~\ref{ass:main} and~\ref{ass:types}, and up to equality almost everywhere, its Bayesian Nash equilibria are the pure profiles \(\alpha^m\) with \(m\in[0,1]\).
\item For every such profile \(\mu\),
\[
\PP_\mu^p=\{\alpha^m\mid m\in[\min_i\bar\mu_i,\max_i\bar\mu_i]\},\PP_\mu^b=\{\delta_{\alpha^m}\mid m\in[\min_i\bar\mu_i,\max_i\bar\mu_i]\}.
\]
In particular, if \(\bar\mu_1=\bar\mu_2\), then \(\PP_\mu^p=\{\alpha^{\bar\mu_1}\}\), regardless of the relative rates at which the tremble probabilities vanish, and uniform tremble distributions select \(\alpha^{1/2}\).
\item \(\PP_{\mathrm{inc}}^p=\{\alpha^m\mid m\in[0,1]\}\) and \(\PP_{\mathrm{inc}}^b=\{\delta_{\alpha^m}\mid m\in[0,1]\}\). The smallest and largest elements of \(\PP_{\mathrm{inc}}^p\) are \(\alpha^0\) and \(\alpha^1\).
\end{enumerate}
\end{claim}

Without prescribed tremble distributions, the bounds are \(\alpha^0\), in which neither firm invests, and \(\alpha^1\), in which both always invest. Prescribing the distributions narrows these bounds to cutoff profiles \(\alpha^m\) with \(m\) between \(\bar\mu_1\) and \(\bar\mu_2\). For the simplified rules of Section~\ref{sec:definition}, \(\bar\mu_i\) is firm \(i\)'s investment probability conditional on using its rule, averaged over types. More generally, whenever \(\PP_\mu^p\) has smallest and largest elements, the inclusion~\eqref{eq:inclusions} and Theorem~\ref{thm:main} imply
\[
\alpha_{\mathrm{inc}}^-\le \min\PP_\mu^p\le\max\PP_\mu^p\le\alpha_{\mathrm{inc}}^+
,\text{almost everywhere}.
\]

Prescribing tremble distributions can also eliminate largest and smallest elements. In the next game, a manufacturer chooses capacity and inventory. A supplier and a distributor contract at zero but may mistakenly deliver or order too much or too little. Under prescribed uniform trembles, different relative frequencies of these mistakes support incomparable equilibria with no common upper or lower bound. Raising a tremble probability around the contracted action shifts probability both upward and downward, so it need not change behavior monotonically in the stochastic order.

\begin{proposition}\label{prop:failure}
There is a game with three players, complete information, and finite action lattices that satisfies Assumptions~\ref{ass:main} and~\ref{ass:types}. For some prescribed profile \(\mu\) of conditional tremble distributions with full support, the following hold.
\begin{enumerate}[label=(\roman*)]
\item \(\PP_\mu^p\) is nonempty but has neither a largest nor a smallest element, and \(\PP_\mu^b\) has neither a largest nor a smallest element.
\item \(\PP_{\mathrm{inc}}^p\) and \(\PP_{\mathrm{inc}}^b\) have largest and smallest elements, attained by pure strategy profiles.
\end{enumerate}
The same conclusions hold in a Bayesian version of the game with independent, atomless types and payoffs that do not depend on types.
\end{proposition}

The proof at the end of this section specifies the game, verifies the assumptions, and constructs the supporting sequences.

\begin{proof}[Proof of Claim~\ref{claim:private-value-errors}]

\noindent(a) The binary action chains, bounded continuous payoffs, and increasing gain \(t_i+a_j-1\) give the payoff assumptions. Independent uniform types give constant beliefs and null compatibility. The interim payoff \(a_i(t_i+m_j-1)\) is identical to that of Example~\ref{ex:interim-beliefs} against every behavioral strategy. By Remark~\ref{rem:perturbed}, the two games therefore have the same equilibrium sets both without perturbations and under every admissible perturbation. The cutoff argument in Claim~\ref{claim:interim-beliefs} therefore proves~(a).

\par\addvspace{\baselineskip}\noindent(b) Fix \(q=(\varepsilon,\mu)\). Let \(x_i\) be player \(i\)'s realized ex ante investment probability. The cutoff best response has investment probability \(x_j\), so equilibrium requires
\[
x_1=\varepsilon_1\bar\mu_1+(1-\varepsilon_1)x_2,x_2=\varepsilon_2\bar\mu_2+(1-\varepsilon_2)x_1.
\]
The unique solution is
\begin{equation}\label{eq:privatevalueprob}
x_1=\tfrac{\varepsilon_1\bar\mu_1+(1-\varepsilon_1)\varepsilon_2\bar\mu_2}{\varepsilon_1+\varepsilon_2-\varepsilon_1\varepsilon_2},x_2=\tfrac{\varepsilon_2\bar\mu_2+(1-\varepsilon_2)\varepsilon_1\bar\mu_1}{\varepsilon_1+\varepsilon_2-\varepsilon_1\varepsilon_2}.
\end{equation}
Both coordinates lie in \((0,1)\), and their cutoff strategies give the unique behavioral equilibrium up to null sets.

As \(\varepsilon_i^n\to0\), every subsequential limit of
\(w_n=\varepsilon_1^n/(\varepsilon_1^n+\varepsilon_2^n)\)
belongs to \([0,1]\). If \(w_n\to w\), then~\eqref{eq:privatevalueprob} gives \(x_1^n,x_2^n\to m=w\bar\mu_1+(1-w)\bar\mu_2\), and the induced strategies converge to \(\delta_{\alpha^m}\) except at the null cutoff type. Conversely, every \(w\in[0,1]\) is attainable by positive vanishing tremble probabilities. For interior weights, keep the ratio of the tremble probabilities fixed at \(w/(1-w)\). At the endpoints, let the two tremble probabilities vanish at the rates \(n^{-2}\) and \(n^{-1}\). Compactness of \([0,1]\) gives a convergent weight subsequence for every supporting sequence. Thus the possible limits are exactly the stated interval of cutoff equilibria, in pure and behavioral strategies. The conclusions for equal means and uniform distributions follow immediately.

\par\addvspace{\baselineskip}\noindent(c) The proof of Claim~\ref{claim:interim-beliefs}(d) gives both equilibrium sets, including their behavioral characterizations. Its supporting distributions \(\mu_i^n=(1-m_n)\delta_0+m_n\delta_1\) are type independent, with \(m_n=(nm+1)/(n+2)\to m\). Thus the characterization also holds under the type independence restriction, including the extrema at \(m=0\) and \(m=1\).
\end{proof}

\begin{proof}[Proof of Proposition~\ref{prop:failure}]

\noindent(i) Consider a game with singleton type spaces. Player 1 chooses \((x,y)\in\{0,1\}\times\{0,1,2\}\), while players 2 and 3 each choose from \(\{-1,0,1\}\). The payoffs are
\(u_1((x,y),b,c)=x(2c_+-b_-)+yb_+-y^2c_-, u_2=-b^2, u_3=-c^2\), 
where \(z_+=\max\{z,0\}\) and \(z_-=\max\{-z,0\}\) for real \(z\). The economic interpretation is given before Proposition~\ref{prop:failure}. The action spaces are finite lattices, and player 1's payoff is separable, hence modular, in own action. Her gains from increasing \(x\), from increasing \(y\) from zero to one, and from increasing \(y\) from one to two are \(2c_+-b_-\), \(b_+-c_-\), and \(b_+-3c_-\), respectively. These expressions increase in \((b,c)\), verifying strategic increasing differences. The remaining assumptions follow from singleton type spaces, chain action spaces for players 2 and 3, and their payoffs depending only on own action.

We prescribe the uniform distribution \(\mu_i\) on each action space and consider the perturbations \(q=(\varepsilon,\mu)\) with \(0<\varepsilon_i<1\). Players 2 and 3 uniquely choose zero. After we remove the constant and the positive multiplier that her own tremble contributes by Remark~\ref{rem:perturbed}, player 1 maximizes
\(\tfrac{\varepsilon_3}{3} \bigl[x(2-k)+ky-y^2\bigr], k=\tfrac{\varepsilon_2}{\varepsilon_3}>0\). 
At \(k=1/2\) and \(k=4\), the unique maximizers are \((1,0)\) and \((0,2)\), respectively. Keep either ratio fixed while all tremble probabilities vanish. For instance, take \(\varepsilon_1^n=\varepsilon_3^n=1/(10n)\) and \(\varepsilon_2^n=k/(10n)\). This supports the incomparable profiles \(((1,0),0,0)\) and \(((0,2),0,0)\) in \(\PP_\mu^p\).

The upper corner \((1,2)\) can be optimal only if \(2-k\ge0\) and \(k-3\ge0\), which is impossible. The lower corner \((0,0)\) can be optimal only if \(2-k\le0\) and \(k-1\le0\), which is also impossible. Neither corner is therefore used in a perturbed equilibrium. Each receives probability \(\varepsilon_1/6\) in the induced perturbed behavioral strategy and zero in its limit. A distribution stochastically dominating both \(\delta_{(1,0)}\) and \(\delta_{(0,2)}\) must satisfy \(\mathbb E[x]\ge1\) and \(\mathbb E[y]\ge2\), since the coordinate functions are increasing. It must therefore equal \(\delta_{(1,2)}\). A common stochastic lower bound must similarly equal \(\delta_{(0,0)}\). Both are excluded, so \(\PP_\mu^b\), and therefore also \(\PP_\mu^p\), has neither a largest nor a smallest element.

\par\addvspace{\baselineskip}\noindent(ii) The conclusion follows from Theorem~\ref{thm:main}, including its version with type independent tremble distributions, since the maintained assumptions have been verified.

For the atomless extension, give each player an independent uniform type in \([0,1]\) and retain the payoffs and prescribed uniform tremble distributions. The assumptions remain satisfied. The perturbations with fixed ratios in part~(i) support the same two constant profiles, and the exclusion and stochastic order arguments apply almost everywhere. Theorem~\ref{thm:main} again gives part~(ii).
\end{proof}

\end{document}